\documentclass[aps,prl,reprint,superscriptaddress]{revtex4-2}

\pdfoutput=1
\usepackage{graphicx}
\usepackage{amssymb,amsmath,tabularx}
\usepackage{bm}
\usepackage{wrapfig}
\usepackage[center]{subfigure}
\usepackage{soul}
\usepackage{comment}

\usepackage{xcolor}

\newcommand{\RomanNumeralCaps}[1]{\MakeUppercase{\romannumeral #1}}

\allowdisplaybreaks

\begin{document}

\title{Noise-driven odd elastic waves in living chiral active matter}
\author{Sang Hyun Choi}
\affiliation{Department of Physics, University of Illinois at
Urbana-Champaign, Loomis Laboratory of Physics, 1110 West Green
Street, Urbana, Illinois 61801-3080, USA}
\affiliation{Center for Living Systems and Department of Physics, University of Chicago, Chicago, Illinois 60637, USA}
\author{Zhi-Feng Huang}
\affiliation{Department of Physics and Astronomy, Wayne State University, Detroit, Michigan 48201, USA}
\author{Nigel Goldenfeld}
\affiliation{Department of Physics, University of Illinois at
Urbana-Champaign, Loomis Laboratory of Physics, 1110 West Green
Street, Urbana, Illinois 61801-3080, USA}
\affiliation{Department of Physics, University of California, San Diego, 9500 Gilman Drive,
La Jolla, California 92093, USA}

\begin{abstract}
Chiral active matter is predicted to exhibit odd elasticity, with nontraditional elastic response arising from a combination of chirality, being out of equilibrium, and the presence of nonreciprocal interactions. One of the resulting phenomena is the possible occurrence of odd elastic waves in overdamped systems, although its experimental realization still remains elusive. Here we show that in overdamped active systems, noise is required to generate persistent elastic waves. In the chiral crystalline phase of active matter, such as that found recently in populations of swimming starfish embryos, the noise arises from the self-driving of active particles and their mutual collisions, a key factor that has been missing in previous studies. We identify the criterion for the occurrence of noise-driven odd elastic waves and construct the corresponding phase diagram, which is also applicable to general chiral active crystals. Our results can be used to predict the experimental conditions for achieving a transition to self-sustained elastic waves in overdamped active systems.
\end{abstract}
\maketitle

Living systems self-organize in novel ways because they are
necessarily open and process energy input that is eventually dissipated to the environment. When forming ordered structures, their response is not subject to the fluctuation-dissipation theorems for equilibrium passive matter, leading to the possibility of unconventional elastic response, where novel linear response coefficients emerge that break the symmetries required in equilibrium materials, generating odd elasticity (in solids of active or living matter) \cite{scheibner2020odd,banerjee2021active,braverman2021topological,tan2022odd,bililign2022motile,HuangPNAS25}
or odd viscosity (in fluids)
\cite{banerjee2017odd,souslov2019topological,banerjee2021active,han2021fluctuating}.
The resultant dynamics can also reflect nonreciprocal interactions (see e.g., Refs.~\cite{ivlev2015statistical,fruchart2021non} for recent accounts), long understood to be important in fields as diverse as animal behavior (see e.g., Ref.~\cite{couzin2005effective}) and the efficient amplification of ring lasers
\cite{geusic1962unidirectional,koester1964amplification,post1967sagnac,menegozzi1973theory}.
In the latter example, nonreciprocity leads to the existence of self-sustained oscillations that accompany the dissipative nonequilibrium steady state of the laser, itself
an example of a parity-time (PT) symmetric state characteristic of non-Hermitian systems with balanced gain and loss
\cite{bender1998real,bender1999pt,peng2014loss}. Such physical systems can enter the PT-symmetry breaking state through a transition involving exceptional points
\cite{kato1966perturbation,dembowski2001experimental,heiss2012physics,fruchart2021non}, and it is expected that the results are generalizable to odd elastic and viscous materials. An interesting prediction is the possible emergence of odd elastic waves in active crystals with odd elasticity \cite{scheibner2020odd}, but the corresponding experimental realization needs to be carefully examined.

The purpose of this Letter is to predict the conditions for achieving persistent collective odd elastic excitations in overdamped active chiral systems. To accomplish this, we present a framework to describe such systems based on linear response theory \cite{kadanoff1963hydrodynamic,han2021fluctuating,banerjee2022hydrodynamic}, exposing new insights into the dynamics through detailed spectral analysis and modeling. In a deterministic system, damping caused by longitudinal interparticle interactions leads to decay of odd elastic waves \cite{scheibner2020odd}, the experimental observation of which would thus be challenging. Here, we propose a new mechanism for the emergence of persistent collective odd elastic waves in overdamped active crystals. It arises from the interplay between self-propulsion and noise, which are important ingredients of active and living systems that have been neglected in the existing models of odd elasticity \cite{scheibner2020odd,tan2022odd,HuangPNAS25}.

We focus on systems where nonreciprocal interactions and chirality are closely connected \cite{fruchart2021non,han2021fluctuating}, with a remarkable biological realization of chiral active materials reported in the behavior of self-organizing bacteria \cite{petroff2015fast} and starfish embryos \cite{tan2022odd} that form rotating 2D living odd crystals. In these examples, nonreciprocity arises from transverse hydrodynamic interactions between spinning bacterial cells or embryos, which has been elucidated in great detail \cite{petroff2015fast,tan2022odd}. Here we show that noise caused by collisions between agents as a result of self-propulsion can lead to persistent noise-driven odd elastic waves. We estimate the criterion for the existence of such waves and present the phase diagram for chiral active matter as a function of degrees of nonreciprocity and noise strength. Our spectral analysis indicates that small oscillations of agents about their mean positions in the experiment of living crystals of starfish embryos \cite{tan2022odd} correspond to a self-circling mode resulting from self-propulsion of individuals, instead of an elastic wave. Our modeling predicts the conditions under which particle self-driving with intrinsic noise can maintain the self-sustained odd elastic wave that is experimentally realizable in an overdamped environment, a key factor that is missing in the current odd elastodynamics study.

\textit{Spectral method using current correlation functions.---} Our analyses of wave behavior are based on correlation functions which provide important information of materials such as their structures and transport properties \cite{kadanoff1963hydrodynamic}. By extracting the value of frequency $\omega$ that maximizes the correlation function for each wave vector $\mathbf{q}$ in the Fourier space, one can construct the dispersion relation $\omega(\mathbf{q})$. Since the dispersion relation incorporates all possible modes, it can be used to diagnose the wave behavior. The dynamic structure factor, or power spectrum, obtained from the density correlation function gives only the longitudinal mode, while both longitudinal and transverse modes can be identified from the current correlation function which yields the full dispersion relations \cite{boon1980molecular} and is thus used here to identify the elastic wave behavior.

The current correlation function is a tensor with elements $C_{\alpha\beta}(\mathbf{q},\omega)=\frac{1}{N}\langle J_{\alpha}^{*}(\mathbf{q},\omega)J_{\beta}(\mathbf{q},\omega) \rangle$, where $J_{\alpha}(\mathbf{q},\omega)$ is the $\alpha$-component of the Fourier transform of the current density vector $\mathbf{J}(\mathbf{r},t) = \sum_{i}^{N}\mathbf{v}_{i}(t)\delta (\mathbf{r}-\mathbf{r}_{i}(t))$, with $\mathbf{v}_{i}(t)$ and $\mathbf{r}_{i}(t)$ the velocity and position of the $i^{th}$ particle respectively. For both longitudinal (subscript $L$) and transverse (subscript $T$) directions, we calculate not only the diagonal elements of the current correlation $C_{LL}(\mathbf{q},\omega)$ and $C_{TT}(\mathbf{q},\omega)$, but also the real and imaginary parts of the off-diagonal cross correlation $C_{LT}(\mathbf{q},\omega)$ (noting that $C_{LT}^*(\mathbf{q},\omega) = C_{TL}(\mathbf{q},\omega)$). 

Analytical calculation of the current correlation function and the corresponding dispersion relation is non-trivial. In fact, there is no simple analytical solution even for a one-dimensional deterministic simple harmonic oscillator \cite{wierling2010wave,radons1983dynamical} (see Supplemental Material (SM) $\S$\RomanNumeralCaps{1} \cite{SM} for details). 
Inspired by the observation that the spectrum of the velocity autocorrelation function \cite{kadanoff1963hydrodynamic,han2021fluctuating,banerjee2022hydrodynamic} corresponds to the small wave number limit of the longitudinal current correlation related to single particle density \cite{boon1980molecular}, we also calculate the velocity correlation function analytically as a rough approximation for the current correlation function and use it to estimate the condition for the onset of persistent elastic waves.

\textit{Noise-driven odd elastic waves.---}
The criterion for the existence of deterministic odd elastic waves has been predicted to be \cite{scheibner2020odd} 
\begin{equation}
B^{2}/4-K^{o}A-(K^{o})^{2}<0,
\label{eq:deterministic_condition}
\end{equation}
for which the system undergoes an exceptional point transition and exhibits PT symmetry breaking \cite{fruchart2021non}. Here $B$ is the bulk modulus and $K^{o}$ and $A$ are the elastic moduli related to odd elasticity \cite{scheibner2020odd}. However, the linear stability analysis of the overdamped elastodynamic equation for the displacement field shows that an odd wave is always damped and the damping arises from the longitudinal interaction between agents (SM $\S$\RomanNumeralCaps{2} and $\S$\RomanNumeralCaps{3} \cite{SM}).  A deterministic persistent wave could only be possible in the case of marginal instability with zero longitudinal force; but this corresponds to zero bulk and shear moduli, and is thus not physically realistic.

To identify the criterion for persistent odd elastic waves, we note that noise can induce pattern formation \cite{butler2009robust,butler2011fluctuation,BiancalaniPRL17} and traveling waves in e.g., population dynamics \cite{biancalani2011stochastic}. 
Fluctuation-driven patterns or waves have been shown to emerge in regimes of  parameter space where their deterministic counterparts cannot occur \cite{butler2009robust,butler2011fluctuation,BiancalaniPRL17,biancalani2011stochastic}. These results suggest the possibility of exciting noise-induced odd elastic waves in overdamped chiral active systems, with an appropriate noise-driven wave criterion that is different from the above deterministic one. 
 
\begin{figure}[!htb]
   \includegraphics[width=\columnwidth]{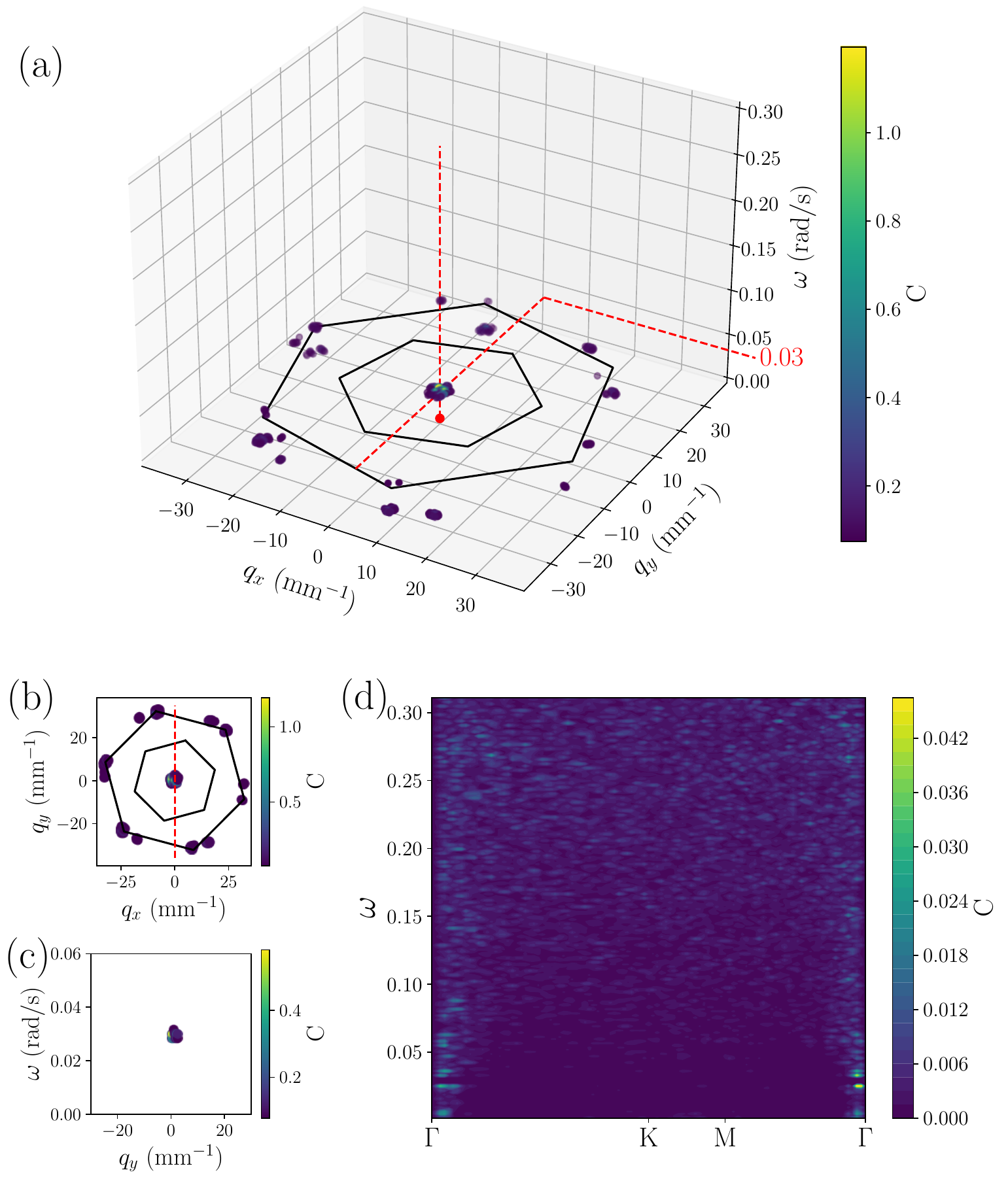}
   \caption{The dispersion result obtained from the current correlation function $C$ for the starfish embryo experimental data of Ref.~\cite{tan2022odd}. (a) Data points with $C$ values exceeding a threshold of 0.075 are shown (to filter out noise), with the full plot given in SM $\S$\RomanNumeralCaps{12} \cite{SM}. The outer and inner hexagons represent the reciprocal lattice and the first Brillouin zone respectively. Red dashed lines are added to indicate the location of the origin (red dot) and the frequency value $\omega = 0.03$ rad/s. Two signals per vertex are detected because the crystal changed its configuration during the experiment and thus slightly rotated in the co-rotating frame \cite{tan2022odd}. (b) The top view of (a), where the coloring represents the maximum $C$ value at each $\mathbf{q}$. (c) Side view of (a) at $q_{x}=0$ (also corresponding to the red dashed line in (b)). (d) The dispersion result in the first Brillouin zone after removing the self-circling signal.}
  \label{fig:expt_dispersion}
\end{figure}

This idea can be verified by adding noise to a 2D toy model of a Hookean spring system with both longitudinal and transverse forces, and results are shown in the End Matter. Here we present how this noise-driven mechanism can be incorporated into a more realistic model describing the living crystal of starfish embryos \cite{tan2022odd}. The same methodology could be generalized to a broader range of chiral active matter \cite{petroff2015fast,hokmabad2022spontaneously,bililign2022motile} including other living crystalline systems.
Specifically, our stochastic odd elastic model is given by
\begin{align}
\begin{split}
\frac{d\mathbf{r}_{i}}{dt}&=\sum_{i\neq j}\biggl[ \overline{\mathbf{v}}_{\text{st}}(\mathbf{r}_{i},\mathbf{r}_{j}) + \frac{1}{\eta R}\mathbf{F}_{\text{rep}}(\lvert \mathbf{r}_{i}-\mathbf{r}_{j} \rvert) \\
&+ R(\omega_{i}+\omega_{j})F_{\text{nf}}(\lvert \mathbf{r}_{i}-\mathbf{r}_{j} \rvert)\hat{\mathbf{r}}_{ij}^{\perp} \biggr] + v_{0}(t)\mathbf{p}_{i},
\end{split} \label{eq:starfish_eom}
\\
\frac{d\theta_{i}}{dt} &= \omega_{i}, \label{eq:theta}
\end{align}
where $R$ is the radius of embryo, $\eta$ is the fluid viscosity, and $\omega_{i}$ represents the self-circling angular frequency of the $i^{th}$ embryo which self-propels with strength $v_0$. $\omega_{i}$ is assumed to be of the same value as the individual embryo self-spinning frequency, based on the experimental observation \cite{tan2022odd} (and also our analysis of the experimental data shown later in Fig.~\ref{fig:expt_dispersion}).
The force terms in Eq.~(\ref{eq:starfish_eom}) are the same as those used in Ref.~\cite{tan2022odd}, including the longitudinal attraction $\overline{\mathbf{v}}_{\text{st}}$ originating from the Stokeslet flow, the longitudinal steric repulsive force $\mathbf{F}_{\text{rep}}$, and the near-field transverse force $F_{\text{nf}}$ between two spinning embryos as obtained from the lubrication theory \cite{o1970asymmetrical,kim2013microhydrodynamics}. Explicit expressions for each term and the parameter values used are given in SM $\S$\RomanNumeralCaps{4} \cite{SM}. 
In contrast to previous modeling of chiral living crystals \cite{petroff2015fast,tan2022odd}, for numerical convenience we use a time-independent constant $\omega_{i} = \bar{\omega}_{0} = 1$ for all particles because time variation of the spinning frequency was found to be negligible here (SM $\S$\RomanNumeralCaps{5} \cite{SM}).

The key factor introduced in this model is
the self-propulsion of embryos, $v_{0}\mathbf{p}_{i}$, where the polarization $\mathbf{p}_{i} = (\cos\theta_{i}, \sin\theta_{i})$ with orientation angle $\theta_{i}$. The contribution from self-propulsion, which is absent in previous studies of odd crystals \cite{scheibner2020odd,tan2022odd}, leads to mutual collision between neighboring embryos, and serves as the main source of noise. Noise is thus incorporated in the self-propulsion strength $v_{0}$ via $v_{0}(t) = \bar{v}_{0} + \xi_{v_{0}}(t)$, with $\bar{v}_{0}$ a constant average and $\xi_{v_{0}}$ a Gaussian white noise determined by $\langle \xi_{v_{0}}(t)\xi_{v_{0}}(t') \rangle =v_{\sigma}^{2}\delta(t-t')$. The overall chirality is retained by keeping $\bar{v}_{0}>0$, but $v_{0}(t)$ of an embryo at a given instant is allowed to be negative (due to noise), leading to local deformation of the living crystal. Physically, the reverse sign stems from the sudden change of direction of the self-propelling motion due to collision between embryos. 
The simulation is performed using the Euler-Maruyama algorithm \cite{maruyama1953markov,maruyama1955continuous} on a $30 \times 30$ 2D triangular lattice with periodic boundary conditions for $10^{4}$ time steps with interval $dt=0.01$ over 1000 realizations.

\begin{figure}[t]
   \includegraphics[width=\columnwidth]{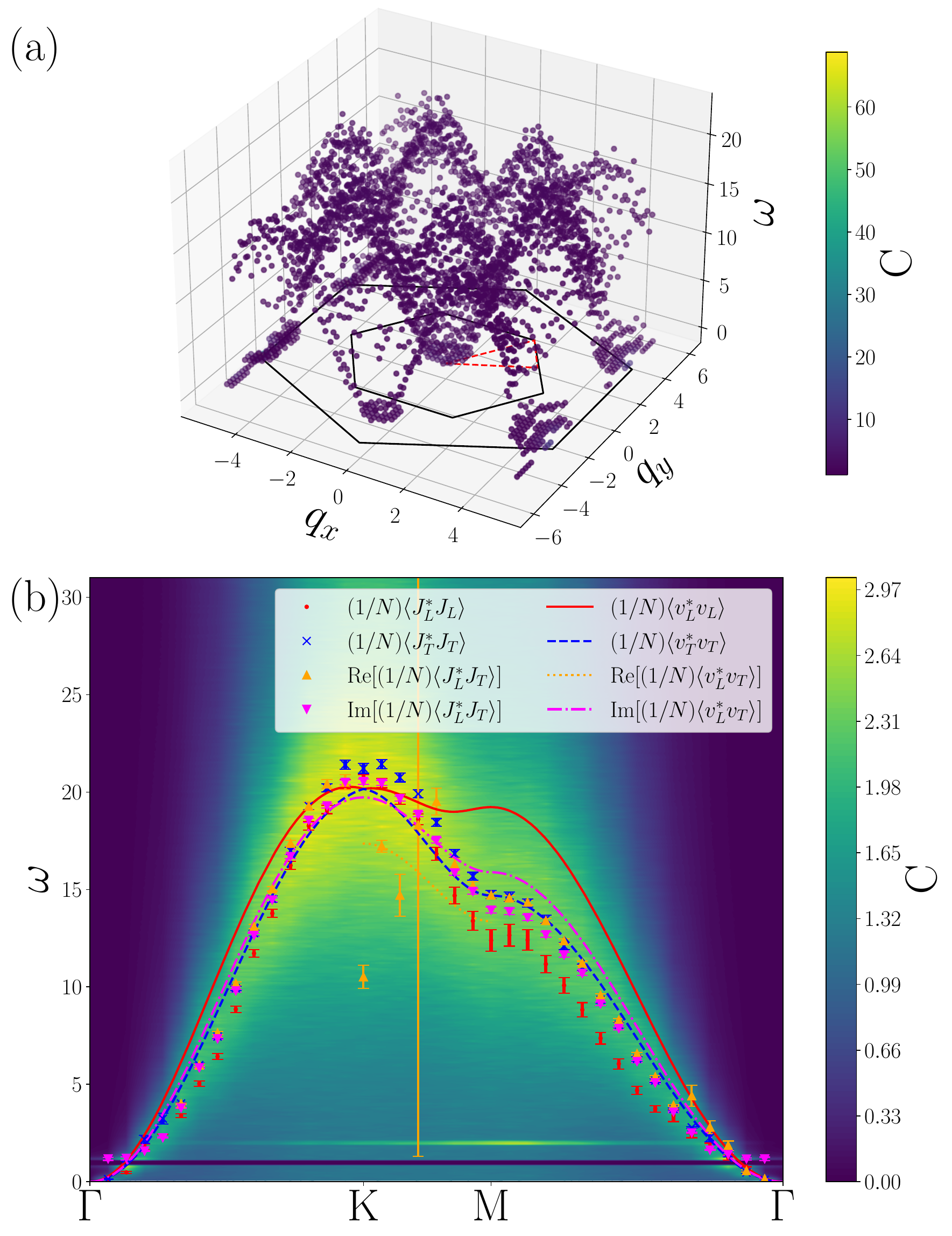}
   \caption{(a) The dispersion results obtained from the current correlation function $C = C_{LL}+C_{TT}$ for the starfish embryo model in the presence of noise. The solid outer hexagon on the $(q_{x}, q_{y}$) plane at $\omega=0$ represents the reciprocal lattice, and the inner hexagon is the first Brillouin zone. (b) The dispersion relations along the red dashed line in the first Brillouin zone. Symbols correspond to the simulation results, while results obtained from the analytically calculated velocity correlation function are shown as solid or dashed curves. Large data variations around the $K$ point are due to the large uncertainties of data fitting for the near-zero noisy values of Re[$(1/N)\langle J_{L}^{*}J_{T} \rangle$] (see SM $\S$\RomanNumeralCaps{7} \cite{SM}), which also leads to a big error bar showing as the vertical line in the middle of (b).
  \label{fig:bandcomp}}
\end{figure}

Figure~\ref{fig:bandcomp}(a) shows the results of dispersion relations for $C = C_{LL}(\mathbf{q},\omega)+C_{TT}(\mathbf{q},\omega) = \frac{1}{N}\langle \mathbf{J}^{*}(\mathbf{q},\omega)\cdot\mathbf{J}(\mathbf{q},\omega) \rangle$, and Fig.~\ref{fig:bandcomp}(b) shows the results for all elements of the current correlation function along a path on the first Brillouin zone, indicating that even in the presence of damping caused by longitudinal force, the wave property still survives as a result of noise excitation. In the figure the self-circling signal at $\omega = \bar{\omega}_{0} = 1$ [indicated by the dark horizontal line in Fig.~\ref{fig:bandcomp}(b)] has been removed to make the dispersion curves visible, since the value of the current correlation corresponding to self-circling motion is much larger than that of the wave behavior even when we set $\bar{v}_{0}$ to be as small as 0.01 (with $v_{\sigma} = 0.1$). The self-circling signal is expected to be detected at integer multiples of the self-circling frequency, i.e., at $\omega = n\bar{\omega}_{0}$, with decreasing magnitude as $n$ increases (see SM $\S$\RomanNumeralCaps{1} \cite{SM} for the derivation to the lowest order of $qv_{0}/\bar{\omega}_{0}$, which gives the analytic results for $n \leq 2$). The $n=2$ signal is shown as the bright horizontal line in Fig.~\ref{fig:bandcomp}(b), while those for $n>2$ are too weak to be visible. In Fig.~\ref{fig:bandcomp}(a) the signals at $\omega=\bar{\omega}_{0}$ and $2\bar{\omega}_{0}$ have also been removed to make the dispersion visible.

We have analytically calculated the dispersion relations from the velocity correlation functions, with results shown in Fig.~\ref{fig:bandcomp}(b) as solid or dashed lines. They are somewhat different from the current correlation functions but capture the overall behavior of the dispersion relations that are consistent with the simulation results. Note that the dispersion curve from the real part of the cross velocity correlation is shown only from $K$ to $M$ point of the Brillouin zone. It is zero from $\Gamma$ to $K$ point and from $M$ to $\Gamma$ point because of the underlying geometry of the lattice (see SM $\S$\RomanNumeralCaps{9} \cite{SM} for derivation).

\textit{The criterion for noise-driven odd elastic waves.---}
An elastic wave induced by noise is expected to emerge under a different condition from the deterministic one \cite{butler2009robust,butler2011fluctuation,BiancalaniPRL17,biancalani2011stochastic}. The wave behavior can be identified from the dispersion relation, when there exist real values of $\omega(\mathbf{q})$ maximizing the current correlation function, as demonstrated in Fig.~\ref{fig:bandcomp}. Since the analytical expression for the current correlation function is not available, we use the velocity correlation function instead to predict the approximate condition for the occurrence of noise-driven elastic waves.
For simplicity, here we use the correlation element $\langle \mathbf{v}^{*}(\mathbf{q},\omega)\cdot\mathbf{v}(\mathbf{q},\omega) \rangle$, which is calculated analytically as
\begin{equation}
\begin{split}
\langle &\mathbf{v}^{*}(\mathbf{q},\omega)\cdot\mathbf{v}(\mathbf{q},\omega) \rangle = \\
&\omega^{2}\frac{(\omega^{2}+M_{21}^{2}+M_{22}^{2})\langle \xi_{1}^{*}\xi_{1} \rangle + (\omega^{2}+M_{11}^{2}+M_{12}^{2})\langle \xi_{2}^{*}\xi_{2} \rangle}{\lvert \det{(-i\omega\mathbb{I}-M)} \rvert^{2}},
\end{split}
\label{eq:generalvv}
\end{equation}
where $M(\mathbf{q})$ is the dynamic matrix and $\xi_i(\mathbf{q},\omega)$ are noise components satisfying $\langle \xi_{1}^{*}\xi_{2} \rangle = \langle \xi_{2}^{*}\xi_{1} \rangle = 0$. Details of the derivation and the results for all other elements of velocity correlation are given in SM $\S$\RomanNumeralCaps{10} \cite{SM}. 
Here we consider the case of a Gaussian white noise and work in the continuum limit such that $M_{11} = -{q^{2}}(B+\mu)/{\gamma}$, $M_{12} = -{q^{2}}K^{\rm o}/{\gamma}$, $M_{21} = {q^{2}}(K^{\rm o}+A)/{\gamma}$, and $M_{22} = -{q^{2}}\mu/{\gamma}$, with $\mu$ the shear modulus and $\gamma$ the friction coefficient. We can then identify the approximate criterion for the existence of real $\omega(\mathbf{q})$ that maximizes Eq.~(\ref{eq:generalvv}), i.e.,
\begin{equation}
(K^{\rm o})^{2}+(K^{\rm o}+A)^{2}+4K^{\rm o}(K^{\rm o}+A)-\mu^{2}-(B+\mu)^{2}>0.
\label{eq:wavecondi}
\end{equation} 
Using the relation between the longitudinal and transverse effective spring constants $k_{L}$ and $k_{T}$ and the elastic moduli for a triangular lattice \cite{braverman2021topological}, we can rewrite the condition of Eq.~(\ref{eq:wavecondi}) as $\alpha \equiv k_{T}/k_{L} > \sqrt{5/11}$ after substituting the specific parameter values chosen in our study, where the ratio $\alpha$ is used to represent the degree of nonreciprocity induced by transverse interaction. 

\begin{figure}[t]
   \includegraphics[width=\columnwidth]{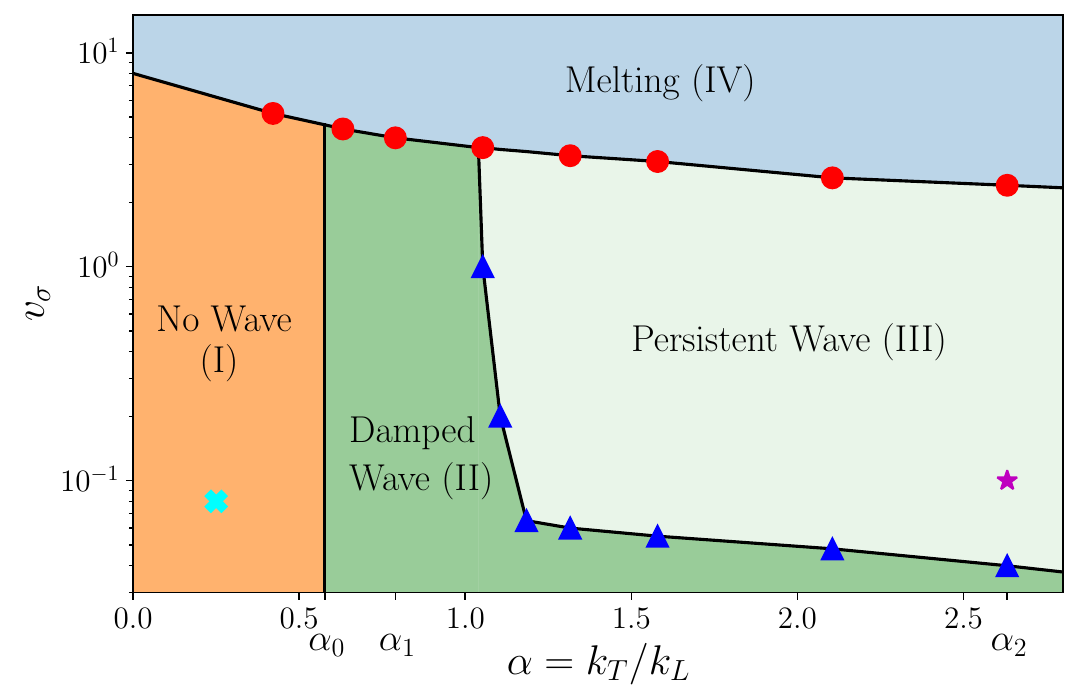}
   \caption{Phase diagram of overdamped odd elastic systems as a function of noise strength $v_{\sigma}$ and the degree of nonreciprocity that is represented by the ratio $\alpha = k_{T}/k_{L}$. Induced by the noise, a new state of persistent noise-driven elastic waves appears at large enough $\alpha$ (regime \RomanNumeralCaps{3}). 
   The purple star point denotes where the simulation for Fig.~\ref{fig:bandcomp} is conducted, and the cyan cross point indicates the estimated location of the experimental starfish embryo system \cite{tan2022odd} [with $\alpha$ calculated from experimental parameters and $v_{\sigma}$ deduced from the experimental data (SM $\S$\RomanNumeralCaps{16} \cite{SM})]. Symbols of red circles and blue triangles represent results at the phase boundaries, as determined by simulations of the stochastic odd elastic model. 
   The boundary at $\alpha_{0}=\sqrt{1/3}$ has been predicted by the deterministic theory \cite{scheibner2020odd}.
   Sample cross sections of the phase diagram at $\alpha_{1}=0.79$ and $\alpha_{2}=2.63$ are provided in SM $\S$\RomanNumeralCaps{11} \cite{SM}, showing transitions between different phases. 
  \label{fig:phase}}
\end{figure}

Based on calculations from the stochastic odd elastic model Eqs.~(\ref{eq:starfish_eom}) and (\ref{eq:theta}),
we present in Fig.~\ref{fig:phase} a phase diagram for odd elastic systems, which captures the overall behavior in different parameter regimes and is applicable to general odd crystalline systems. The phase boundaries are determined via the dynamic Lindemann parameter \cite{lindemann1910ueber,bedanov1985modified,zahn1999two,zahn2000dynamic} and the current correlation at the M point of first Brillouin zone (see SM $\S$\RomanNumeralCaps{11} \cite{SM}). In the absence of noise or at low noise strength, there are two regimes, the no-wave regime \RomanNumeralCaps{1} and the damped wave regime \RomanNumeralCaps{2}, separated by a threshold at $\alpha_{0} = \sqrt{1/3}$ determined by odd elasticity \cite{scheibner2020odd}. At high enough noise strength, a new phase of persistent elastic waves appears (regime \RomanNumeralCaps{3}). However, at a given value of $\alpha$, if the noise is too strong, the crystal becomes unstable and melts (regime \RomanNumeralCaps{4}). Although not depicted in Fig.~\ref{fig:phase}, the phase boundaries between \RomanNumeralCaps{2} and \RomanNumeralCaps{3} and between \RomanNumeralCaps{3} and \RomanNumeralCaps{4} are expected to approach $v_\sigma=0$ as $\alpha \rightarrow \infty$, because the system becomes unstable regardless of noise strength in the absence of damping due to the longitudinal force (i.e., when $k_L=0$).
The appearance of \RomanNumeralCaps{3} is analogous to previous works in population dynamics \cite{butler2009robust,butler2011fluctuation} and reaction-diffusion systems \cite{biancalani2011stochastic,BiancalaniPRL17}, which show how noise can generate and stabilize an ordered phase that occupies a finite region of phase diagram; such phases can include the occurrence of persistent waves induced by stochasticity \cite{biancalani2011stochastic}.

\textit{Oscillatory behavior in the experimental data.---}
We apply the above analysis and theory to the experimental results of starfish embryo living crystals which showed an oscillatory behavior, as identified from the periodically oscillating displacement fields and mode chirality analysis \cite{tan2022odd}. To clarify the nature of this observed oscillatory behavior (i.e., whether being caused by elastic wave), we first conduct a spectral analysis of the corresponding experimental data based on current correlation functions. Results of dispersion $\omega(\mathbf{q})$ obtained from $C = C_{LL}(\mathbf{q},\omega) + C_{TT}(\mathbf{q},\omega)$ are given in Fig.~\ref{fig:expt_dispersion}, and the same conclusion can be drawn from those of each element of the current correlation function (SM $\S$\RomanNumeralCaps{12} \cite{SM}). 
Figure \ref{fig:expt_dispersion} shows local correlation maxima at $\mathbf{q}=\mathbf{0}$ and the vertices of the reciprocal lattice at $\omega \approx 0.03 \text{ rad/s}$. This frequency can be rewritten as $f = {\omega}/{2\pi} \approx 0.29 \text{ min}^{-1}$, very similar to the self-spinning frequency of the embryos inside the living crystal, i.e., $(0.33\pm 0.13) \text{ min}^{-1}$ as measured experimentally \cite{tan2022odd}. Having signals only around $\mathbf{q}=\mathbf{0}$ and the reciprocal lattice vertices indicates a coordinated self-circling of each embryo around its center of mass rather than an elastic wave behavior, as a result of self-driven motion of each embryo during its spinning. In SM $\S$\RomanNumeralCaps{1} and $\S$\RomanNumeralCaps{13} \cite{SM}, through both numerical simulations and analytical calculations of non-interacting self-circling particles, we demonstrate that pure self-circling leads to the dispersion result resembling Fig.~\ref{fig:expt_dispersion}.
As shown in Fig.~\ref{fig:expt_dispersion}(d), cutting off the self-circling signal just yields a noisy result, fundamentally different from Fig.~\ref{fig:bandcomp}(b) (see also SM $\S$\RomanNumeralCaps{14} \cite{SM}). This confirms that the data are dominated by the self-circling signal and do not contain any signature of elastic wave behavior. Emulation of the experimental system based on our model also supports this conclusion (SM $\S$\RomanNumeralCaps{15} and $\S$\RomanNumeralCaps{16} \cite{SM}).

It is worth pointing out that whether or not the starfish embryo experimental system satisfies the criterion for odd elastic waves was inconclusive in Ref.~\cite{tan2022odd}. The elastic moduli calculated from microscopic experimental parameters \cite{tan2022odd} do not satisfy both deterministic and noise-driven criteria for odd elastic waves (Eqs.~(\ref{eq:deterministic_condition}) and (\ref{eq:wavecondi})). In addition, experimental parameters give $k_{T}\approx2.0 \text{ s}^{-1}$ and $k_{L}\approx8.1\text{ s}^{-1}$ \cite{tan2022odd}, and thus $\alpha\approx0.25$ (or if using instead the self-spinning frequency inside a cluster, $\alpha\approx0.0024$), which belongs to regime \RomanNumeralCaps{1} in Fig.~\ref{fig:phase}.
On the other hand, the elastic moduli inferred from local strains caused by topological defects in the living crystal with the use of linear elasticity \cite{tan2022odd} satisfy both conditions. (See SM $\S$\RomanNumeralCaps{17} \cite{SM} for details).
Note that linear-elasticity models \cite{scheibner2020odd,braverman2021topological} break down near topological defect cores where nonlinear elasticity effects prevail. Thus the calculation based on microscopic experimental parameters is more convincing.

Our method of explicitly detecting the existence of persistent odd elastic waves via current correlation functions is significant, as we do not need to estimate whether or not a criterion for the occurrence of elastic wave is satisfied, nor do we need to be able to infer indirectly the measured values of elastic moduli. Also, this method can directly distinguish between behaviors of elastic wave and simple circling motion, as demonstrated above in Figs.~\ref{fig:bandcomp} and \ref{fig:expt_dispersion}. Our results demonstrate that the odd elastic modes that are overdamped in deterministic systems can be excited by noise originating from self-propulsion and collisions of active particles, with conditions identified by our stochastic model and the predicted phase diagram. 

\begin{acknowledgments}
We thank Tzer Han Tan and Nikta Fakhri for providing the experimental
data of Ref.~\cite{tan2022odd} for the trajectories of starfish embryos.
We gratefully acknowledge valuable discussions with Tzer Han Tan, Alexander Mietke, J{\"o}rn Dunkel, Vincenzo Vitelli, and Ken Elder.
Z.-F.H. acknowledges support from the National Science Foundation
under Grant No. DMR-2006446.
\end{acknowledgments}

%

\clearpage

\begin{center}
\textbf{\large End Matter}
\end{center}
\vskip -10pt
\begin{figure}[!h]
   \includegraphics[width=\columnwidth]{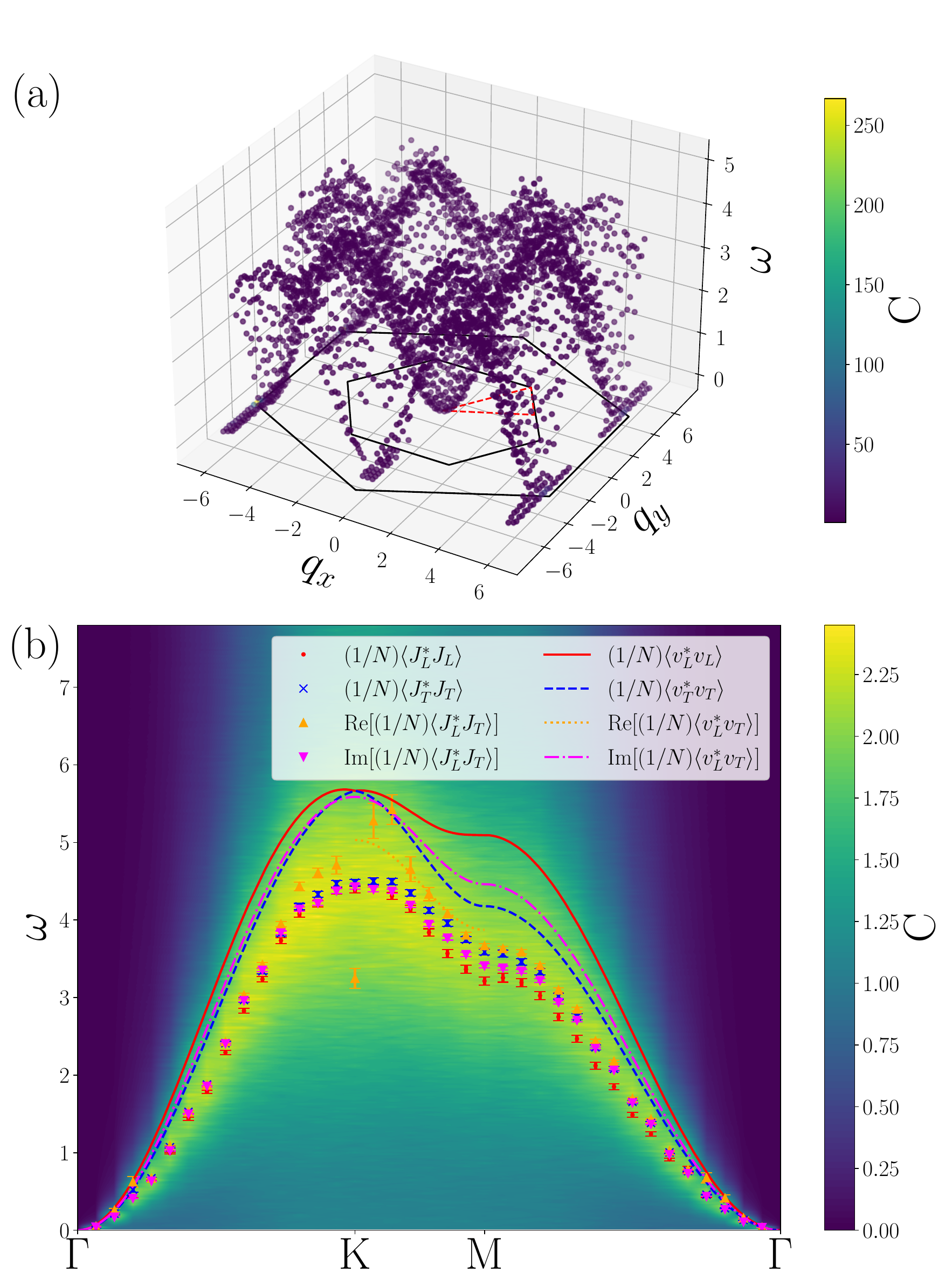}
   \caption{(a) The dispersion results obtained from the current correlation function $C = C_{LL}+C_{TT}$ for the toy model in the presence of noise. The solid outer hexagon on the $(q_{x}, q_{y}$) plane at $\omega=0$ represents the reciprocal lattice, and the inner hexagon is the first Brillouin zone. (b) The dispersion relations along the red dashed line in the first Brillouin zone. Symbols correspond to the simulation results, while results obtained from the analytically calculated velocity correlation function are shown as solid or dashed curves. Large data variations around the $K$ point are due to the large uncertainties of data fitting for the near-zero noisy values of Re[$(1/N)\langle J_{L}^{*}J_{T} \rangle$] (see SM $\S$ \RomanNumeralCaps{7} \cite{SM}).}
  \label{fig:bandcomp_toy}
\end{figure}

In this End Matter, we present a stochastic version of a 2D toy model of a Hookean spring system with both longitudinal and transverse forces. This toy model, in the absence of noise dynamics, has served as a minimal model to demonstrate the exceptional point transition in the non-Hermitian, odd-elastic overdamped system leading to the wave behavior (although damped) \cite{scheibner2020odd}, and is thus extended here. In this stochastic model, each particle $i$ is governed by
\begin{equation}
\frac{d\mathbf{r}_{i}}{dt}=\sum_{j}\bigg[-(k_{L}\hat{\mathbf{r}}_{ij}+k_{T}\hat{\mathbf{r}}^{\perp}_{ij})(r_{ij}-r_{ij}^{0})\bigg] + \bm{\xi}_{i},
\label{eq:sto_toy}
\end{equation}
where $r_{ij}$ is the distance between particles $i$ and $j$ with equilibrium spacing $r_{ij}^{0}$, $\hat{\mathbf{r}}_{ij}$ is the unit vector in the direction of $\mathbf{r}_{ij} \equiv \mathbf{r}_{i}-\mathbf{r}_{j}$, and $(\hat{\mathbf{r}}^{\perp}_{ij})_{\alpha} = \epsilon_{\alpha\beta}(\hat{\mathbf{r}}_{ij})_{\beta}$ with $\alpha, \beta = x, y$ and $\epsilon_{\alpha\beta}$ the 2D Levi-Civita symbol. The spring constants in the longitudinal and transverse directions are denoted as $k_{L}$ and $k_{T}$ respectively. We only consider the nearest-neighbor interactions and assume a Gaussian white noise with correlation $\langle \xi_{i}^\alpha(t)\xi_{j}^\beta(t') \rangle = 2D\delta_{ij}\delta_{\alpha\beta}\delta(t-t')$ for simplicity. As done for the starfish embryo model in the main text, the simulation is conducted via the Euler-Maruyama algorithm \cite{maruyama1953markov,maruyama1955continuous} on a $30 \times 30$ 2D triangular lattice with periodic boundary conditions, for $4\times10^{4}$ time steps (of step interval $dt = 0.01$) with 1000 realizations. The parameter values are chosen as $r_{ij}^{0} = 1$, $k_{L}=0.5$, $k_{T}=1$, and $D=10^{-4}$. 

Figure \ref{fig:bandcomp_toy}(a) shows the results of dispersion relations for $C = C_{LL}(\mathbf{q},\omega) + C_{TT}(\mathbf{q},\omega) = \frac{1}{N}\langle \mathbf{J}^{*}(\mathbf{q},\omega)\cdot\mathbf{J}(\mathbf{q},\omega) \rangle$, while Fig.~\ref{fig:bandcomp_toy}(b) shows the results for all elements of the current correlation function along a path in the first Brillouin zone. These dispersion results verify that noise-driven odd elastic waves are observable even in this overdamped spring system. The results presented in Fig.~\ref{fig:bandcomp_toy} are very similar to those in Fig.~\ref{fig:bandcomp}, which demonstrates that our framework is not unique to the starfish embryo system but can be generally applied to other chiral active systems. Note that, while in this toy model the noise is simply added to the equation of motion as shown in Eq.~(\ref{eq:sto_toy}), the noise is incorporated into the newly introduced self-propulsion term in the starfish embryo model (Eq.~(\ref{eq:starfish_eom})). The latter is constructed in this way such that it explains the physical origin of the intrinsic noise in living crystals that is strong enough to drive agents as large as the starfish embryos, i.e., the noise originates from the collision between embryos due to self-propulsion.

\end{document}


\title{SUPPLEMENTAL MATERIAL \\ \vskip 0.5cm Noise-driven odd elastic waves in living chiral active matter}

\author{Sang Hyun Choi}
\affiliation{Department of Physics, University of Illinois at
Urbana-Champaign, Loomis Laboratory of Physics, 1110 West Green Street, Urbana, Illinois 61801-3080, USA}
\affiliation{Center for Living Systems and Department of Physics, University of Chicago, Chicago, Illinois 60637, USA}
\author{Zhi-Feng Huang}
\affiliation{Department of Physics and Astronomy, Wayne State University, Detroit, Michigan 48201, USA}
\author{Nigel Goldenfeld}
\affiliation{Department of Physics, University of Illinois at
Urbana-Champaign, Loomis Laboratory of Physics, 1110 West Green
Street, Urbana, Illinois 61801-3080, USA}
\affiliation{Department of Physics, University of California, San Diego, 9500 Gilman Drive,
La Jolla, California 92093, USA}

\begin{abstract}
This Supplemental Material contains additional numerical analyses, analytical derivations and discussions for odd elastic waves in active chiral crystalline systems.
\end{abstract}

\maketitle
\tableofcontents

\section{Derivation of the dispersion curve for self-circling motion}
\label{sec:flat}

This section demonstrates analytically that the self-circling motion of particles gives rise to dispersion around the $\Gamma$ point at frequency values equal to the integer multiples of the self-circling frequency $\omega_{0}$. 

The current density is written as $\mathbf{J}(\mathbf{r},t)=\sum_{i=1}^{N}\mathbf{v}_{i}(t)\delta (\mathbf{r}-\mathbf{r}_{i}(t))$ where the particles are labeled with subscript $i$ and the time-dependent position and velocity of the $i^{th}$ particle are $\mathbf{r}_{i}(t)$ and $\mathbf{v}_{i}(t)$ respectively. After the spatial Fourier transform, it becomes $\mathbf{J}(\mathbf{q},t)=\sum_{i=1}^{N}\mathbf{v}_{i}(t)e^{i\mathbf{q}\cdot\mathbf{r}_{i}(t)}$.
Then the current correlation function in Fourier space is given by
\begin{equation}
\begin{split}
C_{\alpha\beta}(\mathbf{q},\omega)&=\frac{1}{N}\langle J_{\alpha}^{*}(\mathbf{q},\omega)J_{\beta}(\mathbf{q},\omega) \rangle \\
&=\frac{1}{N}\int dt ~e^{-i\omega t} \langle J_{\alpha}(-\mathbf{q},0)J_{\beta}(\mathbf{q},t) \rangle,
\end{split}
\end{equation}
where the subscript $\alpha, \beta$ represents the component of the current vector. The correlation inside the integral is written as
\begin{equation}
\begin{split}
\langle J_{\alpha}(-\mathbf{q},0)J_{\beta}(\mathbf{q},t) \rangle &= \bigg\langle \sum_{i,j} v_{j}^{\alpha}(0) v_{i}^{\beta}(t) e^{i\mathbf{q}\cdot (\mathbf{r}_{i}(t)-\mathbf{r}_{j}(0))} \bigg\rangle \\
&= \sum_{i,j} e^{i\mathbf{q}\cdot (\mathbf{r}_{i}^{c}-\mathbf{r}_{j}^{c})} \bigg\langle v_{j}^{\alpha}(0) v_{i}^{\beta}(t) e^{i\mathbf{q}\cdot (\mathbf{u}_{i}(t)-\mathbf{u}_{j}(0))}\bigg\rangle.
\end{split}
\label{eq:JJinqt}
\end{equation}
Here $\mathbf{r}_{i}(t) = \mathbf{r}_{i}^{c} + \mathbf{u}_{i}(t)$ where $\mathbf{r}_{i}^{c}$ is the position of the $i^{th}$ particle in the undeformed perfect lattice and $\mathbf{u}_{i}(t)$ is its time-dependent displacement. 

Equation (\ref{eq:JJinqt}) is nontrivial to calculate and does not have a simple analytical result. In fact, the calculation of current correlation functions is nontrivial even for much simpler systems such as one-dimensional simple harmonic oscillators \cite{wierling2010wave, radons1983dynamical}.
Although in principle one can generalize the one-dimensional calculations done in Refs.~\cite{wierling2010wave} and \cite{radons1983dynamical} to two-dimensional cases, our stochastic odd elastic systems add extra complexities. For the calculation of current correlation functions of one-dimensional simple harmonic oscillators, the quadratic form of the Hamiltonian is utilized to obtain the expectation value from the averaging $\langle \cdot\cdot\cdot\rangle$, with the use of normal-mode coordinates as well as the harmonic approximation \cite{wierling2010wave,radons1983dynamical}. However, the existence of transverse force or non-potential force as well as the stochastic noise in our case makes it difficult to directly apply this formalism, as we cannot use the Hamiltonian as done in conventional equilibrium systems.
Therefore, we generally cannot calculate the analytical expression for the current correlation functions. 

However, we can calculate the special case of $N$ non-interacting self-circling particles to obtain the exact solution in a small scale limit.
The velocity of the $i^{th}$ non-interacting self-circling particle is expressed as $\mathbf{v}_{i}(t)=(v_{x}^{i}(t),v_{y}^{i}(t))=v_{0}(\text{cos}(\omega_{0}t+\theta_{i}),\text{sin}(\omega_{0}t+\theta_{i}))$ where we have assumed all the particles have the same constant self-circling frequency $\omega_{0}$, and $\theta_{i}$ is the initial orientation angle. Integrating $\mathbf{v}_{i}(t)$ gives $\mathbf{r}_{i}(t) = \mathbf{r}_{i}^{c} + l_{0}(\text{sin}(\omega_{0}t+\theta_{i}),-\text{cos}(\omega_{0}t+\theta_{i}))$ where $\mathbf{r}_{i}^{c}$ is the vector coordinate of the $i^{th}$ particle in the undeformed perfect lattice and $l_{0} = {v_{0}}/{\omega_{0}}$. The current density vector is written as $\mathbf{J}(\mathbf{q},\omega)=(J_{x}(\mathbf{q},\omega),J_{y}(\mathbf{q},\omega))$ where $J_{x}(\mathbf{q},\omega)=\sum_{i} J_{x}^{i}(\mathbf{q},\omega)$ is the Fourier transform of the $x$-component of the current density $J_{x}(\mathbf{r},t)$. For a self-circling particle, $J_{x}^{i}(\mathbf{q},\omega)$ is given by
\begin{equation}
\begin{split}
J_{x}^{i}(\mathbf{q},\omega)&=\int dt e^{-i\omega t} v_{x}^{i}(t)e^{i\mathbf{q}\cdot\mathbf{r}_{i}(t)} \\
&=\int dt e^{-i\omega t}v_{0}\cos{(\omega_{0}t+\theta_{i})}e^{i\mathbf{q}\cdot\mathbf{r}_{i}^{c}}e^{il_{0}(q_{x}\sin{(\omega_{0}t+\theta_{i})}-q_{y}\cos{(\omega_{0}t+\theta_{i})})} \\
&\simeq v_{0}e^{i\mathbf{q}\cdot\mathbf{r}_{i}^{c}}\int dt e^{-i\omega t}\cos{(\omega_{0}t+\theta_{i})}\bigg\{ 1+il_{0}q_{x}\sin{(\omega_{0}t+\theta_{i})} -il_{0}q_{y}\cos{(\omega_{0}t+\theta_{i})} \bigg\} \\
&= v_{0}e^{i\mathbf{q}\cdot\mathbf{r}_{i}^{c}}\int dt e^{-i\omega t} \bigg\{ \cos{(\omega_{0}t+\theta_{i})} +i\frac{l_{0}q_{x}}{2}\sin{(2\omega_{0}t+2\theta_{i})}-i\frac{l_{0}q_{y}}{2}\big(1+\cos{(2\omega_{0}t+2\theta_{i})}\big) \bigg\},
\end{split}
\end{equation}
in the limit of small $l_{0}q_{x}$ and $l_{0}q_{y}$. The justification for small $l_{0}= {v_{0}}/{\omega_{0}}$ is that the embryos cannot move much inside a crowded cluster. As we are only interested in finite values of $q_{x}$ and $q_{y}$ inside the first Brillouin zone, our assumption gives small $l_{0}q_{x}$ and $l_{0}q_{y}$. Here the expansion is done only up to the first order of $l_0 q$, although in principle higher order terms proportional to $\cos(n\omega_{0}t+n\theta_{i})$ with integer $n>2$ can be obtained through higher-order expansions.

Calculating the above integral gives
\begin{equation}
\begin{split}
J_{x}^{i}(\mathbf{q},\omega) \simeq&
\pi v_{0}e^{i\mathbf{q}\cdot\mathbf{r}_{i}^{c}}\bigg\{ e^{i\theta_{i}}\delta(\omega-\omega_{0}) + e^{-i\theta_{i}}\delta(\omega+\omega_{0}) -il_{0}q_{y}\delta(\omega) + \frac{l_{0}}{2}e^{2i\theta_{i}}(q_{x}-iq_{y})\delta(\omega-2\omega_{0}) \\
&- \frac{l_{0}}{2}e^{-2i\theta_{i}}(q_{x}+iq_{y})\delta(\omega+2\omega_{0}) \bigg\}.
\end{split}
\end{equation}
Then we obtain
\begin{equation}
\begin{split}
J_{x}^{i}(\mathbf{q},\omega)J_{x}^{j}(\mathbf{q},\omega)^{*} \simeq& 
\pi^{2}v_{0}^{2}e^{i\mathbf{q}\cdot(\mathbf{r}_{i}^{c}-\mathbf{r}_{j}^{c})}
\bigg\{ e^{i(\theta_{i}-\theta_{j})}\delta^{2}(\omega-\omega_{0}) + e^{-i(\theta_{i}-\theta_{j})}\delta^{2}(\omega+\omega_{0}) + l_{0}^{2}q_{y}^{2}\delta^{2}(\omega) \\
&+ \frac{l_{0}^{2}}{4}e^{2i(\theta_{i}-\theta_{j})}q^{2}\delta^{2}(\omega-2\omega_{0}) + \frac{l_{0}^{2}}{4}e^{-2i(\theta_{i}-\theta_{j})}q^{2}\delta^{2}(\omega+2\omega_{0}) \bigg\},
\end{split}
\end{equation}
where $q^{2}\equiv q_{x}^{2}+q_{y}^{2}$ and we have neglected cross-terms of Dirac delta functions such as $\delta(\omega-\omega_{0})\delta(\omega+\omega_{0})$ because they are equal to zero. 
For the whole population, we then get
\begin{equation}
\begin{split}
J_{x}(\mathbf{q},\omega)J_{x}(\mathbf{q},\omega)^{*} =& \sum_{i,j}J_{x}^{i}(\mathbf{q},\omega)J_{x}^{j}(\mathbf{q},\omega)^{*} \\
\simeq& \pi^{2}v_{0}^{2}\sum_{i}\bigg[ \delta^{2}(\omega-\omega_{0}) + \delta^{2}(\omega+\omega_{0}) + \frac{l_{0}^{2}}{4}q^{2}\delta^{2}(\omega-2\omega_{0}) + \frac{l_{0}^{2}}{4}q^{2}\delta^{2}(\omega+2\omega_{0}) + l_{0}^{2}q_{y}^{2}\delta^{2}(\omega) \bigg] \\
&+ 2\pi^{2}v_{0}^{2}\sum_{i<j}\bigg[ \cos{(\mathbf{q}\cdot(\mathbf{r}_{i}^{c}-\mathbf{r}_{j}^{c})+\theta_{i}-\theta_{j})}\delta^{2}(\omega-\omega_{0}) + \cos{(\mathbf{q}\cdot(\mathbf{r}_{i}^{c}-\mathbf{r}_{j}^{c})-\theta_{i}+\theta_{j})}\delta^{2}(\omega+\omega_{0}) \\
&+ \frac{l_{0}^{2}}{4}q^{2}\cos{(\mathbf{q}\cdot(\mathbf{r}_{i}^{c}-\mathbf{r}_{j}^{c})+2\theta_{i}-2\theta_{j})}\delta^{2}(\omega-2\omega_{0}) \\
&+ \frac{l_{0}^{2}}{4}q^{2}\cos{(\mathbf{q}\cdot(\mathbf{r}_{i}^{c}-\mathbf{r}_{j}^{c})-2\theta_{i}+2\theta_{j})}\delta^{2}(\omega+2\omega_{0}) + l_{0}^{2}q_{y}^{2}\cos{(\mathbf{q}\cdot(\mathbf{r}_{i}^{c}-\mathbf{r}_{j}^{c}))}\delta^{2}(\omega) \bigg].
\end{split}
\label{eq:JxJx_flat}
\end{equation}
Similarly, for the $y$-component, we calculate
\begin{equation}
\begin{split}
J_{y}^{i}(\mathbf{q},\omega) &= \int dt e^{-i\omega t}v_{y}^{i}(t)e^{i\mathbf{q}\cdot\mathbf{r}_{i}(t)} \\
&\simeq \pi v_{0}e^{i\mathbf{q}\cdot\mathbf{r}_{i}^{c}}\int dt e^{-i\omega t}\bigg\{ \sin(\omega_{0}t+\theta_{i}) + i\frac{l_{0}q_{x}}{2}\big( 1-\cos(2\omega_{0}t+2\theta_{i}) \big) -i\frac{l_{0}q_{y}}{2}\sin(2\omega_{0}t+2\theta_{i}) \bigg\},
\end{split}
\end{equation}
with small $l_{0} q$ approximation as before. Then we have
\begin{equation}
\begin{split}
J_{y}^{i}(\mathbf{q},\omega) \simeq& 
\frac{\pi v_{0}}{i}e^{i\mathbf{q}\cdot\mathbf{r}_{i}^{c}}\bigg\{ e^{i\theta_{i}}\delta(\omega-\omega_{0}) - e^{-i\theta_{i}}\delta(\omega+\omega_{0}) - l_{0}q_{x}\delta(\omega) + \frac{l_{0}}{2}e^{2i\theta_{i}}(q_{x}-iq_{y})\delta(\omega-2\omega_{0})  \\
&+ \frac{l_{0}}{2}e^{-2i\theta_{i}}(q_{x}+iq_{y})\delta(\omega+2\omega_{0}) \bigg\},
\end{split}
\end{equation}
from which we get
\begin{equation}
\begin{split}
J_{y}(\mathbf{q},\omega)J_{y}(\mathbf{q},\omega)^{*} =& \sum_{i,j}J_{y}^{i}(\mathbf{q},\omega)J_{y}^{j}(\mathbf{q},\omega)^{*} \\
\simeq& \pi^{2}v_{0}^{2}\sum_{i}\bigg[ \delta^{2}(\omega-\omega_{0}) + \delta^{2}(\omega+\omega_{0}) + \frac{l_{0}^{2}}{4}q^{2}\delta^{2}(\omega-2\omega_{0}) + \frac{l_{0}^{2}}{4}q^{2}\delta^{2}(\omega+2\omega_{0}) + l_{0}^{2}q_{x}^{2}\delta^{2}(\omega) \bigg] \\
&+ 2\pi^{2}v_{0}^{2}\sum_{i<j}\bigg[ \cos{(\mathbf{q}\cdot(\mathbf{r}_{i}^{c}-\mathbf{r}_{j}^{c})+\theta_{i}-\theta_{j})}\delta^{2}(\omega-\omega_{0}) + \cos{(\mathbf{q}\cdot(\mathbf{r}_{i}^{c}-\mathbf{r}_{j}^{c})-\theta_{i}+\theta_{j})}\delta^{2}(\omega+\omega_{0}) \\
&+ \frac{l_{0}^{2}}{4}q^{2}\cos{(\mathbf{q}\cdot(\mathbf{r}_{i}^{c}-\mathbf{r}_{j}^{c})+2\theta_{i}-2\theta_{j})}\delta^{2}(\omega-2\omega_{0}) \\
&+ \frac{l_{0}^{2}}{4}q^{2}\cos{(\mathbf{q}\cdot(\mathbf{r}_{i}^{c}-\mathbf{r}_{j}^{c})-2\theta_{i}+2\theta_{j})}\delta^{2}(\omega+2\omega_{0}) + l_{0}^{2}q_{x}^{2}\cos{(\mathbf{q}\cdot(\mathbf{r}_{i}^{c}-\mathbf{r}_{j}^{c}))}\delta^{2}(\omega) \bigg].
\end{split}
\label{eq:JyJy_flat}
\end{equation}
We calculate the cross current correlation terms as
\begin{equation}
\begin{split}
J_{y}(\mathbf{q},\omega)J_{x}(\mathbf{q},\omega)^{*} =& \sum_{i,j}J_{y}^{i}(\mathbf{q},\omega)J_{x}^{j}(\mathbf{q},\omega)^{*} \\
\simeq& \pi^{2}v_{0}^{2}\sum_{i}\bigg[ -i\delta^{2}(\omega-\omega_{0})+i\delta^{2}(\omega+\omega_{0}) - l_{0}^{2}q_{x}q_{y}\delta^{2}(\omega) -i\frac{l_{0}^{2}}{4}q^{2}\delta^{2}(\omega-2\omega_{0})+i\frac{l_{0}^{2}}{4}q^{2}\delta^{2}(\omega+2\omega_{0}) \bigg] \\
&+ 2\pi^{2}v_{0}^{2}\sum_{i<j}\bigg[ -i\cos{(\mathbf{q}\cdot(\mathbf{r}_{i}^{c}-\mathbf{r}_{j}^{c})+\theta_{i}-\theta_{j})}\delta^{2}(\omega-\omega_{0}) + i\cos{(\mathbf{q}\cdot(\mathbf{r}_{i}^{c}-\mathbf{r}_{j}^{c})-\theta_{i}+\theta_{j})}\delta^{2}(\omega+\omega_{0}) \\
&-l_{0}^{2}q_{x}q_{y}\cos{(\mathbf{q}\cdot(\mathbf{r}_{i}^{c}-\mathbf{r}_{j}^{c}))}\delta^{2}(\omega) -i\frac{l_{0}^{2}}{4}q^{2}\cos{(\mathbf{q}\cdot(\mathbf{r}_{i}^{c}-\mathbf{r}_{j}^{c})+2\theta_{i}-2\theta_{j})}\delta^{2}(\omega-2\omega_{0}) \\
&+ i\frac{l_{0}^{2}}{4}q^{2}\cos{(\mathbf{q}\cdot(\mathbf{r}_{i}^{c}-\mathbf{r}_{j}^{c})-2\theta_{i}+2\theta_{j})}\delta^{2}(\omega+2\omega_{0}) \bigg].
\end{split}
\label{eq:JyJx_flat}
\end{equation}

The current density we work with is in the basis of longitudinal and transverse directions. The longitudinal and transverse currents are defined as 
\begin{align}
J_{L}(\mathbf{q},\omega) &= J_{x}(\mathbf{q},\omega)\frac{q_{x}}{q} + J_{y}(\mathbf{q},\omega)\frac{q_{y}}{q} \label{eq:jldef}, \\
J_{T}(\mathbf{q},\omega) &= J_{y}(\mathbf{q},\omega)\frac{q_{x}}{q} - J_{x}(\mathbf{q},\omega)\frac{q_{y}}{q}, \label{eq:jtdef}
\end{align}
from which we calculate each element of the current correlation function $C_{\alpha\beta}(\mathbf{q},\omega) = \frac{1}{N}\langle J_{\alpha}^{*}(\mathbf{q},\omega)J_{\beta}(\mathbf{q},\omega) \rangle$ with the subscripts $\alpha$ and $\beta$ being either $L$ or $T$ representing the longitudinal or transverse direction.

First, let's look at the longitudinal current correlation function $C_{LL}(\mathbf{q},\omega) = \frac{1}{N}\langle J_{L}^{*}(\mathbf{q},\omega)J_{L}(\mathbf{q},\omega) \rangle$. From Eq.~(\ref{eq:jldef}) we get
\begin{equation}
J_{L}^{*}(\mathbf{q},\omega)J_{L}(\mathbf{q},\omega) = J_{x}^{*}(\mathbf{q},\omega)J_{x}(\mathbf{q},\omega)\frac{q_{x}^{2}}{q^{2}} + J_{y}^{*}(\mathbf{q},\omega)J_{y}(\mathbf{q},\omega)\frac{q_{y}^{2}}{q^{2}} + (J_{y}^{*}(\mathbf{q},\omega)J_{x}(\mathbf{q},\omega) + J_{x}^{*}(\mathbf{q},\omega)J_{y}(\mathbf{q},\omega))\frac{q_{x}q_{y}}{q^{2}}. 
\label{eq:jljl}
\end{equation}
Using Eqs.~(\ref{eq:JxJx_flat}), (\ref{eq:JyJy_flat}), and (\ref{eq:JyJx_flat}), we calculate Eq.~(\ref{eq:jljl}) as
\begin{equation}
\begin{split}
J_{L}(\mathbf{q},\omega)J_{L}(\mathbf{q},\omega)^{*} 
\simeq& \pi^{2}v_{0}^{2}\sum_{i}\bigg[ \delta^{2}(\omega-\omega_{0}) + \delta^{2}(\omega+\omega_{0}) + \frac{l_{0}^{2}}{4}q^{2}\delta^{2}(\omega-2\omega_{0}) + \frac{l_{0}^{2}}{4}q^{2}\delta^{2}(\omega+2\omega_{0}) \bigg] \\
&+ 2\pi^{2}v_{0}^{2}\sum_{i<j}\bigg[ \cos{(\mathbf{q}\cdot(\mathbf{r}_{i}^{c}-\mathbf{r}_{j}^{c})+\theta_{i}-\theta_{j})}\delta^{2}(\omega-\omega_{0}) + \cos{(\mathbf{q}\cdot(\mathbf{r}_{i}^{c}-\mathbf{r}_{j}^{c})-\theta_{i}+\theta_{j})}\delta^{2}(\omega+\omega_{0}) \\
&+ \frac{l_{0}^{2}}{4}q^{2}\cos{(\mathbf{q}\cdot(\mathbf{r}_{i}^{c}-\mathbf{r}_{j}^{c})+2\theta_{i}-2\theta_{j})}\delta^{2}(\omega-2\omega_{0}) \\
&+ \frac{l_{0}^{2}}{4}q^{2}\cos{(\mathbf{q}\cdot(\mathbf{r}_{i}^{c}-\mathbf{r}_{j}^{c})-2\theta_{i}+2\theta_{j})}\delta^{2}(\omega+2\omega_{0}) \bigg].
\end{split}
\label{eq:JlJl_flat}
\end{equation}
Although Eq.~(\ref{eq:JlJl_flat}) looks complicated, we focus on its dependence on $\omega$ and notice that the $\omega$-dependence is only through the Dirac delta functions $\delta(\omega-n\omega_{0})$ with $n$ being 0, $\pm 1$ and $\pm 2$. Equation (\ref{eq:JlJl_flat}) and all the other equations of correlations calculated in this section are from the expansion up to first order of $l_0 q$. Higher-order expansions will give terms such as $\sin^{2}(\omega_{0}t+\theta_{i})$, $\sin^{3}(\omega_{0}t+\theta_{i})$, ... and $\cos^{2}(\omega_{0}t+\theta_{i})$, $\cos^{3}(\omega_{0}t+\theta_{i})$, ..., which lead to $\sin(3\omega_{0}t+3\theta_{i})$, $\cos(3\omega_{0}t+3\theta_{i})$, ... , etc. in the expansion of current density $\mathbf{J}(\mathbf{q}, \omega)$. This results in higher-order terms of $\delta^2(\omega-n\omega_{0})$ (with $n$ a general integer) with dependence on $(l_{0}q)^{n}\cos (\mathbf{q}\cdot(\mathbf{r}_{i}^{c}-\mathbf{r}_{j}^{c})+n\theta_{i}-n\theta_{j})$ in the current correlation functions.
Then a plot of the dispersion relation obtained from Eq.~(\ref{eq:JlJl_flat}) would consist of layers of flat planes located at $\omega=n\omega_{0}$, with the $\mathbf{q}$-dependence shown on each plane. 
Note that Eq.~(\ref{eq:JlJl_flat}) is the result in the small $l_{0}q$ approximation. Therefore, the strongest contribution comes from $\delta(\omega-\omega_{0})$ terms (with $n=1$), as we measure positive frequencies, and the signal decreases as $n$ increases because of higher-order dependence on $l_{0}q$. This result agrees with the simulation result of our starfish embryo model with self-circling motion where dispersion signals are visible only at frequency values $\omega = \omega_{0}$ and $\omega = 2\omega_{0}$. The $q^{2}$ dependence for the $n=2$ term agrees with the observation that the bright horizontal line at $\omega = 2\omega_{0}$ in Fig.~1(b) of the main text has stronger signal around the $K$ or $M$ point compared to the $\Gamma$ point in the first Brillouin zone.

For the transverse current correlation function $C_{TT}(\mathbf{q},\omega) = \frac{1}{N}\langle J_{T}^{*}(\mathbf{q},\omega)J_{T}(\mathbf{q},\omega) \rangle$, from Eq.~(\ref{eq:jtdef}) we get
\begin{equation}
J_{T}^{*}(\mathbf{q},\omega)J_{T}(\mathbf{q},\omega) = J_{y}^{*}(\mathbf{q},\omega)J_{y}(\mathbf{q},\omega)\frac{q_{x}^{2}}{q^{2}} + J_{x}^{*}(\mathbf{q},\omega)J_{x}(\mathbf{q},\omega)\frac{q_{y}^{2}}{q^{2}} - (J_{y}^{*}(\mathbf{q},\omega)J_{x}(\mathbf{q},\omega) + J_{x}^{*}(\mathbf{q},\omega)J_{y}(\mathbf{q},\omega))\frac{q_{x}q_{y}}{q^{2}},
\label{eq:jtjt}
\end{equation}
which leads to
\begin{equation}
\begin{split}
J_{T}(\mathbf{q},\omega)J_{T}(\mathbf{q},\omega)^{*} 
\simeq& \pi^{2}v_{0}^{2}\sum_{i}\bigg[ \delta^{2}(\omega-\omega_{0}) + \delta^{2}(\omega+\omega_{0}) + \frac{l_{0}^{2}}{4}q^{2}\delta^{2}(\omega-2\omega_{0}) + \frac{l_{0}^{2}}{4}q^{2}\delta^{2}(\omega+2\omega_{0}) + l_{0}^{2}q^{2}\delta^{2}(\omega) \bigg] \\
&+ 2\pi^{2}v_{0}^{2}\sum_{i<j}\bigg[ \cos{(\mathbf{q}\cdot(\mathbf{r}_{i}^{c}-\mathbf{r}_{j}^{c})+\theta_{i}-\theta_{j})}\delta^{2}(\omega-\omega_{0}) + \cos{(\mathbf{q}\cdot(\mathbf{r}_{i}^{c}-\mathbf{r}_{j}^{c})-\theta_{i}+\theta_{j})}\delta^{2}(\omega+\omega_{0}) \\
&+ \frac{l_{0}^{2}}{4}q^{2}\cos{(\mathbf{q}\cdot(\mathbf{r}_{i}^{c}-\mathbf{r}_{j}^{c})+2\theta_{i}-2\theta_{j})}\delta^{2}(\omega-2\omega_{0}) \\
&+ \frac{l_{0}^{2}}{4}q^{2}\cos{(\mathbf{q}\cdot(\mathbf{r}_{i}^{c}-\mathbf{r}_{j}^{c})-2\theta_{i}+2\theta_{j})}\delta^{2}(\omega+2\omega_{0})
+ l_{0}^{2}q^{2}\cos{(\mathbf{q}\cdot(\mathbf{r}_{i}^{c}-\mathbf{r}_{j}^{c}))}\delta^{2}(\omega) \bigg].
\end{split}
\label{eq:JtJt_flat}
\end{equation}
Equation (\ref{eq:JtJt_flat}) is the same as Eq.~(\ref{eq:JlJl_flat}) other than the additional $\delta(\omega)$ terms. Since the $\delta(\omega)$ terms are only related to $\omega = 0$ cases, the behavior of $C_{TT}(\mathbf{q},\omega)$ is the same as $C_{LL}(\mathbf{q},\omega)$ for finite values of $\omega$.

For the cross current correlation functions between the longitudinal and transverse currents, since $C_{LT}(\mathbf{q},\omega) = \frac{1}{N}\langle J_{L}^{*}(\mathbf{q},\omega)J_{T}(\mathbf{q},\omega) \rangle$ and $C_{TL}(\mathbf{q},\omega) = \frac{1}{N}\langle J_{T}^{*}(\mathbf{q},\omega)J_{L}(\mathbf{q},\omega) \rangle$ which are complex conjugates of each other, we need to consider the real and imaginary parts of $C_{LT}(\mathbf{q},\omega)$.
Using Eq.~(\ref{eq:jldef}) and Eq.~(\ref{eq:jtdef}), we have 
\begin{align}
J_{L}^{*}(\mathbf{q},\omega)J_{T}(\mathbf{q},\omega) &= J_{x}^{*}(\mathbf{q},\omega)J_{y}(\mathbf{q},\omega)\frac{q_{x}^{2}}{q^{2}} - J_{y}^{*}(\mathbf{q},\omega)J_{x}(\mathbf{q},\omega)\frac{q_{y}^{2}}{q^{2}} + (J_{y}^{*}(\mathbf{q},\omega)J_{y}(\mathbf{q},\omega) - J_{x}^{*}(\mathbf{q},\omega)J_{x}(\mathbf{q},\omega))\frac{q_{x}q_{y}}{q^{2}}, \\
J_{T}^{*}(\mathbf{q},\omega)J_{L}(\mathbf{q},\omega) &= J_{y}^{*}(\mathbf{q},\omega)J_{x}(\mathbf{q},\omega)\frac{q_{x}^{2}}{q^{2}} - J_{x}^{*}(\mathbf{q},\omega)J_{y}(\mathbf{q},\omega)\frac{q_{y}^{2}}{q^{2}} + (J_{y}^{*}(\mathbf{q},\omega)J_{y}(\mathbf{q},\omega) - J_{x}^{*}(\mathbf{q},\omega)J_{x}(\mathbf{q},\omega))\frac{q_{x}q_{y}}{q^{2}}, 
\end{align}
and thus
\begin{equation}
\begin{split}
\text{Re}[J_{L}^{*}(\mathbf{q},\omega)J_{T}(\mathbf{q},\omega)] =& \frac{q_{x}^{2}-q_{y}^{2}}{q^{2}}\text{Re}[J_{x}^{*}(\mathbf{q},\omega)J_{y}(\mathbf{q},\omega)] + \frac{q_{x}q_{y}}{q^{2}}(J_{y}^{*}(\mathbf{q},\omega)J_{y}(\mathbf{q},\omega)-J_{x}^{*}(\mathbf{q},\omega)J_{x}(\mathbf{q},\omega)) \\
\simeq& \frac{q_{x}^{2}-q_{y}^{2}}{q^{2}}\bigg\{ \pi^{2}v_{0}^{2}\sum_{i}\bigg[ -l_{0}^{2}q_{x}q_{y}\delta^{2}(\omega) \bigg] + 2\pi^{2}v_{0}^{2}\sum_{i<j}\bigg[ -l_{0}^{2}q_{x}q_{y}\cos{(\mathbf{q}\cdot(\mathbf{r}_{i}^{c}-\mathbf{r}_{j}^{c}))}\delta^{2}(\omega) \bigg] \bigg\} \\
&+ \frac{q_{x}q_{y}}{q^{2}}\bigg\{ \pi^{2}v_{0}^{2}\sum_{i}l_{0}^{2}(q_{x}^{2}-q_{y}^{2})\delta^{2}(\omega) + 2\pi^{2}v_{0}^{2}\sum_{i<j}l_{0}^{2}(q_{x}^{2}-q_{y}^{2})\cos{(\mathbf{q}\cdot(\mathbf{r}_{i}^{c}-\mathbf{r}_{j}^{c}))}\delta^{2}(\omega) \bigg\} \\
=& 0,
\end{split}
\end{equation}
when $\omega$ is nonzero.
Since the real part of $C_{LT}(\mathbf{q},\omega)$ is zero for nonzero frequency, it does not affect the dispersion curve shown in Fig.~1(b) of the main paper. The imaginary part of $C_{LT}(\mathbf{q},\omega)$ is calculated as
\begin{equation}
\begin{split}
\text{Im}[J_{L}^{*}(\mathbf{q},\omega)J_{T}(\mathbf{q},\omega)] =& \text{Im}[J_{x}^{*}(\mathbf{q},\omega)J_{y}(\mathbf{q},\omega)] \\
\simeq& \pi^{2}v_{0}^{2}\sum_{i}\bigg[ -\delta^{2}(\omega-\omega_{0}) + \delta^{2}(\omega+\omega_{0}) - \frac{l_{0}^{2}}{4}q^{2}\delta^{2}(\omega-2\omega_{0}) + \frac{l_{0}^{2}}{4}q^{2}\delta^{2}(\omega+2\omega_{0}) \bigg] \\
&+ 2\pi^{2}v_{0}^{2}\sum_{i<j}\bigg[ -\cos{(\mathbf{q}\cdot(\mathbf{r}_{i}^{c}-\mathbf{r}_{j}^{c})+\theta_{i}-\theta_{j})}\delta^{2}(\omega-\omega_{0}) + \cos{(\mathbf{q}\cdot(\mathbf{r}_{i}^{c}-\mathbf{r}_{j}^{c})-\theta_{i}+\theta_{j})}\delta^{2}(\omega+\omega_{0}) \\
&- \frac{l_{0}^{2}}{4}q^{2}\cos{(\mathbf{q}\cdot(\mathbf{r}_{i}^{c}-\mathbf{r}_{j}^{c})+2\theta_{i}-2\theta_{j})}\delta^{2}(\omega-2\omega_{0}) \\
&+ \frac{l_{0}^{2}}{4}q^{2}\cos{(\mathbf{q}\cdot(\mathbf{r}_{i}^{c}-\mathbf{r}_{j}^{c})-2\theta_{i}+2\theta_{j})}\delta^{2}(\omega+2\omega_{0}) \bigg].
\end{split}
\label{eq:ImJlJt}
\end{equation}
Equation (\ref{eq:ImJlJt}) looks very similar to Eq.~(\ref{eq:JlJl_flat}) except that the signs of the $\delta(\omega-\omega_{0})$ term and the $\delta(\omega-2\omega_{0})$ term are flipped.
Since the $\omega$ dependence is the same, this difference does not change the shape of the dispersion curve. The $\delta$-function dependence of $\omega$ gives rise to the dispersion at integer multiples of self-circling frequency values like the cases for $C_{LL}(\mathbf{q},\omega)$ and $C_{TT}(\mathbf{q},\omega)$.

Since we show the experimental dispersion relations obtained from the current correlation $\frac{1}{N}\langle \mathbf{J}^{*}(\mathbf{q},\omega)\cdot\mathbf{J}(\mathbf{q},\omega) \rangle = C_{LL}(\mathbf{q},\omega) + C_{TT}(\mathbf{q},\omega)$ in the main text (Fig.~1), we calculate this quantity as well, which is
\begin{equation}
\begin{split}
\mathbf{J}^{*}(\mathbf{q},\omega)\cdot\mathbf{J}(\mathbf{q},\omega) 
=& J_{L}^{*}(\mathbf{q},\omega)J_{L}(\mathbf{q},\omega) + J_{T}^{*}(\mathbf{q},\omega)J_{T}(\mathbf{q},\omega) \\
=& J_{x}^{*}(\mathbf{q},\omega)J_{x}(\mathbf{q},\omega) + J_{y}^{*}(\mathbf{q},\omega)J_{y}(\mathbf{q},\omega) \\
\simeq& 2\pi^{2}v_{0}^{2}\sum_{i}\bigg[ \delta^{2}(\omega-\omega_{0}) + \delta^{2}(\omega+\omega_{0}) + \frac{l_{0}^{2}}{2}q^{2}\delta^{2}(\omega) + \frac{l_{0}^{2}}{4}q^{2}\delta^{2}(\omega-2\omega_{0}) + \frac{l_{0}^{2}}{4}q^{2}\delta^{2}(\omega+2\omega_{0}) \bigg] \\
&+ 4\pi^{2}v_{0}^{2}\sum_{i<j}\bigg[ \cos{(\mathbf{q}\cdot(\mathbf{r}_{i}^{c}-\mathbf{r}_{j}^{c})+\theta_{i}-\theta_{j})}\delta^{2}(\omega-\omega_{0}) + \cos{(\mathbf{q}\cdot(\mathbf{r}_{i}^{c}-\mathbf{r}_{j}^{c})-\theta_{i}+\theta_{j})}\delta^{2}(\omega+\omega_{0}) \\
&+ \frac{l_{0}^{2}}{4}q^{2}\cos{(\mathbf{q}\cdot(\mathbf{r}_{i}^{c}-\mathbf{r}_{j}^{c})+2\theta_{i}-2\theta_{j})}\delta^{2}(\omega-2\omega_{0}) \\
&+ \frac{l_{0}^{2}}{4}q^{2}\cos{(\mathbf{q}\cdot(\mathbf{r}_{i}^{c}-\mathbf{r}_{j}^{c})-2\theta_{i}+2\theta_{j})}\delta^{2}(\omega+2\omega_{0}) + \frac{l_{0}^{2}}{2}q^{2}\cos{(\mathbf{q}\cdot(\mathbf{r}_{i}^{c}-\mathbf{r}_{j}^{c}))}\delta^{2}(\omega) \bigg].
\label{eq:JJ_flat}
\end{split}
\end{equation}
Equation (\ref{eq:JJ_flat}) has a form similar to Eqs.~(\ref{eq:JlJl_flat}) and (\ref{eq:JtJt_flat}), thus leading to the same conclusion that the self-circling motion causes dispersion at integer multiples of the self-circling frequency. 

We note that the analytical calculations given in this section are not intended to be an accurate first-principles representation of the starfish embryo system. We make a number of simplifications to enable the calculations to be performed.  The starfish embryos interact with each other, but here we assume non-interacting particles that are making circular trajectories. We have used the first-order expansion of the exponential in the integral expression of the current density, which assumes the limit of small $l_{0}q$. We justify our approximation by assuming that the embryos cannot move too much inside a dense cluster, which leads to small $l_{0}= {v_{0}}/{\omega_{0}}$; also we only look at the property in the first Brillouin zone with finite $q$. Even so, this approximation would work better in the proximity to the $\Gamma$ point than the boundary of the Brillouin zone. 

Despite the limitations given above, our analytical calculation for the non-interacting self-circling particles successfully describes the dispersion due to self-circling observed in the simulation (Fig.~1(b) of the main text) as well as the experiment (Fig.~3 of the main text). Note that the $\delta$-function dependence of $\omega$ in the current correlation functions signifies the dispersion signals at integer multiples of the self-circling frequency $\omega_{0}$, while the $q$ dependence of each term depicts where in the $q$-space one will see strong signals. The $\cos{(\mathbf{q}\cdot(\mathbf{r}_{i}^{c}-\mathbf{r}_{j}^{c}))}$ terms predicts that $\mathbf{q}=\mathbf{0}$ and the reciprocal lattice vectors, all of them being represented by the $\Gamma$ point, are the values of $\mathbf{q}$ that yield the strongest signal. This prediction agrees with the results obtained from the experimental data of Ref.~\cite{tan2022odd} (Fig.~3 of the main text) as well as the simulation data (Fig.~1 of the main text). In the experiment, the starfish embryos are circling in phase with their neighbors, which allows us to use $\theta_{i}-\theta_{j} \approx 0$. In the simulation we use the random initial condition for particle orientations, and the phase differences between particles are averaged out when calculating the current correlation function, giving the strongest signal around the $\Gamma$ point.

\section{Damping of odd elastic waves}
\label{sec:damping}

This section shows that the deterministic odd elastic waves are always damped, based on the results of odd elasticity given in Refs.~\cite{scheibner2020odd} and \cite{braverman2021topological}.

The linearized equation of motion for the displacement field $\mathbf{u}$ in an overdamped system is given by \cite{scheibner2020odd}
\begin{equation}
\gamma\dot{u}_{j} = C_{ijmn}\partial_{i}\partial_{m}u_{n},
\end{equation}
where $\gamma$ is the friction coefficient and $C_{ijmn}$ is the element of elastic modulus tensor. In Fourier space it becomes
\begin{equation}
-i\omega\gamma\tilde{u}_{j}=-q_{i}q_{m}C_{ijmn}\tilde{u}_{n}.
\label{eq:u_eom}
\end{equation}
We can rewrite Eq.~(\ref{eq:u_eom}) in terms of the 2D representation of the elastic moduli tensor $C^{\alpha\beta} = \frac{1}{2}\tau_{ij}^{\beta}C_{ijmn}\tau_{mn}^{\alpha}$, where $\tau^{\alpha}$ is defined by \cite{braverman2021topological}
\begin{align}
\tau^{0} &= 
\begin{pmatrix}
1 & 0 \\
0 & 1
\end{pmatrix},
&
\tau^{1} &=
\begin{pmatrix}
0 & -1 \\
1 & 0
\end{pmatrix},
&
\tau^{2} &=
\begin{pmatrix}
1 & 0 \\
0 & -1
\end{pmatrix},
&
\tau^{3} &=
\begin{pmatrix}
0 & 1 \\
1 & 0
\end{pmatrix}.
\end{align}
Then in the basis of the longitudinal and transverse displacement vector, Eq.~(\ref{eq:u_eom}) becomes
\begin{equation}
-i\omega\gamma 
\begin{pmatrix}
\tilde{u}_{L} \\
\tilde{u}_{T}
\end{pmatrix} 
= -q^{2}
\begin{pmatrix}
B+\mu & K^{o} \\
-K^{o}-A & \mu
\end{pmatrix}
\begin{pmatrix}
\tilde{u}_{L} \\
\tilde{u}_{T}
\end{pmatrix},
\label{eq:u_eigen}
\end{equation}
where $\tilde{u}_{L}$ is the longitudinal displacement field, $\tilde{u}_{T}$ is the transverse displacement field, $B$ is the bulk modulus, $\mu$ is the shear modulus, $K^{o}$ is the odd modulus coupling the shears anti-symmetrically, and $A$ is the odd modulus coupling the compression to the internal torque density \cite{scheibner2020odd}. 

Eigenvalues of the dynamic matrix in Eq.~(\ref{eq:u_eigen}) are given by 
\begin{equation}
\sigma = -\bigg( \frac{B}{2} + \mu \bigg)\pm\sqrt{\bigg( \frac{B}{2} \bigg)^{2}-K^{o}A-(K^{o})^{2}},
\label{eq:eigen}
\end{equation}
which provide the information on the stability of the solution and the condition of wave behavior. As discussed in Ref.~\cite{scheibner2020odd}, if $\big( \frac{B}{2} \big)^{2}-K^{o}A-(K^{o})^{2}<0$, the eigenvalue has a nonzero imaginary part and thus the system exhibits an oscillatory behavior. Therefore, this inequality provides the condition for the onset of odd elastic waves. In addition to the imaginary part, we need to consider the real part of $\sigma$, which is $-\big( \frac{B}{2} + \mu \big)$ and governs the stability of the solution. Since the elastic moduli $B>0$ and $\mu>0$, the real part of the eigenvalue is always negative. Therefore, even if the wave condition is satisfied, the deterministic wave would always become damped. 

To prevent the damping, we need to make the real part of the eigenvalue zero (giving marginal stability), i.e., $\frac{B}{2} + \mu=0$ or $B=\mu=0$. For a two-dimensional triangular lattice, each elastic modulus is expressed in terms of the linearized force as \cite{braverman2021topological}
\begin{align}
B &= \frac{\sqrt{3}}{2}\bigg( k_{L}+\frac{F_{L}^{0}}{r_{0}} \bigg), 
&
\mu &= \frac{\sqrt{3}}{4}\bigg( k_{L}-\frac{3F_{L}^{0}}{r_{0}}\bigg),
&
A &= \frac{\sqrt{3}}{2}\bigg( k_{T}+\frac{F_{T}^{0}}{r_{0}} \bigg),
&
K^{o} &= \frac{\sqrt{3}}{4}\bigg( k_{T}-\frac{3F_{T}^{0}}{r_{0}} \bigg),
\label{eq:elastic_moduli}
\end{align}
where $k_{L}$ and $k_{T}$ are the longitudinal and transverse spring constants of the linearized equation of motion, $F_{L}^{0}$ and $F_{T}^{0}$ are the zeroth order terms of the longitudinal and transverse forces between two neighboring particles, and $r_{0}$ is the lattice spacing. $B$ and $\mu$ only depend on the longitudinal parameters $k_{L}$ and $F_{L}^{0}$. When the particles (such as starfish embryos) form the crystal, $F_{L}^{0}=0$. Therefore, to satisfy $\frac{B}{2} + \mu=0$ we need to make $k_{L}=0$. In a linearized system, this means that the longitudinal force has to be zero to prevent the waves from being damped. It is not the case for the real system of starfish embryos, and in fact, it is unrealistic because the crystal structure cannot be maintained without the longitudinal force or with zero elastic moduli. This result of damping can be verified through simulations, as conducted in the next section (SM $\S$\ref{sec:simul_damping}).

\section{Simulation verification of damping due to longitudinal force}
\label{sec:simul_damping}

This section confirms through simulation results that the deterministic odd elastic waves are damped unless the longitudinal force is absent. 

\begin{figure}[!htb]
   \includegraphics[width=0.72\columnwidth]{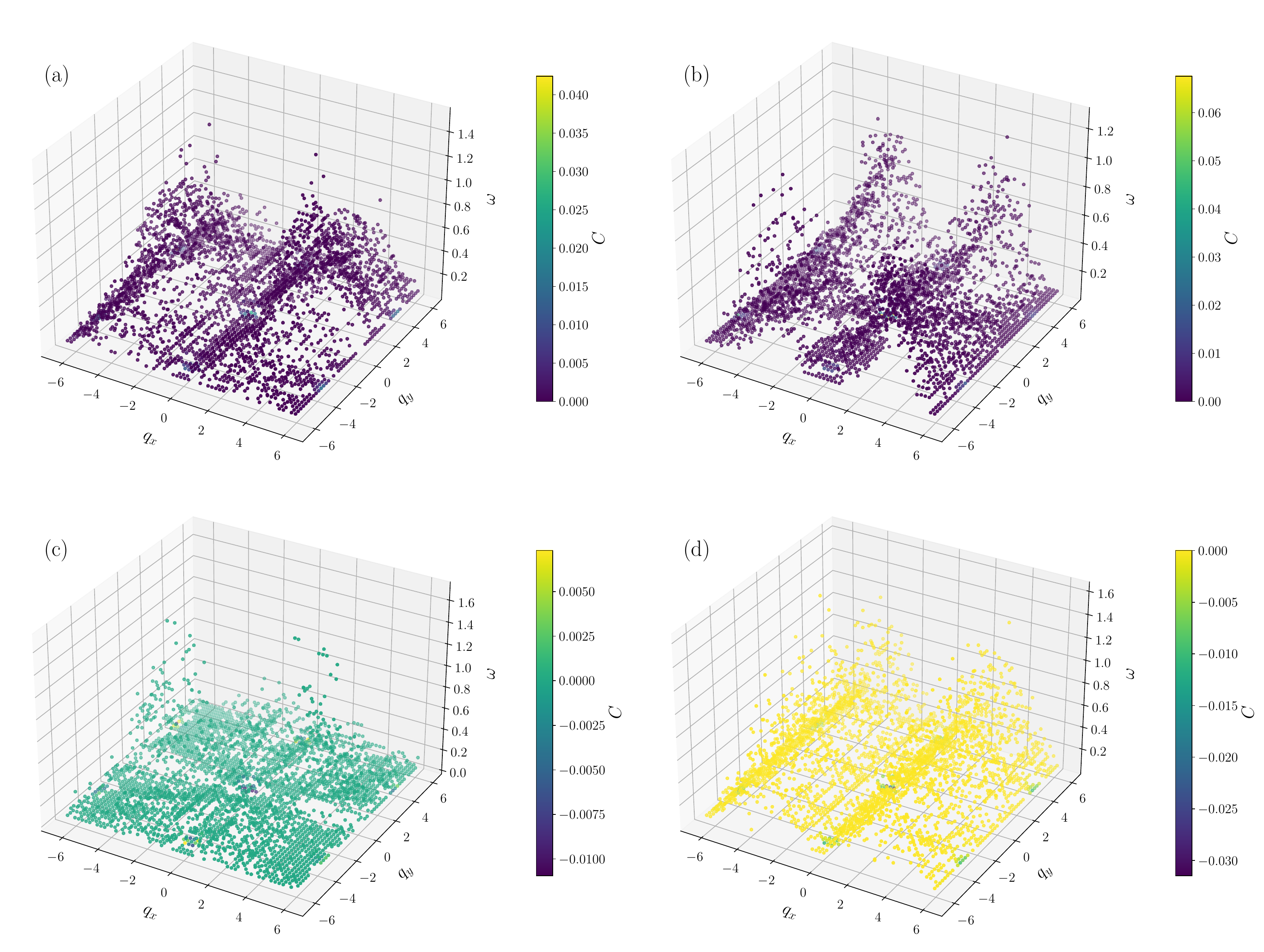}
   \caption{The dispersion results for the toy model in the presence of longitudinal force, as obtained from (a) $C_{LL}(\mathbf{q},\omega)$, (b) $C_{TT}(\mathbf{q},\omega)$, (c) $\text{Re}[C_{LT}(\mathbf{q},\omega)]$, and (d) $\text{Im}[C_{LT}(\mathbf{q},\omega)]$. 
  \label{fig:noNoise_Damp_toy}}
\end{figure}
\begin{figure}[!htb]
   \includegraphics[width=0.72\columnwidth]{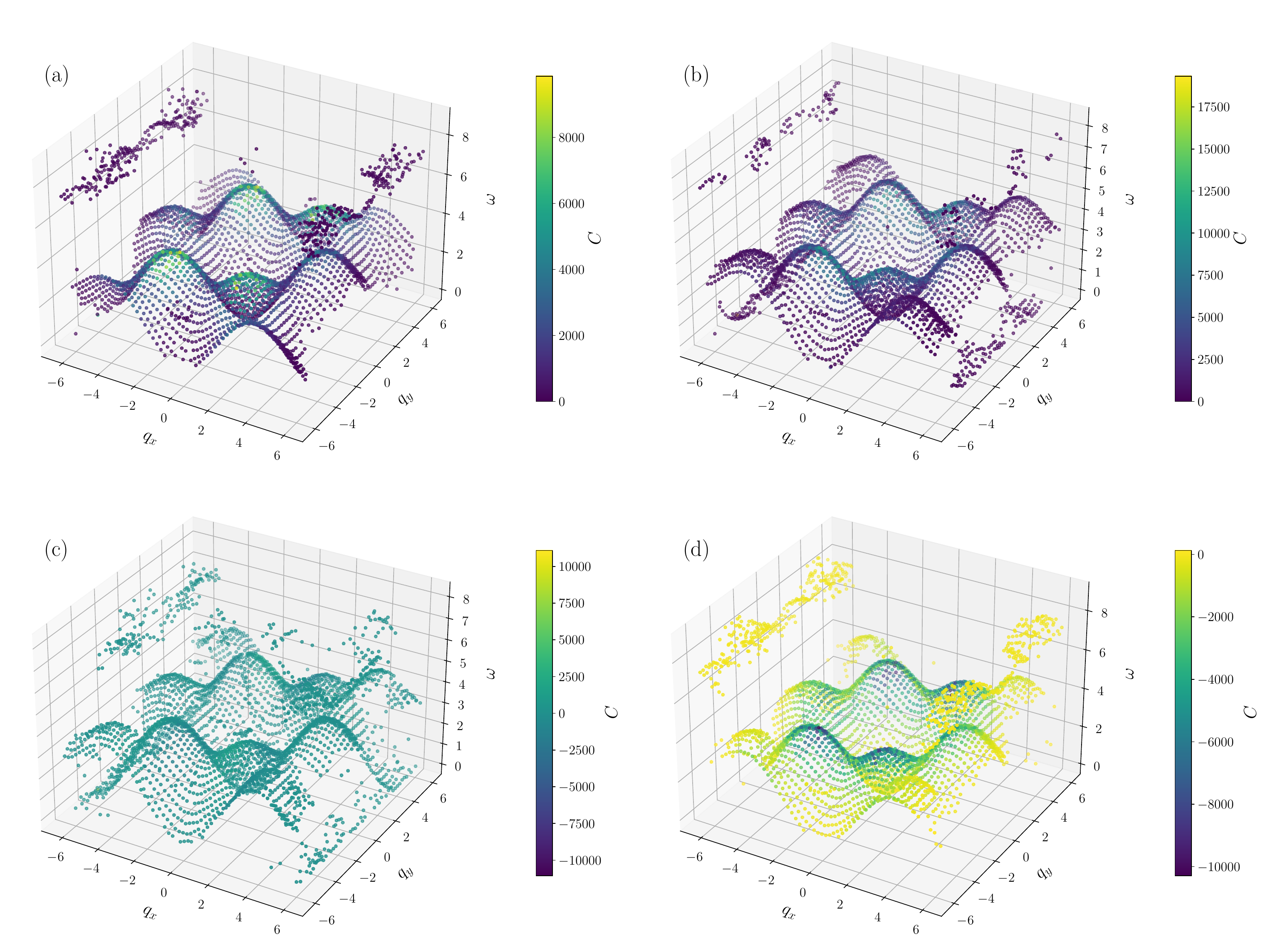}
   \caption{The dispersion relations for the toy model in the absence of longitudinal force, as obtained from (a) $C_{LL}(\mathbf{q},\omega)$, (b) $C_{TT}(\mathbf{q},\omega)$, (c) $\text{Re}[C_{LT}(\mathbf{q},\omega)]$, and (d) $\text{Im}[C_{LT}(\mathbf{q},\omega)]$. 
  \label{fig:noNoise_noDamp_toy}}
\end{figure}

Figures \ref{fig:noNoise_Damp_toy} and \ref{fig:noNoise_noDamp_toy} show the dispersion relations obtained from the toy model with and without the longitudinal force. The toy model with the longitudinal and transverse spring forces is given by \cite{scheibner2020odd}
\begin{equation}
\frac{d\mathbf{r}_{i}}{dt}=\sum_{j}\bigg[-\left (k_{L}\hat{\mathbf{r}}_{ij}+k_{T}\hat{\mathbf{r}}^{\perp}_{ij} \right ) \left (r_{ij}-r_{ij}^{0} \right )\bigg],
\label{eq:det_toy}
\end{equation}
where $r_{ij}$ is the distance between particles $i$ and $j$, $r_{ij}^{0}$ is their equilibrium distance, $\hat{\mathbf{r}}_{ij}$ is the unit vector in the direction of $\mathbf{r}_{ij} \equiv \mathbf{r}_{i}-\mathbf{r}_{j}$, and $(\hat{\mathbf{r}}^{\perp}_{ij})_{\alpha} = \epsilon_{\alpha\beta}(\hat{\mathbf{r}}_{ij})_{\beta}$ with $\epsilon_{\alpha\beta}$ the two-dimensional Levi-Civita symbol. Here $k_{L}$ is the spring constant of the longitudinal force while $k_{T}$ is the spring constant of the transverse force.

The simulations are conducted with 900 particles located on a $30\times 30$ two-dimensional triangular lattice with periodic boundary conditions. We used the Euler algorithm \cite{euler1768foundations} with time step $dt = 0.001$, up to the final time $t_{f} = 100$ ($10^{5}$ time steps), and recorded the data every 100 time steps (with the measured $\Delta t=100\times dt=0.1$). We chose such a small time step and short iteration time because in the absence of damping (Fig.~\ref{fig:noNoise_noDamp_toy}), the system is unstable and eventually loses its crystal structure at later time. The results shown here have been averaged over 100 simulation runs.
The parameter values are $r_{ij}^{0} = 1.0$, $k_{T} = 1.0$, and $k_{L} = 0.5$ for Fig.~\ref{fig:noNoise_Damp_toy} and $k_{L} = 0$ for Fig.~\ref{fig:noNoise_noDamp_toy}. The value of $k_{T}$ and $k_{L}$ are chosen such that the deterministic wave condition \cite{scheibner2020odd} is satisfied. 

As seen in Fig.~\ref{fig:noNoise_Damp_toy}, in the presence of nonzero longitudinal force the dispersion results are noisy and do not show any structure for all elements of the current correlation function, although the parameter values satisfy the deterministic wave condition. The value of each current correlation is in the order of $10^{-2}$, which is very small, as all the dynamics is damped.
On the other hand, in the absence of longitudinal force so that there is no damping, as shown in Fig.~\ref{fig:noNoise_noDamp_toy}, we can identify the dispersion relations from all elements of the current correlation function. The value of each current correlation is in the order of $10^{3}$, much larger than that of Fig.~\ref{fig:noNoise_Damp_toy}. 

Now we look at a more realistic model describing the living crystal of starfish embryos \cite{tan2022odd}, of which the equation of motion is written as
\begin{equation}
\frac{d\mathbf{r}_{i}}{dt} = \sum_{i\neq j}\biggl[ \bar{\mathbf{v}}_{\text{st}}(\mathbf{r}_{i};\mathbf{e}_{z},\mathbf{r}_{j}) + \frac{1}{\eta R}\mathbf{F}_{\text{rep}}(\lvert \mathbf{r}_{i}-\mathbf{r}_{j} \rvert) + R(\omega_{i}+\omega_{j})F_{\text{nf}}(\lvert \mathbf{r}_{i}-\mathbf{r}_{j} \rvert)\hat{\mathbf{r}}_{ij}^{\perp} \biggr].
\label{eq:det_starfish}
\end{equation}
The explicit expressions for each force term in Eq.~(\ref{eq:det_starfish}) are shown in the next section (SM $\S$\ref{sec:starfish_model}). 

\begin{figure}[b]
   \includegraphics[width=0.72\columnwidth]{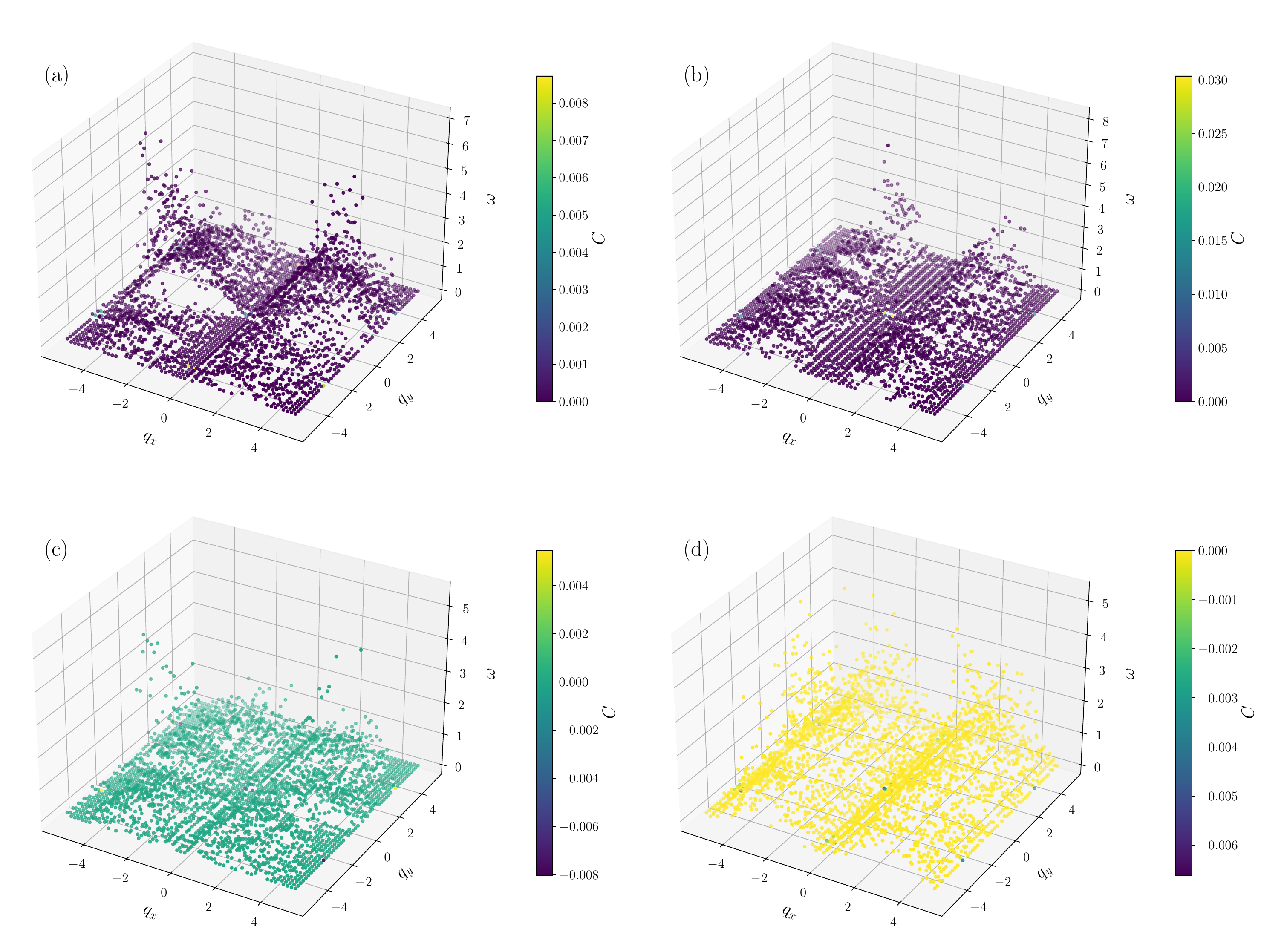}
   \caption{The dispersion results for the starfish embryo model in the presence of longitudinal force, as obtained from (a) $C_{LL}(\mathbf{q},\omega)$, (b) $C_{TT}(\mathbf{q},\omega)$, (c) $\text{Re}[C_{LT}(\mathbf{q},\omega)]$, and (d) $\text{Im}[C_{LT}(\mathbf{q},\omega)]$. 
  \label{fig:noNoise_Damp_starfish}}
\end{figure}
\begin{figure}[!htb]
   \includegraphics[width=0.72\columnwidth]{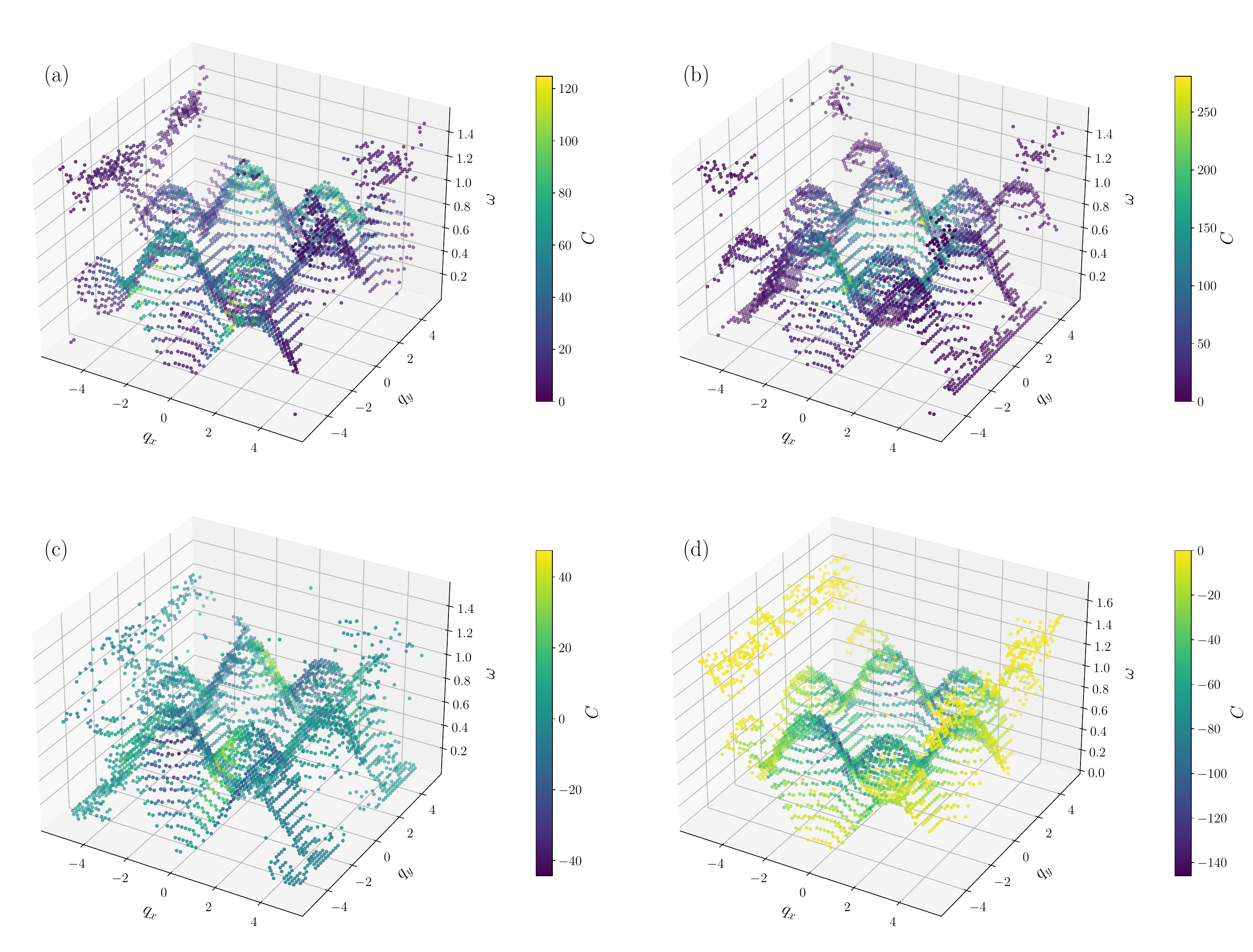}
   \caption{The dispersion relations for the starfish embryo model in the absence of longitudinal force, as obtained from (a) $C_{LL}(\mathbf{q},\omega)$, (b) $C_{TT}(\mathbf{q},\omega)$, (c) $\text{Re}[C_{LT}(\mathbf{q},\omega)]$, and (d) $\text{Im}[C_{LT}(\mathbf{q},\omega)]$. 
  \label{fig:noNoise_noDamp_starfish}}
\end{figure}

As in the case for the toy model, the simulations are done with 900 particles on a $30\times30$ two-dimensional triangular lattice with periodic boundary conditions. We use the Euler algorithm \cite{euler1768foundations} with $dt=0.001$ and the final time $t_{f}=100$ ($10^{5}$ time steps), record the data every 100 time steps (with the measured $\Delta t=0.1$) and average the results over 100 runs of simulations. We non-dimensionalized Eq.~(\ref{eq:det_starfish}), setting the nondimensional radius of the embryos to be $R=0.5$, and adjust the experimental parameter values accordingly. The equilibrium spacing between the embryos is set to be 1.2 so that it is slightly longer than $2R$.
For other parameter values used in our simulations (see SM $\S$\ref{sec:starfish_model} for the detailed definition of each parameter),
to obtain results in Fig.~\ref{fig:noNoise_Damp_starfish} we set the rescaled Stokeslet strength responsible for the longitudinal attraction between embryos to be $\tilde{F}_{\text{st}}=53.7$ and the rescaled amplitude of the longitudinal steric repulsion between embryos to be $\tilde{F}_{\text{rep}}=785.1$. For Fig.~\ref{fig:noNoise_noDamp_starfish}, we use $\tilde{F}_{\text{st}}=0$ and $\tilde{f}_{\text{rep}}=0$ so that the longitudinal force is set to zero. 
For parameters determining the transverse force amplitude, we use $\bar{\omega}_{0}=1.0$ and $f_{0}=1.0$ for Fig.~\ref{fig:noNoise_Damp_starfish}, and $\bar{\omega}_{0}=0.1$ and $f_{0}=0.5$ for Fig.~\ref{fig:noNoise_noDamp_starfish}. These values are all different from those used in Ref.~\cite{tan2022odd} for the starfish embryo experiment. For Fig.~\ref{fig:noNoise_Damp_starfish}, we have increased the transverse force amplitude to make the system satisfy the deterministic wave condition. However, the undamped case depicted in Fig.~\ref{fig:noNoise_noDamp_starfish} is so unstable that we have to use smaller parameter values to prevent the system from melting within the simulation time. When the longitudinal force is zero, the wave condition is still satisfied even if we use smaller transverse force.

Similar to Fig.~\ref{fig:noNoise_Damp_toy}, Fig.~\ref{fig:noNoise_Damp_starfish} does not show meaningful dispersion relations in the presence of longitudinal force. The value of the current correlation function is mostly in the order of $10^{-3}$, as the dynamics is all damped.
On the other hand, Fig.~\ref{fig:noNoise_noDamp_starfish} shows that, similar to Fig.~\ref{fig:noNoise_noDamp_toy}, in the absence of longitudinal force and thus damping, the dispersion relations appear from the calculation of current correlation functions. The values of current correlation functions are in the order of 10 or $10^{2}$, much larger than the damped case. 

These results confirm that the longitudinal force is responsible for the damping of odd elastic waves, and the deterministic waves could be retained only if the longitudinal force is removed. However, it is known that the longitudinal interaction always exists in real systems, such as the starfish embryos \cite{tan2022odd}. In fact, without the longitudinal force the crystal structure cannot be maintained. Therefore, the odd elastic wave is not expected to be observable according to the existing model, and we need a new model to generate persistent elastic waves in this overdamped system.

\section{Explicit expressions for the starfish embryo model}
\label{sec:starfish_model}

In this section, we give the explicit expressions of the force terms used in the starfish embryo model. It follows the original model given in Ref.~\cite{tan2022odd}, while we add a new self-propulsion term with the important noise-driven effect.
The equations of motion for the self-propelling starfish embryo model are given by (as reproduced from the main text)
\begin{align}
\frac{d\mathbf{r}_{i}}{dt} &= \sum_{i\neq j}\biggl[ \bar{\mathbf{v}}_{\text{st}}(\mathbf{r}_{i},\mathbf{r}_{j}) + \frac{1}{\eta R}\mathbf{F}_{\text{rep}}(\lvert \mathbf{r}_{i}-\mathbf{r}_{j} \rvert) + R(\omega_{i}+\omega_{j})F_{\text{nf}}(\lvert \mathbf{r}_{i}-\mathbf{r}_{j} \rvert)\hat{\mathbf{r}}_{ij}^{\perp} \biggr] + v_{0}\mathbf{p}_{i},
\label{eq:starfish_eom}
\\
\frac{d\theta_{i}}{dt} &= \omega_{i},
\end{align}
where the polarization vector $\mathbf{p}_{i}= (\cos\theta_{i}, \sin\theta_{i})$, $\theta_{i}(t)$ is the time-dependent orientation angle of the $i^{th}$ embryo, and $v_{0}$ is the self-propulsion strength. 
The first term in Eq.~(\ref{eq:starfish_eom}) represents the longitudinal attraction between the embryos due to the Stokeslet flow \cite{tan2022odd}, written explicitly as
\begin{equation}
\bar{\mathbf{v}}_{\text{st}}(\mathbf{r}_{i};\mathbf{e}_{z},\mathbf{r}_{j}) = -\frac{F_{\text{st}}}{8\pi\eta}\frac{2h(\mathbf{r}_{i}-\mathbf{r}_{j})}{(\lvert \mathbf{r}_{i}-\mathbf{r}_{j} \rvert^{2}+(2h)^{2})^{3/2}},
\label{eq:vst}
\end{equation}
where $F_{\text{st}}$ is the Stokeslet strength, $\eta$ is the fluid viscosity, and $h$ is the distance below the fluid surface of the suspended embryo. We assume $h=R$ where $R$ is the radius of the embryo. Equation (\ref{eq:vst}) is the sum of two mirrored Stokeslet flow fields below and above the fluid surface at distance $h$. The unit vector in the $z$ direction, $\mathbf{e}_{z}$, is from the expression for the original Stokeslet flow field but does not appear in the explicit expression for the mirror sum that gives Eq.~(\ref{eq:vst}).
In our simulations, we use the non-dimensionalized version of Eq.~(\ref{eq:starfish_eom}), and Eq.~(\ref{eq:vst}) then becomes
\begin{equation}
\bar{\mathbf{v}}_{\text{st}}(\mathbf{r}_{i};\mathbf{e}_{z},\mathbf{r}_{j}) = -\frac{\tilde{F}_{\text{st}}}{8\pi}\frac{\tilde{\mathbf{r}}_{i}-\tilde{\mathbf{r}}_{j}}{(\lvert \tilde{\mathbf{r}}_{i}-\tilde{\mathbf{r}}_{j} \rvert^{2} + 1)^{3/2}},
\end{equation}
where we have used $\mathbf{r}_{i} = 2R\tilde{\mathbf{r}}_{i}$ and $F_{\text{st}}/\eta = (2R)^{2}\tilde{F}_{\text{st}}$ for rescaling.

The second term in Eq.~(\ref{eq:starfish_eom}) represents the steric repulsion between embryos along the longitudinal direction, i.e.,
\begin{equation}
\mathbf{F}_{\text{rep}}(\lvert \mathbf{r}_{i}-\mathbf{r}_{j} \rvert) = 12f_{\text{rep}}\frac{R^{13}(\mathbf{r}_{i}-\mathbf{r}_{j})}{\lvert \mathbf{r}_{i}-\mathbf{r}_{j} \rvert^{14}}.
\end{equation}
Following the same non-dimensionalization for the Stokeslet flow, the second term in Eq.~\ref{eq:starfish_eom} becomes
\begin{equation}
\frac{1}{\eta R}\mathbf{F}_{\text{rep}}(\lvert \mathbf{r}_{i}-\mathbf{r}_{j} \rvert) = \frac{3}{2^{10}}\tilde{f}_{\text{rep}}\frac{\tilde{\mathbf{r}}_{i}-\tilde{\mathbf{r}}_{j}}{\lvert \tilde{\mathbf{r}}_{i}-\tilde{\mathbf{r}}_{j} \rvert^{14}},
\end{equation}
where $f_{\text{rep}}/\eta = (2R)^{2}\tilde{f}_{\text{rep}}$ has been applied for rescaling.

Lastly, the third term in Eq.~(\ref{eq:starfish_eom}) represents the transverse hydrodynamic near-field force whose amplitude is given by the lubrication theory \cite{o1970asymmetrical,kim2013microhydrodynamics}, leading to \cite{tan2022odd}
\begin{equation}
F_{\text{nf}}(\lvert \mathbf{r}_{i}-\mathbf{r}_{j} \rvert) = 
\begin{cases}
f_{0}\ln{\frac{d_{c}}{d_{ij}}}, & \text{if } d_{ij} < d_{c}\\
0, & \text{otherwise}
\end{cases}
\label{eq:Fnf}
\end{equation}
where $d_{ij} = \lvert \mathbf{r}_{i}-\mathbf{r}_{j} \rvert - 2R$, and $d_{c}$ sets the range within which the near-field transverse force applies. In Ref.~\cite{tan2022odd}, $d_{c} = 0.5R$ was used. However, in our simulations we set $d_{c}=R$ because the previous value was too short for the embryos to feel the effect of the transverse force. After the non-dimensionalization, the third term in Eq.~(\ref{eq:starfish_eom}) becomes
\begin{equation}
R(\omega_{i}+\omega_{j})F_{\text{nf}}(\lvert \mathbf{r}_{i}-\mathbf{r}_{j} \rvert)\hat{\mathbf{r}}_{ij}^{\perp}  = \frac{\omega_{i}+\omega_{j}}{2}f_{0}\ln\frac{d_{c}/2R}{\lvert \tilde{\mathbf{r}}_{i}-\tilde{\mathbf{r}}_{j} \rvert - 1}\hat{\mathbf{r}}_{ij}^{\perp},
\end{equation}
for $d_{ij}<d_{c}$, or in other words, $\lvert \tilde{\mathbf{r}}_{i}-\tilde{\mathbf{r}}_{j} \rvert - 1 < d_{c}/2R$.

Using the experimental values \cite{tan2022odd} of $R=0.11 \text{ mm}$, $F_{\text{st}}/\eta = 2.6 \text{ mm}^{2}/\text{s}$, and $f_{\text{rep}}/\eta = 38 \text{ mm}^{2}/\text{s}$, we get $\tilde{F}_{\text{st}} = 53.7$ and $\tilde{f}_{\text{rep}} = 785.1$ for our simulations. The dimensionless parameter $f_{0}$ is given as 0.06 in Ref.~\cite{tan2022odd}, but we use a larger value of $f_{0} = 1.0$ to make the system satisfy the wave condition by increasing the strength of transverse force. In Ref.~\cite{tan2022odd}, each embryo is initially set to self-spin at the average angular frequency $\omega_{0}/2\pi = 0.72$ Hz, but at $t>0$ the value of $\omega_{i}$ decreases and keeps varying inside the living crystal as a result of torque balance. In our simulations we fix the self-spinning frequency to be $\omega_{i} = \bar{\omega}_{0} = 1.0$ for all $i$, with $\bar{\omega}_{0}$ the emergent collective self-spinning frequency inside the crystal, as the modulation in $\omega_{i}$ is found to be very small and have negligible effect (see SM $\S$\ref{sec:wupdate}). Here we have rescaled the time to be dimensionless, via $\tilde{t} = t / \tau $ where $t$ is the real time and $\tau = 1$ s.

\section{Justification for constant self-circling frequency in the modeling}
\label{sec:wupdate}

In the previous models of chiral active living systems such as starfish embryos \cite{tan2022odd} and swimming bacteria \cite{petroff2015fast}, the self-spinning frequency of each individual agent varies with time and is determined by the balance between the torque exerted by the agent's cilia or flagella and the viscous torque. However, our model neglects the time evolution of the self-spinning frequency because the modulation in the frequency is too small. In this section, we present the time series of the self-spinning frequency, which has the same value as the self-circling frequency if self-propulsion is included, in the presence of torque balance to support the use of constant frequency in our model.

We follow the temporal dynamics of self-spinning frequency given in Refs.~\cite{tan2022odd} and \cite{petroff2015fast}. The system is overdamped, and thus the time derivative of the self-spinning frequency of the $i^{th}$ particle $d \omega_{i}/dt$ is neglected, as it is the second-order time derivative of the orientation angle of the agent. The temporal variation of $\omega_{i}$ is then governed by an instantaneous update due to torque balance instead of a time-differential equation, shown as \cite{tan2022odd}
\begin{equation}
\omega_{i} = \omega_{0}-\sum_{j \neq i}(\omega_{i}+\omega_{j})T_{\text{nf}}(\lvert \mathbf{r}_{i}-\mathbf{r}_{j} \rvert),
\label{eq:wupdate}
\end{equation}
where $i$ and $j$ label individual particles and the summation is over the neighbors of the $i^{th}$ particle. $\omega_{0}$ is the average self-spinning frequency of a single particle, which is a constant, $\mathbf{r}_{i}$ is the position of the $i^{th}$ particle, and $T_{\text{nf}}(\lvert \mathbf{r}_{i}-\mathbf{r}_{j} \rvert)$ is a function of the distance between two particles whose mathematical form is determined by the lubrication theory \cite{o1970asymmetrical,kim2013microhydrodynamics} and given as \cite{tan2022odd}
\begin{equation}
T_{\text{nf}}(\lvert \mathbf{r}_{i}-\mathbf{r}_{j} \rvert) = 
\begin{cases}
\tau_{0}\ln{\frac{d_{c}}{d_{ij}}}, & \text{if } d_{ij} < d_{c}\\
0, & \text{otherwise}
\end{cases}
\label{eq:Tnf}
\end{equation}
where $\tau_{0}$ is a constant coefficient, $d_{ij}=\lvert \mathbf{r}_{i}-\mathbf{r}_{j} \rvert -2R$, and $d_{c}$ sets the range within which the near-field interaction between spinning applies. As for the transverse force in SM $\S$\ref{sec:starfish_model}, we set $d_{c}=R$ instead of $d_{c}=0.5R$ used in Ref.~\cite{tan2022odd} because this previous value was too short for the embryos to feel the near-field contribution. 
Equation (\ref{eq:wupdate}) can be rewritten as a matrix form $\bm{\omega} = \bm{\omega}_{0} - M\bm{\omega}$, where $\bm{\omega}$ is an $N$-dimensional vector with elements $\omega_{i}$ for each particle from $i=1$ to $i=N$ ($N$ being the particle number), $\bm{\omega}_{0}$ is an $N$-dimensional vector with all of its elements being $\omega_{0}$, and $M$ is an $N \times N$ matrix with elements $M_{ij} = T_{\text{nf}}(\lvert \mathbf{r}_{i}-\mathbf{r}_{j} \rvert)$ for $i \neq j$ and $M_{ii} = \sum_{j \neq i}M_{ij}$. Then the self-spinning frequency at a given time is determined as
\begin{equation}
\bm{\omega}(t) = (\mathbb{I}+M(t))^{-1}\bm{\omega}_{0},
\label{eq:wupdate_invM}
\end{equation}
where $\mathbb{I}$ is an identity matrix. Because the position of each particle depends on time, the elements of the matrix $M$ are time dependent as well, which leads to the change of the particle self-circling frequency $\omega_{i}$ with time.

\begin{figure}[!htb]
   \includegraphics[width=\columnwidth]{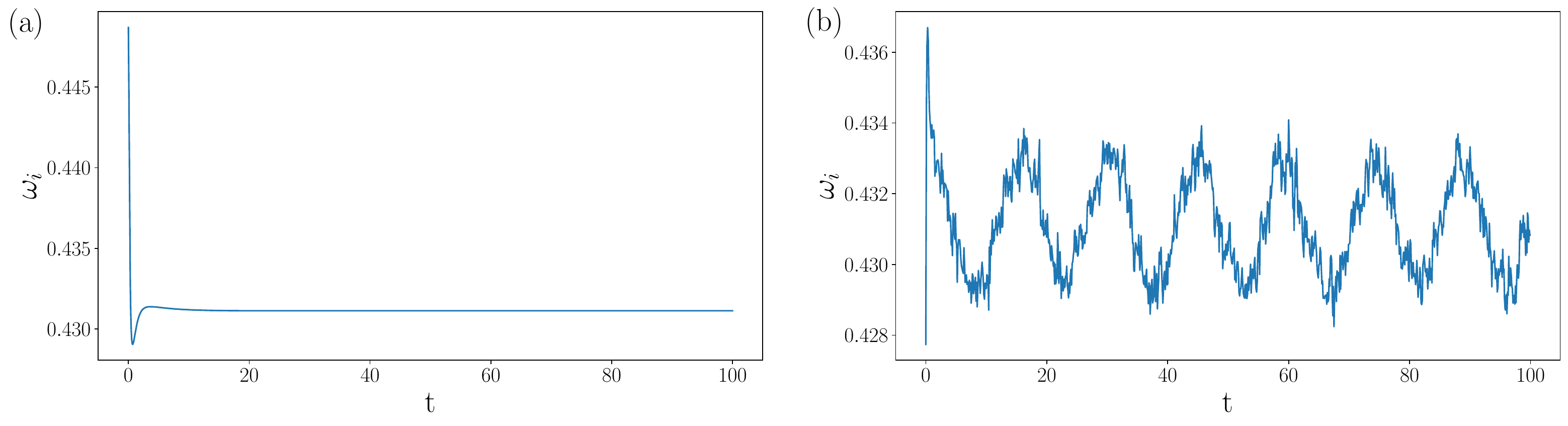}
   \caption{Time series of self-spinning frequency $\omega_{i}$ of a particle in the starfish embryo model (a) without self-propulsion and (b) with self-propulsion at $\bar{v}_0=0.01$ and $v_{\sigma} = 0.1$. 
  \label{fig:wupdate}}
\end{figure}

Figure~\ref{fig:wupdate} shows the value of self-spinning frequency of one particle as time progresses. The simulations of the starfish embryo model (SM \S\ref{sec:starfish_model}) were conducted on a $20 \times 20$ two-dimensional triangular lattice with periodic boundary conditions for $10^{5}$ time steps with $dt = 0.001$ (recorded every 100 time steps so that the measured $\Delta t=0.1$). The parameter values used are the same as those given in SM \S\ref{sec:starfish_model}: $\tilde{F}_{\text{st}} = 53.7$ for the longitudinal attractive force, $\tilde{f}_{\text{rep}} = 785.1$ for the longitudinal repulsive force, $f_{0} = 1.0$ for the transverse force, and $\omega_{0} = 1.0$ for the initial self-spinning frequency for all particles. For the spinning frequency update, $\tau_{0}=0.12$ was used in Eq.~(\ref{eq:Tnf}), which is the same value used in Ref.~\cite{tan2022odd}. The results presented are from one run of the simulation for one of the particles, but the results are qualitatively the same for different runs of the simulation for all particles. 

Figure~\ref{fig:wupdate}(a) gives the time evolution of $\omega_{i}$ without self-propulsion, for which the Euler algorithm has been used in solving the model equations. Since self-propulsion is not included in this particular simulation, the corresponding model is identical to that in Ref.~\cite{tan2022odd} although with different parameter values. Like the result presented in Ref.~\cite{tan2022odd}, although the initial value was $\omega_{i}=\omega_{0}=1.0$, it first rapidly decreases because of the interaction between neighboring particles governed by Eq.~(\ref{eq:wupdate}), before saturating to a constant value. Here our focus is not on the early-time decrease in $\omega_{i}$ but on the observation that the value of $\omega_{i}$ does not change in later time. 

Figure~\ref{fig:wupdate}(b) shows the result with the presence of self-propulsion and hence self-circling, thus corresponding to our new model presented in this Letter. Given the effect of particle self-propulsion, $\omega_{i}$ can now be called self-circling frequency, which has the same value as the self-spinning frequency. As in the simulations presented in the main text, the mean self-propulsion strength $\bar{v}_{0} = 0.01$ and the noise strength $v_{\sigma} = 0.1$, and the Euler-Maruyama algorithm \cite{maruyama1953markov,maruyama1955continuous} has been used. As in the deterministic case shown in Fig.~\ref{fig:wupdate}(a), $\omega_{i}$ first decreases from its initial value 1.0. Because of the self-propulsion, the particles keep changing their positions and collide with each other. This leads to constant modulation of relative distances between the particles, and hence $\omega_{i}$ does not converge to a constant value but oscillates with time around a mean value. As shown in Fig.~\ref{fig:wupdate}(b), the amplitude of the self-spinning frequency oscillation is at the order of $10^{-3}$ and is about $1\%$ of the mean value. In other words, the time variations of $\omega_{i}$ are very small.

Thus results in Fig.~\ref{fig:wupdate} demonstrate that, in both the absence and the presence of self-propulsion, the change in $\omega_{i}$ due to torque balance is very small and not significant at later time. As depicted in Eq.~(\ref{eq:wupdate_invM}), the instantaneous update of $\omega_{i}$ at every time step involves the calculation of an $N \times N$ matrix inversion (where $N$ is the number of particles), which is computationally demanding. Therefore, in this work, instead of performing computationally expensive tasks without significant effect, we assume $\omega_{i}$ to be constant in our modeling and simulations.

\section{Fitting procedure to identify the peak locations of current correlation functions for the dispersion relations}
\label{sec:fitting}

This section illustrates how we identify the frequency $\omega$ that maximizes the current correlation function for each wave vector $\mathbf{q}$, as used to plot the dispersion curves.

\begin{figure}[!htb]
   \includegraphics[width=0.72\columnwidth]{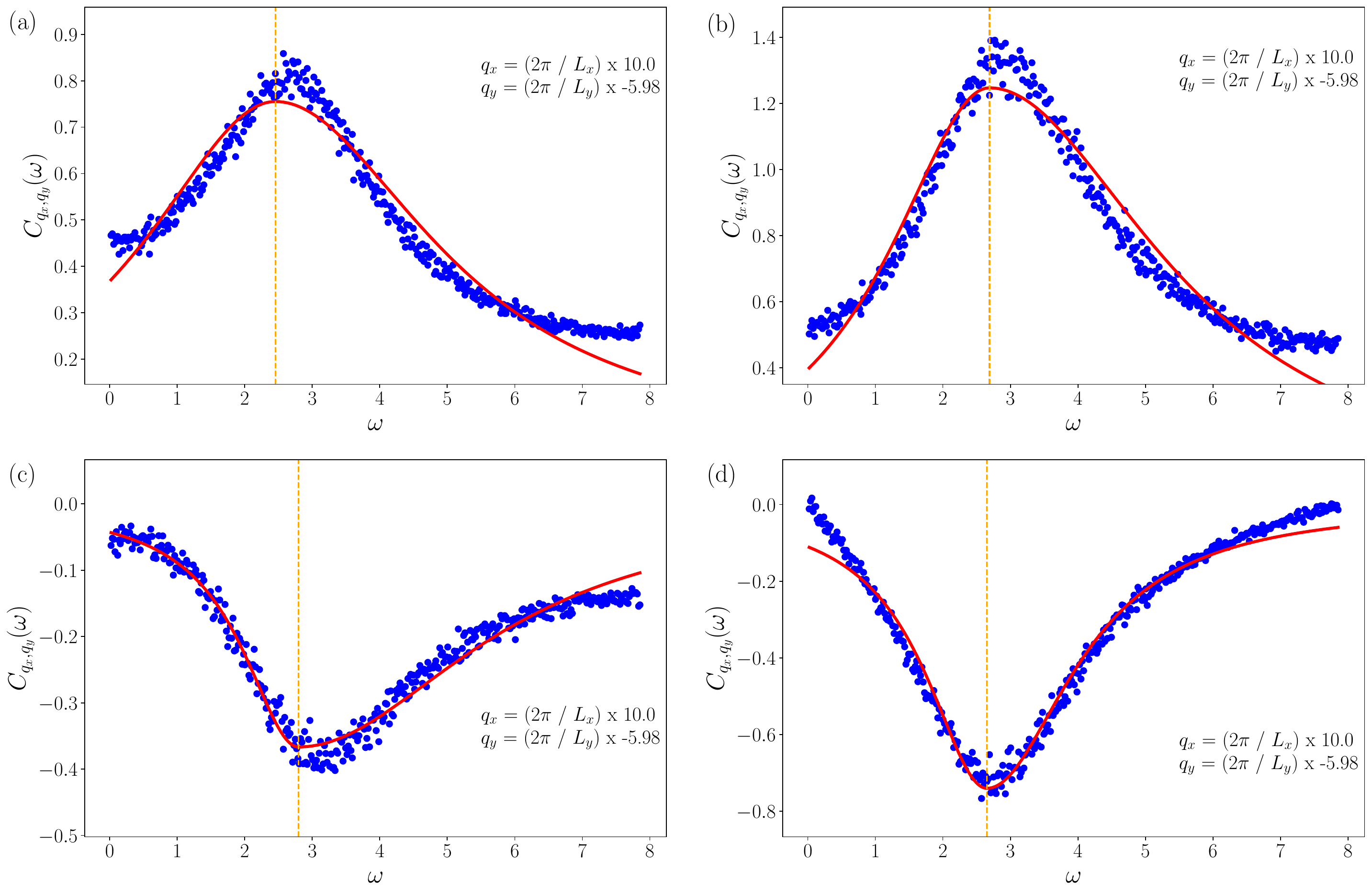}
   \caption{Current correlation function as a function of $\omega$ at a given wave vector $\mathbf{q}$, as calculated from the  simulation data of the toy model with $k_{T}=1.0$ and $k_{L}=0.5$, for (a) $C_{LL}(\mathbf{q},\omega)$, (b) $C_{TT}(\mathbf{q},\omega)$, (c) Re[$C_{LT}(\mathbf{q},\omega)$], and (d) Im[$C_{LT}(\mathbf{q},\omega)$]. Each current correlation function is fitted with the skewed Cauchy distribution (red solid line) to identify the peak location (orange dashed line) of the distribution.
  \label{fig:Cfit_toy}}
\end{figure}

\begin{figure}[!htb]
   \includegraphics[width=0.72\columnwidth]{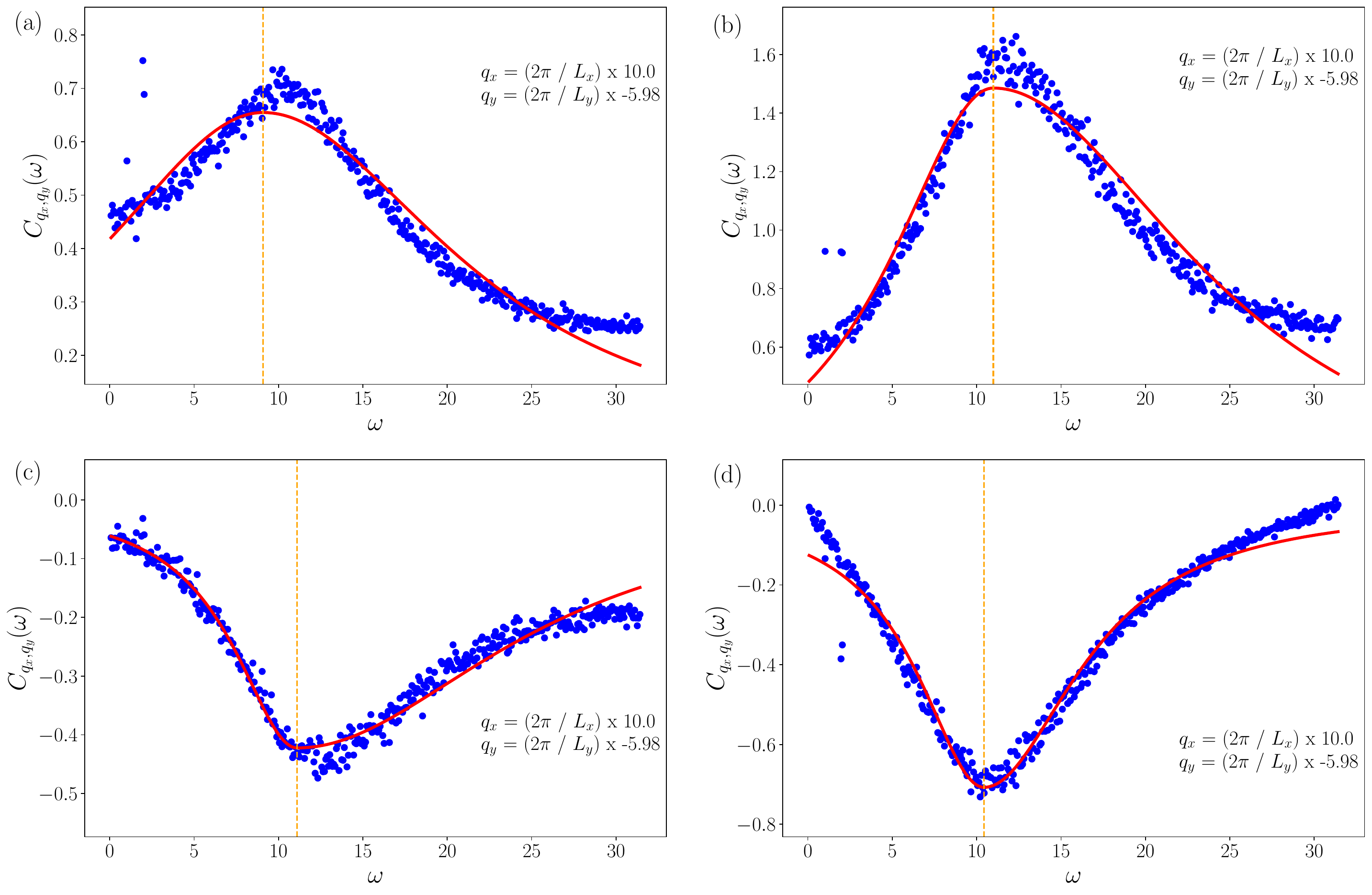}
   \caption{Current correlation function as a function of $\omega$ at a given wave vector $\mathbf{q}$, as calculated from the simulation data of the starfish embryo model with the same parameter values used to generate Fig.~1 of the main text, for (a) $C_{LL}(\mathbf{q},\omega)$, (b) $C_{TT}(\mathbf{q},\omega)$, (c) Re[$C_{LT}(\mathbf{q},\omega)$], and (d) Im[$C_{LT}(\mathbf{q},\omega)$]. Each current correlation function is fitted with the skewed Cauchy distribution (red solid line) to identify the peak location (orange dashed line) of the distribution.
  \label{fig:Cfit_starfish}}
\end{figure}

Each current correlation function, $C_{LL}(\mathbf{q},\omega)$, $C_{TT}(\mathbf{q},\omega)$, Re[$C_{LT}(\mathbf{q},\omega)$], or Im[$C_{LT}(\mathbf{q},\omega)$] at each $\mathbf{q}$, shows a skewed distribution as a function of $\omega$ as seen in Figs.~\ref{fig:Cfit_toy} and \ref{fig:Cfit_starfish}. We fit these peaks with a skewed Cauchy distribution 
\begin{equation}
f(\omega) = \frac{1}{\pi\sigma}\Bigg[ 1+\frac{(\omega-\omega_{m})^{2}}{\sigma^{2}(1+\lambda \text{ sgn}(\omega-\omega_{m}))^{2}} \Bigg]^{-1},
\label{eq:skewedCauchy}
\end{equation}
where $\omega_{m}$ is the peak location of the distribution, $\sigma > 0$ is a scale parameter, and $-1< \lambda <1$ is the skewness parameter. We use this skewed Cauchy distribution because the Cauchy distribution, usually called the Lorentzian distribution in physics, is commonly used to depict the spectral distribution, and our distributions here are skewed. Note that finding the physical origin or the rigorous derivation of the skewed Cauchy form of the current correlation function is beyond the scope of this work. We are interested in extracting the frequency value at the maximum or peak of each current correlation function. 

The fitting is done with a Python function \texttt{scipy.optimize.curve\char`_fit} after manually defining the skewed Cauchy distribution according to Eq.~(\ref{eq:skewedCauchy}). Then we collect the value of fitting parameter $\omega_{m}$ to identify the peak location, which is the frequency value maximizing the corresponding current correlation function for a given $\mathbf{q}$. These collected frequency values of the peaks make up the dispersion relations $\omega(\mathbf{q})$.

\section{The structure of the real part of cross current correlation}
\label{sec:reClt}

This section discusses the reason of large deviations seen in the data points of the dispersion curves obtained from Re[$C_{LT}(\mathbf{q},\omega)$] in Fig.~1 of the main text. 

\begin{figure}[!htb]
   \includegraphics[width=0.8\columnwidth]{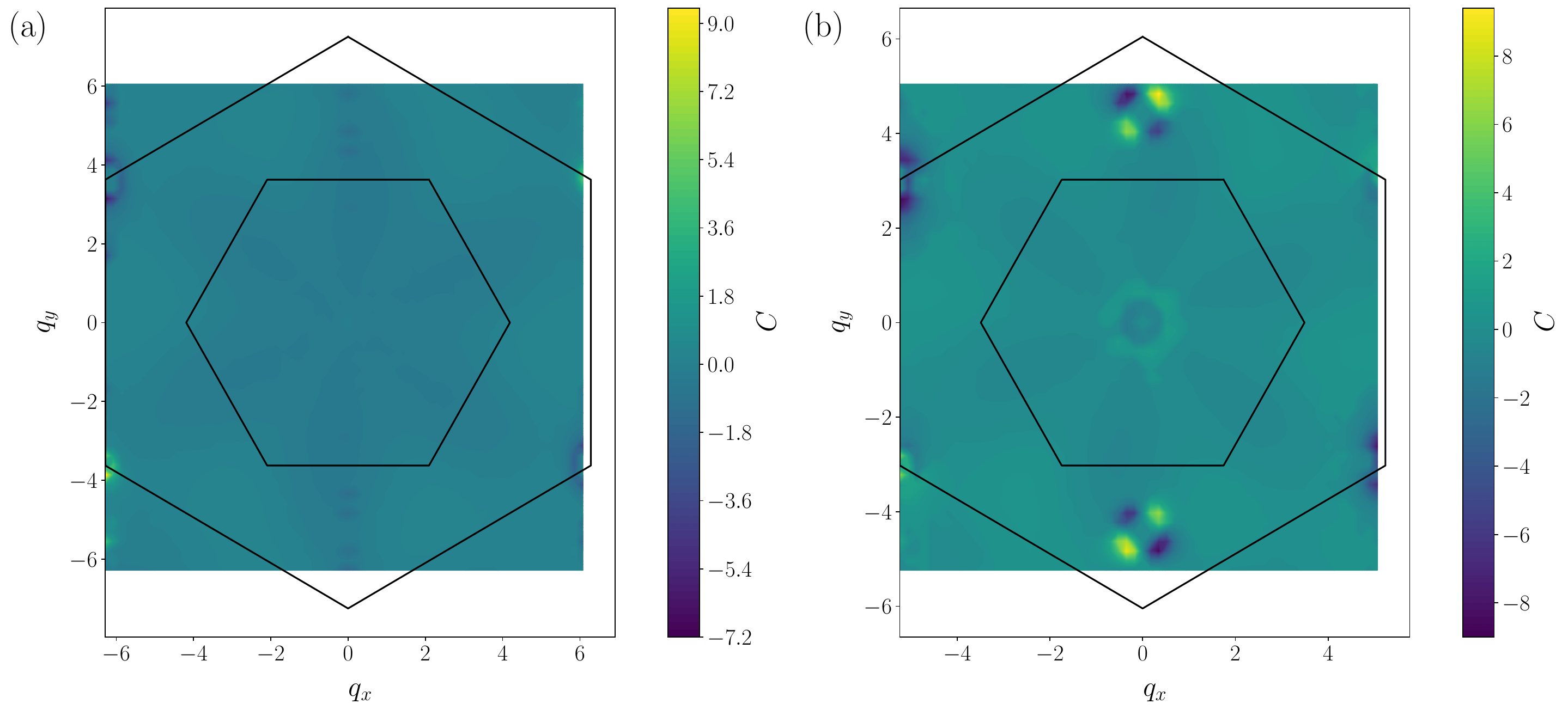}
   \caption{Contour plot showing the maximum peak value of Re[$C_{LT}(\mathbf{q},\omega)$] at each $\mathbf{q}$ for (a) the toy model and (b) the starfish embryo model. The outer hexagon is the reciprocal lattice, and the inner hexagon is the first Brillouin zone. Because of large dominant values of Re[$C_{LT}(\mathbf{q},\omega)$] in some regions, the structure inside the Brillouin zone is blurry.
  \label{fig:contour_Cltr_unclear}}
\end{figure}

To see the overall structure of Re[$C_{LT}(\mathbf{q},\omega)$], we draw its contour plot in the Fourier space, as shown in Fig.~\ref{fig:contour_Cltr_unclear} for both toy model and starfish embryo model. At each wave vector $\mathbf{q}$, we choose the value of Re[$C_{LT}(\mathbf{q},\omega)$] with the largest magnitude, which corresponds to the height of the peak shown in Fig.~\ref{fig:Cfit_toy} or Fig.~\ref{fig:Cfit_starfish}, for example. Then we plot these peak values of Re[$C_{LT}(\mathbf{q},\omega)$] as the contour plots in Fig.~\ref{fig:contour_Cltr_unclear}, where 
the inner hexagon depicts the boundary of the first Brillouin zone while the outer hexagon shows the reciprocal lattice. Because of large dominant values of the current correlation in some regions of the $q$-space, the structure inside the Brillouin zone looks blurry.

\begin{figure}[!htb]
   \includegraphics[width=0.8\columnwidth]{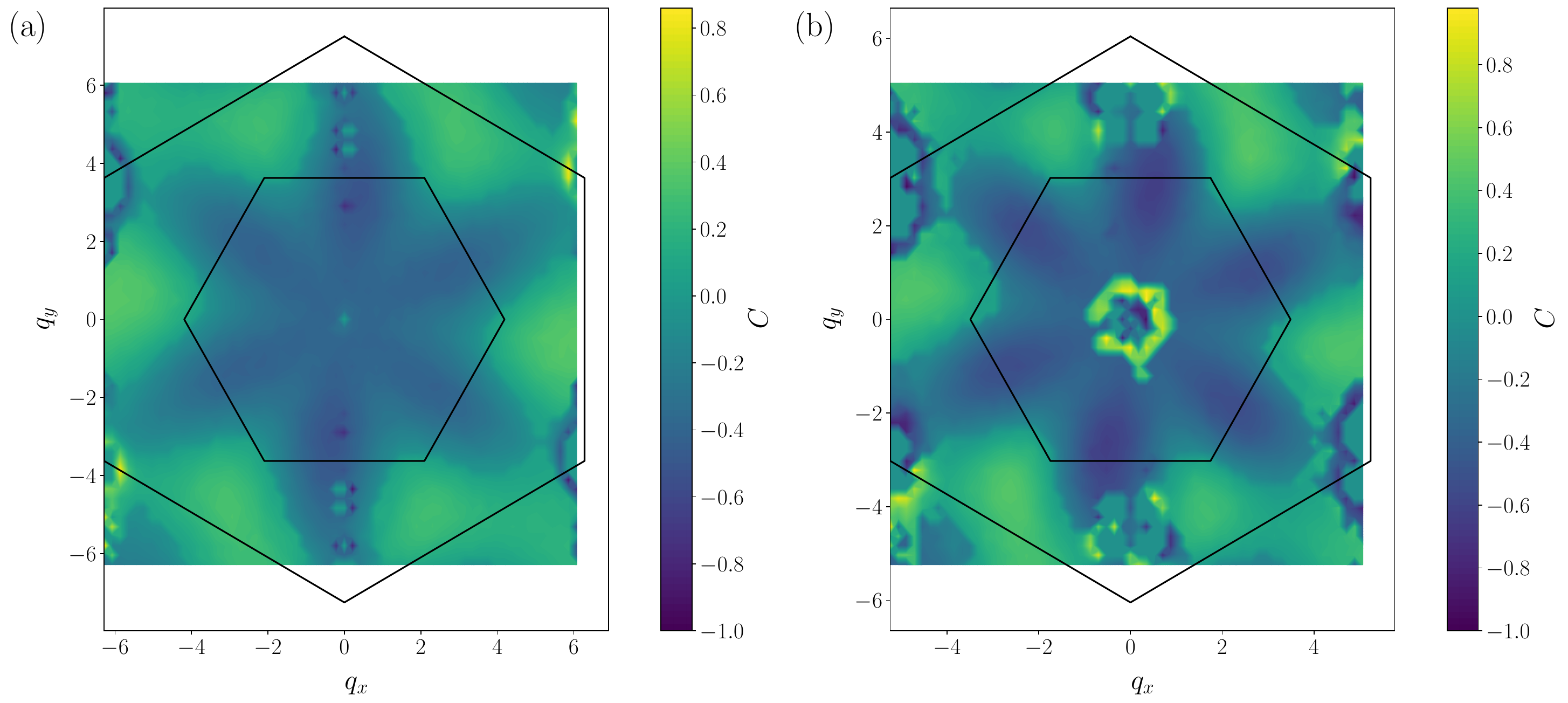}
   \caption{Contour plot showing the maximum peak value of Re[$C_{LT}(\mathbf{q},\omega)$] at each $\mathbf{q}$ for (a) the toy model and (b) the starfish embryo model. The outer hexagon is the reciprocal lattice, and the inner hexagon is the first Brillouin zone. The Re[$C_{LT}(\mathbf{q},\omega)$] values that are larger than 1 or smaller than -1 have been set to zero to make the structure inside the Brillouin zone more visible. 
  \label{fig:contour_Cltr}}
\end{figure}

To get a better illustration for the structure of the current correlation inside the first Brillouin zone, we set the value of Re[$C_{LT}(\mathbf{q},\omega)$] to be zero if it is smaller than -1 or bigger than 1. The resulting Fig.~\ref{fig:contour_Cltr} shows a clear radial region of hexagonal symmetry with negative values of Re[$C_{LT}(\mathbf{q},\omega)$] and the other regions with positive values of Re[$C_{LT}(\mathbf{q},\omega)$]. The hexagonal symmetry shown in this structure stems from the underlying triangular lattice of the crystal. In the regions around the $K$ points, which are the vertices of the first Brillouin zone (the inner hexagon in Fig.~\ref{fig:contour_Cltr}), the sign of Re[$C_{LT}(\mathbf{q},\omega)$] changes, i.e., values of Re[$C_{LT}(\mathbf{q},\omega)$] are close to zero around the $K$ points.  

\begin{figure}[!htb]
   \includegraphics[width=\columnwidth]{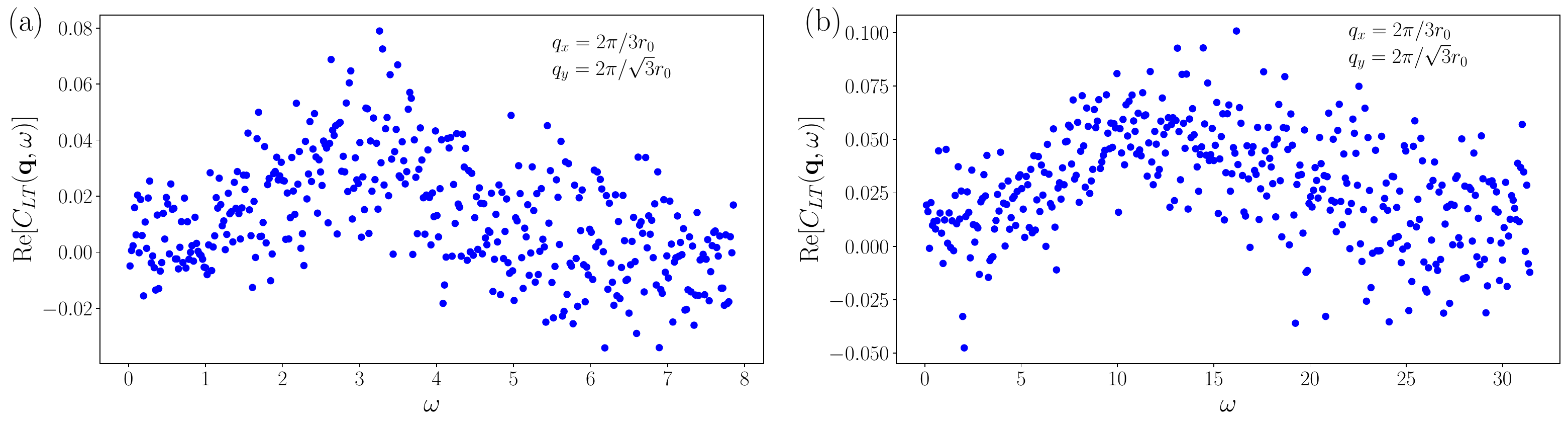}
   \caption{Re[$C_{LT}(\mathbf{q},\omega)$] as a function of $\omega$ at a given $\mathbf{q}$ of the $K$ point, for (a) the toy model and (b) the starfish embryo model. $\mathbf{q}$ represents the location of the $K$ point on the $q$-space with $r_{0}$ being the lattice spacing. For both models, the values of Re[$C_{LT}(\mathbf{q},\omega)$] are near zero for all values of $\omega$, and the data are noisy, which makes it difficult to identify a peak location. 
  \label{fig:Cw_Cltr_K}}
\end{figure}

Figure \ref{fig:Cw_Cltr_K} shows Re[$C_{LT}(\mathbf{q},\omega)$] as a function of $\omega$ at the $K$ point. For both models, the values of this current correlation function are close to zero for all values of $\omega$, and the large fluctuation makes it difficult to pinpoint the location of a peak. The comparison of Fig.~\ref{fig:Cw_Cltr_K} with Fig.~\ref{fig:Cfit_toy}(c) and Fig.~\ref{fig:Cfit_starfish}(c) shows a clear contrast, as clear peaks of Re[$C_{LT}(\mathbf{q},\omega)$] are present in the latter cases. This observation explains why the result of $\omega$ maximizing Re[$C_{LT}(\mathbf{q},\omega)$] around the $K$ point gives large deviations for the corresponding dispersion relations shown in Fig.~1 of the main text.

\section{Derivation of the stochastic velocity correlation functions}
\label{sec:analytical_vv}

In this section, we derive the general equation for the velocity correlation function in the presence of noise. We use the results here to obtain the approximate analytical result for the dispersion relations, as the analytical calculation for the current correlation function is nontrivial (see SM $\S$\ref{sec:flat}).

We start with the stochastic linearized equation of motion for the displacement field. In Fourier space the Langevin equation becomes
\begin{equation}
    -i\omega \mathbf{u}(\mathbf{q},\omega)=M(\mathbf{q})\mathbf{u}(\mathbf{q},\omega) + \bm{\xi}(\mathbf{q},\omega),
    \label{eq:u}
\end{equation}
where $M(\mathbf{q})$ is the dynamic matrix and $\bm{\xi}(\mathbf{q},\omega)$ is the Fourier-transformed noise. Each element of the dynamic matrix is calculated as
\begin{equation}
\begin{aligned}
M_{11}(\mathbf{q}) &= -k_{L}g_{1}(\mathbf{q})-k_{T}g_{2}(\mathbf{q}),
&
M_{12}(\mathbf{q}) &= -k_{L}g_{2}(\mathbf{q})-k_{T}g_{3}(\mathbf{q}),
\\
M_{21}(\mathbf{q}) &= -k_{L}g_{2}(\mathbf{q})+k_{T}g_{1}(\mathbf{q}),
& 
M_{22}(\mathbf{q}) &= -k_{L}g_{3}(\mathbf{q})+k_{T}g_{2}(\mathbf{q}),
\end{aligned}
\label{eq:Mab}
\end{equation}
where $k_{L}$ is the effective longitudinal spring constant and $k_{T}$ is the effective transverse spring constant, and
\begin{equation}
\begin{aligned}
g_{1}(\mathbf{q}) &= 3-2\cos{(q_{x}r_{0})} - \cos{\bigg(\frac{q_{x}r_{0}}{2}\bigg)}\cos{\bigg(\frac{\sqrt{3}q_{y}r_{0}}{2}\bigg)}, \\
g_{2}(\mathbf{q}) &= \sqrt{3}\sin{\bigg( \frac{q_{x}r_{0}}{2} \bigg)}\sin{\bigg( \frac{\sqrt{3}q_{y}r_{0}}{2} \bigg)}, \\
g_{3}(\mathbf{q}) &= 3-3\cos{\bigg(\frac{q_{x}r_{0}}{2}\bigg)}\cos{\bigg(\frac{\sqrt{3}q_{y}r_{0}}{2}\bigg)},
\end{aligned}
\label{eq:gi}
\end{equation}
for a two-dimensional triangular lattice. Here $r_{0}$ is the equilibrium lattice spacing.
Note that setting $k_{T}=0$ in Eq.~(\ref{eq:Mab}) gives the dynamic matrix for a regular spring system with a triangular lattice. Equation (\ref{eq:gi}) is the result in the $xy$-basis. However, we are working in the basis of the longitudinal and transverse directions. The change of basis gives each element of the dynamic matrix as
\begin{equation}
\begin{aligned}
M_{LL}(\mathbf{q}) &= M_{11}(\mathbf{q})\frac{q_{x}^{2}}{q^{2}}+\big(M_{12}(\mathbf{q})+M_{21}(\mathbf{q})\big)\frac{q_{x}q_{y}}{q^{2}}+M_{22}(\mathbf{q})\frac{q_{y}^{2}}{q^{2}},
\\
M_{LT}(\mathbf{q}) &= M_{12}(\mathbf{q})\frac{q_{x}^{2}}{q^{2}}+\big(-M_{11}(\mathbf{q})+M_{22}(\mathbf{q})\big)\frac{q_{x}q_{y}}{q^{2}}-M_{21}(\mathbf{q})\frac{q_{y}^{2}}{q^{2}},
\\
M_{TL}(\mathbf{q}) &= M_{21}(\mathbf{q})\frac{q_{x}^{2}}{q^{2}}+\big(-M_{11}(\mathbf{q})+M_{22}(\mathbf{q})\big)\frac{q_{x}q_{y}}{q^{2}}-M_{12}(\mathbf{q})\frac{q_{y}^{2}}{q^{2}},
\\
M_{TT}(\mathbf{q}) &= M_{22}(\mathbf{q})\frac{q_{x}^{2}}{q^{2}}-\big(M_{12}(\mathbf{q})+M_{21}(\mathbf{q})\big)\frac{q_{x}q_{y}}{q^{2}}+M_{11}(\mathbf{q})\frac{q_{y}^{2}}{q^{2}},
\end{aligned}
\label{eq:Mltab}
\end{equation}
with the longitudinal and transverse displacement fields
\begin{align}
u_{L}(\mathbf{q},\omega) &= u_{x}(\mathbf{q},\omega)\frac{q_{x}}{q} + u_{y}(\mathbf{q},\omega)\frac{q_{y}}{q} \label{eq:uldef}, \\
u_{T}(\mathbf{q},\omega) &= u_{y}(\mathbf{q},\omega)\frac{q_{x}}{q} - u_{x}(\mathbf{q},\omega)\frac{q_{y}}{q} \label{eq:utdef},
\end{align}
where $u_{x}(\mathbf{q},\omega)=u_{1}(\mathbf{q},\omega)$ and $u_{y}(\mathbf{q},\omega)=u_{2}(\mathbf{q},\omega)$ following the matrix index notation.

In the elastodynamical equation (\ref{eq:u}), the noise vector undergoes the same change of basis as Eqs.~(\ref{eq:uldef}) and (\ref{eq:utdef}), with longitudinal and transverse noise components $\xi_L(\mathbf{q},\omega)$ and $\xi_T(\mathbf{q},\omega)$. To solve this equation of motion, we follow the procedure similar to that in Ref.~\cite{mckane2005predator} for population dynamics, giving
\begin{equation}
\begin{pmatrix}
u_{L} \\
u_{T}
\end{pmatrix}
=
\frac{1}{\det(-i\omega\mathbb{I}-M)}
\begin{pmatrix}
-i\omega-M_{TT} & M_{LT} \\
M_{TL} & -i\omega-M_{LL}
\end{pmatrix}
\begin{pmatrix}
\xi_{L} \\
\xi_{T}
\end{pmatrix},
\label{eq:usol}
\end{equation}
where $\mathbb{I}$ is the identity matrix. 

Since the velocity is the time derivative of the displacement, we get the relation between the velocity correlation function and the displacement correlation function as
\begin{equation}
\langle v_{\alpha}^{*}(\mathbf{q},\omega)v_{\beta}(\mathbf{q},\omega) \rangle = \omega^{2}\langle u_{\alpha}^{*}(\mathbf{q},\omega)u_{\beta}(\mathbf{q},\omega) \rangle.
\end{equation}
Therefore, from Eq.~(\ref{eq:usol}) for the solution of displacement fields, we can obtain the analytic results for the velocity correlation function, i.e.,
\begin{equation}
\begin{split}
\langle v_{L}^{*}(\mathbf{q},\omega)v_{L}(\mathbf{q},\omega) \rangle =& \frac{\omega^{2}}{\lvert \det(-i\omega\mathbb{I}-M(\mathbf{q})) \rvert^{2}}
\Big\{\big(\omega^{2}+M_{TT}^{2}(\mathbf{q})\big)\langle \xi_{L}^{*}(\mathbf{q},\omega)\xi_{L}(\mathbf{q},\omega) \rangle + M_{LT}^{2}(\mathbf{q})\langle \xi_{T}^{*}(\mathbf{q},\omega)\xi_{T}(\mathbf{q},\omega) \rangle \\
&+ 2\text{Re}\big[ \big(i\omega-M_{TT}(\mathbf{q})\big)M_{LT}(\mathbf{q})\langle \xi_{L}^{*}(\mathbf{q},\omega)\xi_{T}(\mathbf{q},\omega) \rangle \big] \Big\},
\end{split}
\label{eq:vlvl}
\end{equation}
\begin{equation}
\begin{split}
\langle v_{T}^{*}(\mathbf{q},\omega)v_{T}(\mathbf{q},\omega) \rangle =& \frac{\omega^{2}}{\lvert \det(-\omega\mathbb{I}-M(\mathbf{q})) \rvert^{2}}
\Big\{M_{TL}^{2}(\mathbf{q}) \langle \xi_{L}^{*}(\mathbf{q},\omega)\xi_{L}(\mathbf{q},\omega) \rangle + \big(\omega^{2}+M_{LL}^{2}(\mathbf{q})\big)\langle \xi_{T}^{*}(\mathbf{q},\omega)\xi_{T}(\mathbf{q},\omega) \rangle \\
&+ 2\text{Re}\big[ \big(-i\omega-M_{LL}(\mathbf{q})\big)M_{TL}(\mathbf{q})\langle \xi_{L}^{*}(\mathbf{q},\omega)\xi_{T}(\mathbf{q},\omega) \rangle \big] \Big\},
\end{split}
\label{eq:vtvt}
\end{equation}
\begin{equation}
\begin{split}
\langle v_{L}^{*}(\mathbf{q},\omega)v_{T}(\mathbf{q},\omega) \rangle =& \frac{\omega^{2}}{\lvert \det(-i\omega\mathbb{I}-M(\mathbf{q})) \rvert^{2}}
\Big\{ \big(i\omega-M_{TT}(\mathbf{q})\big)M_{TL}(\mathbf{q})\langle \xi_{L}^{*}(\mathbf{q},\omega)\xi_{L}(\mathbf{q},\omega) \rangle \\
&+ \big( -i\omega-M_{LL}(\mathbf{q}) \big)M_{LT}(\mathbf{q})\langle \xi_{T}^{*}(\mathbf{q},\omega)\xi_{T}(\mathbf{q},\omega) \rangle
+ M_{LT}(\mathbf{q})M_{TL}(\mathbf{q})\langle \xi_{T}^{*}(\mathbf{q},\omega)\xi_{L}(\mathbf{q},\omega) \rangle \\
&+ \big( \omega^{2} + M_{LL}(\mathbf{q})M_{TT}(\mathbf{q}) - i\omega \big( M_{LL}(\mathbf{q}) - M_{TT}(\mathbf{q}) \big) \big)\langle \xi_{L}^{*}(\mathbf{q},\omega)\xi_{T}(\mathbf{q},\omega) \rangle \Big\},
\end{split}
\label{eq:vlvt}
\end{equation}
\begin{equation}
\begin{split}
\langle v_{T}^{*}(\mathbf{q},\omega)v_{L}(\mathbf{q},\omega) \rangle =& \frac{\omega^{2}}{\lvert \det(-i\omega\mathbb{I}-M(\mathbf{q})) \rvert^{2}}
\Big\{ \big(-i\omega-M_{TT}(\mathbf{q})\big)M_{TL}(\mathbf{q})\langle \xi_{L}^{*}(\mathbf{q},\omega)\xi_{L}(\mathbf{q},\omega) \rangle \\
&+ \big( i\omega-M_{LL}(\mathbf{q}) \big)M_{LT}(\mathbf{q})\langle \xi_{T}^{*}(\mathbf{q},\omega)\xi_{T}(\mathbf{q},\omega) \rangle 
+ M_{LT}(\mathbf{q})M_{TL}(\mathbf{q})\langle \xi_{L}^{*}(\mathbf{q},\omega)\xi_{T}(\mathbf{q},\omega) \rangle \\
&+ \big( \omega^{2} + M_{LL}(\mathbf{q})M_{TT}(\mathbf{q}) + i\omega \big( M_{LL}(\mathbf{q}) - M_{TT}(\mathbf{q}) \big) \big)\langle \xi_{T}^{*}(\mathbf{q},\omega)\xi_{L}(\mathbf{q},\omega) \rangle \Big\}.
\end{split}
\label{eq:vtvl}
\end{equation}
These results of velocity correlation functions are general, and apply to systems with different forces or lattice structures and different types of noise $\bm{\xi}(\mathbf{q},\omega)$. For the dispersion results given in Fig.~1 of the main text, we assume Gaussian white noise for simplicity, but in principle other types of colored noise depending on $\mathbf{q}$ or $\omega$ can be used.
Like the current correlation functions, the cross velocity correlation functions shown in Eqs.~(\ref{eq:vlvt}) and (\ref{eq:vtvl}) are complex conjugates of each other. We thus use the real and imaginary parts of $\langle v_{L}^{*}(\mathbf{q},\omega)v_{T}(\mathbf{q},\omega) \rangle$ to calculate the corresponding dispersion relations.

\section{Derivation of zero velocity correlation function for certain paths}
\label{sec:revlvt_0}

Figure 1 in the main text shows the analytical results of dispersion relations obtained from each element of the velocity correlation function. However, for the real part of the cross velocity correlation, we only show the results along the path from $K$ to $M$ point of the first Brillouin zone. As will be derived in this section, the real part of the cross velocity correlation is zero from $\Gamma$ to $K$ point and from $M$ to $\Gamma$ point in the first Brillouin zone, and thus no corresponding dispersion relations exist.

The expression of $\text{Re}[\langle v_{L}^{*}(\mathbf{q},\omega)v_{T}(\mathbf{q},\omega) \rangle]$ is obtained from Eqs.~(\ref{eq:vlvt}) and (\ref{eq:vtvl}). For simplicity, Gaussian white noise is assumed, such that $\langle \xi_{L}^{*}(\mathbf{q},\omega)\xi_{T}(\mathbf{q},\omega) \rangle = \langle \xi_{T}^{*}(\mathbf{q},\omega)\xi_{L}(\mathbf{q},\omega) \rangle = 0$. This assumption allows us to set $\langle \xi_{L}^{*}(\mathbf{q},\omega)\xi_{L}(\mathbf{q},\omega) \rangle$ and $\langle \xi_{T}^{*}(\mathbf{q},\omega)\xi_{T}(\mathbf{q},\omega) \rangle$ to be a constant, chosen to be 1 without loss of generality, as it is just an overall constant factor that does not change the qualitative behavior of the velocity correlation. Then $\text{Re}[\langle v_{L}^{*}(\mathbf{q},\omega)v_{T}(\mathbf{q},\omega) \rangle]$ is simplified as
\begin{equation}
\text{Re}[\langle v_{L}^{*}(\mathbf{q},\omega)v_{T}(\mathbf{q},\omega) \rangle] = \frac{\omega^{2}}{\lvert \det(-i\omega\mathbb{I}-M(\mathbf{q})) \rvert^{2}}\Big\{ -M_{LL}(\mathbf{q})M_{LT}(\mathbf{q}) - M_{TL}(\mathbf{q})M_{TT}(\mathbf{q}) \Big\}.
\end{equation}
Using Eqs.~(\ref{eq:Mab}) and (\ref{eq:Mltab}) we have
\begin{equation}
\begin{split}
M_{LL}(\mathbf{q})M_{LT}(\mathbf{q}) + M_{TL}(\mathbf{q})M_{TT}(\mathbf{q}) =& \big( M_{11}(\mathbf{q})M_{12}(\mathbf{q})+M_{21}(\mathbf{q})M_{22}(\mathbf{q}) \big)\frac{q_{x}^{2}-q_{y}^{2}}{q^{2}} \\
&+ \big( -M_{11}^{2}(\mathbf{q})+M_{12}^{2}(\mathbf{q})-M_{21}^{2}(\mathbf{q})+M_{22}^{2}(\mathbf{q}) \big)\frac{q_{x}q_{y}}{q^{2}} \\
=& (k_{L}^{2} + k_{T}^{2})\big( g_{1}(\mathbf{q}) + g_{3}(\mathbf{q}) \big)\Bigg\{ g_{2}(\mathbf{q})\frac{q_{x}^{2}-q_{y}^{2}}{q^{2}} + \big( -g_{1}(\mathbf{q})+g_{3}(\mathbf{q}) \big)\frac{q_{x}q_{y}}{q^{2}} \Bigg\},
\end{split}
\label{eq:Revlvt_qdep}
\end{equation}
where the expressions of $g_{i}(\mathbf{q})$ are given in Eq.~(\ref{eq:gi}).

To evaluate Eq.~(\ref{eq:Revlvt_qdep}) in the first Brillouin zone, we choose the path connecting the $\Gamma$ point $(q_{x},q_{y})=(0,0)$, $K$ point $(q_{x},q_{y}) = \big(\frac{4\pi}{3r_{0}},0\big)$, and $M$ point $(q_{x},q_{y})=\big( \frac{\pi}{r_{0}},\frac{\pi}{\sqrt{3}r_{0}} \big)$, where $r_{0}$ is the lattice spacing in real space.
Along the path connecting the $\Gamma$ and $K$ points where $q_{y}=0$, the second term inside the bracket of Eq.~(\ref{eq:Revlvt_qdep}) is zero due to its linear dependence on $q_{y}$, and the first term is also zero because $g_{2}(\mathbf{q}) = \sqrt{3} \sin(q_xr_0/2)\sin{(\sqrt{3}q_{y}r_{0}/2)} =0$  (see Eq.~(\ref{eq:gi})). Thus $\text{Re}[\langle v_{L}^{*}(\mathbf{q},\omega)v_{T}(\mathbf{q},\omega) \rangle]=0$ on the path from $\Gamma$ to $K$ point.
On the path from $M$ to $\Gamma$ point, $q_{x} = \frac{\sqrt{3}}{2}q$ and $q_{y}=\frac{1}{2}q$. Substituting these values of $q_{x}$ and $q_{y}$ into Eq.~(\ref{eq:gi}), we calculate the terms inside the bracket of Eq.~(\ref{eq:Revlvt_qdep}) as
\begin{equation}
\begin{split}
g_{2}(\mathbf{q})\frac{q_{x}^{2}-q_{y}^{2}}{q^{2}} + \big( -g_{1}(\mathbf{q})+g_{3}(\mathbf{q}) \big)\frac{q_{x}q_{y}}{q^{2}} &= \sqrt{3}\sin^{2}\Bigg( \frac{\sqrt{3}}{4}qr_{0} \Bigg)\frac{1}{2}q^{2} + \Bigg\{ -5\sin^{2}\Bigg( \frac{\sqrt{3}}{4}qr_{0} \Bigg) + 3\sin^{2}\Bigg( \frac{\sqrt{3}}{4}qr_{0} \Bigg)\Bigg\} \frac{\sqrt{3}}{4}q^{2} \\
&= 0.
\end{split}
\end{equation}
Therefore, $\text{Re}[\langle v_{L}^{*}(\mathbf{q},\omega)v_{T}(\mathbf{q},\omega) \rangle]=0$ on the path from $M$ to $\Gamma$ point as well.

\section{Derivation of the criterion for noise-driven odd elastic waves}
\label{sec:wave_condi}

In this section, we derive the approximate criterion for the occurrence of persistent noise-driven odd elastic waves, by finding the condition for the existence of real $\omega$ that maximizes the velocity correlation functions. We use the velocity correlation functions since the analytical expressions for the current correlation functions are not available (see SM $\S$\ref{sec:flat}). The corresponding result is an approximation in the continuum limit, not meant to be exact. 

We first derive the wave criterion using the velocity correlation function $\langle \mathbf{v}^{*}(\mathbf{q},\omega)\cdot\mathbf{v}(\mathbf{q},\omega) \rangle = \langle v_{L}^{*}(\mathbf{q},\omega)v_{L}(\mathbf{q},\omega) \rangle + \langle v_{T}^{*}(\mathbf{q},\omega)v_{T}(\mathbf{q},\omega) \rangle$, as presented in the main text. As in the other calculations conducted in this work, we assume a Gaussian white noise. This leads to
\begin{equation}
\langle \mathbf{v}^{*}(\mathbf{q},\omega)\cdot\mathbf{v}(\mathbf{q},\omega) \rangle 
= \frac{\omega^{2}\Big\{ 2\omega^{2} + M_{LL}^{2}(\mathbf{q}) + M_{LT}^{2}(\mathbf{q}) + M_{TL}^{2}(\mathbf{q}) + M_{TT}^{2}(\mathbf{q}) \Big\}}{\lvert \det(-i\omega\mathbb{I}-M(\mathbf{q})) \rvert^{2}},
\label{eq:rep_vv}
\end{equation}
where we have set $\langle \xi_{L}^{*}(\mathbf{q},\omega)\xi_{L}(\mathbf{q},\omega) \rangle = \langle \xi_{T}^{*}(\mathbf{q},\omega)\xi_{T}(\mathbf{q},\omega) \rangle = 1$ without loss of generality.

For simplicity, we calculate the approximate criterion for the appearance of a persistent noise-driven wave in the continuum limit. Taking this limit also allows us to directly compare the criterion with the deterministic one given in Ref.~\cite{scheibner2020odd} in terms of odd elasticity.
In this limit, we express the elements of the dynamic matrix via elastic moduli, such that the stochastic elastodynamical equation is written as
\begin{equation}
-i\omega \begin{pmatrix} \tilde{u}_{L} \\ \tilde{u}_{T} \end{pmatrix} 
= -\frac{q^{2}}{\gamma}
\begin{pmatrix} 
B+\mu & K^{o} \\
-K^{o}-A & \mu
\end{pmatrix}
\begin{pmatrix} \tilde{u}_{L} \\ \tilde{u}_{T} \end{pmatrix} 
+ \begin{pmatrix} \xi_{L} \\ \xi_{T} \end{pmatrix},
\label{eq:Langevin_displacement}
\end{equation}
which is the equation of motion for the longitudinal and transverse displacement fields introduced in Ref.~\cite{scheibner2020odd} (see also SM $\S$\ref{sec:damping}) with the addition of noise. Substituting the elements of the dynamic matrix in Eq.~(\ref{eq:Langevin_displacement}) into Eq.~(\ref{eq:rep_vv}), we get
\begin{equation}
\langle \mathbf{v}^{*}(\mathbf{q},\omega)\cdot\mathbf{v}(\mathbf{q},\omega) \rangle =
\frac{2\omega^{4}+c_{1}\omega^{2}q^{4}}{(\omega^{2}-c_{2}q^{4})^{2}+c_{3}\omega^{2}q^{4}},
\label{eq:rep_vv_simp}
\end{equation}
where
\begin{equation}
\begin{aligned}
c_{1} &= \frac{1}{\gamma^{2}}\big\{ \mu^{2} + (B+\mu)^{2} + (K^{o})^{2} + (K^{o}+A)^{2} \big\}, \\
c_{2} &= \frac{1}{\gamma^{2}}\big\{ \mu(B+\mu) + K^{o}(K^{o}+A) \big\}, \\
c_{3} &= \frac{1}{\gamma^{2}}(B+2\mu)^{2}.
\end{aligned}
\label{eq:ci}
\end{equation}
To find the condition of $\omega$ that maximizes Eq.~(\ref{eq:rep_vv_simp}) at a given $\mathbf{q}$, we need (with $s\equiv \omega^{2}$)
\begin{align}
\frac{\partial}{\partial s}\bigg[ \langle \mathbf{v}^{*}(\mathbf{q},\omega)\cdot\mathbf{v}(\mathbf{q},\omega) \rangle \bigg] &= 0,
\label{eq:dds_repvv}
\\
\frac{\partial^{2}}{\partial^{2} s}\bigg[ \langle \mathbf{v}^{*}(\mathbf{q},\omega)\cdot\mathbf{v}(\mathbf{q},\omega) \rangle \bigg] &< 0,
\label{eq:d2d2s_repvv}
\end{align}
which yield the solution
\begin{equation}
s = \omega^{2} = \frac{2c_{2}^{2}+\sqrt{4c_{2}^{4}+(c_{1}+4c_{2}-2c_{3})c_{1}c_{2}^{2}}}{c_{1}+4c_{2}-2c_{3}}q^{4}.
\label{eq:ssol_repvv}
\end{equation}
This solution exists (i.e., $s$ is real and positive) if $c_{1}+4c_{2}-2c_{3}>0$, or equivalently
\begin{equation}
(K^{o})^{2}+(K^{o}+A)^{2}+4K^{o}(K^{o}+A)-\mu^{2}-(B+\mu)^{2}>0,
\label{eq:condi_repvv}
\end{equation}
after using Eq.~(\ref{eq:ci}) for the expressions of $c_{i}$ in terms of elastic moduli. This is Eq.~(5) in the main text.
Given the relation between the effective spring constants and elastic moduli \cite{scheibner2020odd}
\begin{equation}
\begin{aligned}
\mu &= \frac{\sqrt{3}}{4}k_{L},
&
B &= \frac{\sqrt{3}}{2}k_{L},
&
K^{o} &= \frac{\sqrt{3}}{4}k_{T},
&
A &= \frac{\sqrt{3}}{2}k_{T},
\end{aligned}
\label{eq:moduli_simp}
\end{equation}
we rewrite the criterion of Eq.~(\ref{eq:condi_repvv}) as 
\begin{equation}
\frac{k_{T}}{k_{L}} > \sqrt{\frac{5}{11}}.
\end{equation}
This threshold $\sqrt{5/11}$ is larger than $\sqrt{1/3}$ which is the threshold value between the no-wave regime and the damped wave regime calculated in Ref.~\cite{scheibner2020odd} for the deterministic case. This seemingly stricter condition may seem opposite to the results of the previous works \cite{butler2009robust,butler2011fluctuation,biancalani2011stochastic} which showed that the driving by fluctuation makes it possible to observe pattern formation \cite{butler2009robust,butler2011fluctuation,BiancalaniPRL17} or traveling waves \cite{biancalani2011stochastic} in the parameter regime where it was not possible to do so in the deterministic case. However, in our case the elastic wave is not observable in the deterministic case due to damping, and is observable only in the presence of noise. Therefore, our result is consistent with the previous findings \cite{butler2009robust,butler2011fluctuation,biancalani2011stochastic} in the sense that the area of the parameter space to observe a persistent wave increases from zero to a non-zero value.

We next extend our calculation of the criterion for the noise-driven odd elastic waves to all elements of the velocity correlation function. With white noise, the longitudinal velocity correlation function in Eq.~(\ref{eq:vlvl}) simplifies to
\begin{equation}
\langle v_{L}^{*}(\mathbf{q},\omega)v_{L}(\mathbf{q},\omega) \rangle = \frac{\omega^{2}\Big\{ \omega^{2} + M_{LT}^{2}(\mathbf{q}) + M_{TT}^{2}(\mathbf{q}) \Big\}}{\lvert \det(-i\omega\mathbb{I}-M(\mathbf{q})) \rvert^{2}}
= \frac{\omega^{4}+c_{1}\omega^{2}q^{4}}{(\omega^{2}-c_{2}q^{4})^{2}+c_{3}\omega^{2}q^{4}}.
\label{eq:vlvl_simp}
\end{equation}
Note that Eq.~(\ref{eq:vlvl_simp}) has essentially the same functional form as Eq.~(\ref{eq:rep_vv_simp}) except for the term $\omega^{4}$ instead of $2\omega^{4}$ in the numerator. 
Expressions of $c_{2}$ and $c_{3}$ remain the same as Eq.~(\ref{eq:ci}), but $c_{1}$ is now defined as
\begin{equation}
c_{1} = \frac{1}{\gamma^{2}}\big[ \mu^{2} + (K^{o})^{2} \big].
\end{equation}
The frequency that maximizes the longitudinal velocity correlation function in Eq.~(\ref{eq:vlvl_simp}) is given by
\begin{equation}
s = \omega^{2} = \frac{c_{2}^{2}+\sqrt{c_{2}^{4}+(c_{1}+2c_{2}-c_{3})c_{1}c_{2}^{2}}}{c_{1}+2c_{2}-c_{3}}q^{4},
\label{eq:ssol_vlvl}
\end{equation}
which is real and positive if $c_{1}+2c_{2}-c_{3}>0$, or equivalently
\begin{equation}
(K^{o})^{2}+2K^{o}(K^{o}+A)-(B+\mu)^{2}>0.
\end{equation}
Expressed in terms of the effective spring constants, this condition becomes
\begin{equation}
\frac{k_{T}}{k_{L}} > \frac{3}{\sqrt{7}}.
\label{eq:alpha_LL}
\end{equation}

In the presence of white noise, the transverse velocity correlation function in Eq.~(\ref{eq:vtvt}) simplifies to
\begin{equation}
\langle v_{T}^{*}(\mathbf{q},\omega)v_{T}(\mathbf{q},\omega) \rangle = \frac{\omega^{2}\Big\{ \omega^{2} + M_{LL}^{2}(\mathbf{q}) + M_{TL}^{2}(\mathbf{q}) \Big\}}{\lvert \det(-i\omega\mathbb{I}-M(\mathbf{q})) \rvert^{2}}
= \frac{\omega^{4}+c_{1}\omega^{2}q^{4}}{(\omega^{2}-c_{2}q^{4})^{2}+c_{3}\omega^{2}q^{4}},
\label{eq:vtvt_simp}
\end{equation}
with
\begin{equation}
c_{1} = \frac{1}{\gamma^{2}}\big\{ (B+\mu)^{2} + (K^{o}+A)^{2} \big\}.
\end{equation}
Going through the same procedure, we can obtain the condition for a persistent wave as
\begin{equation}
(K^{o}+A)^{2}+2K^{o}(K^{o}+A)-\mu^{2}>0.
\end{equation}
In terms of the effective spring constants, it becomes
\begin{equation}
\frac{k_{T}}{k_{L}} > \frac{1}{\sqrt{15}}.
\label{eq:alpha_TT}
\end{equation}

For the real part of the cross velocity correlation function in Eq.~(\ref{eq:vlvt}) with white noise, we have 
\begin{equation}
\text{Re}[\langle v_{L}^{*}(\mathbf{q},\omega)v_{T}(\mathbf{q},\omega) \rangle] = -\frac{\omega^{2}\Big\{ M_{LL}(\mathbf{q})M_{LT}(\mathbf{q}) + M_{TL}(\mathbf{q})M_{TT}(\mathbf{q}) \Big\}}{\lvert \det(-i\omega\mathbb{I}-M(\mathbf{q})) \rvert^{2}}
= \frac{c_{1}\omega^{2}q^{4}}{(\omega^{2}-c_{2}q^{4})^{2}+c_{3}\omega^{2}q^{4}},
\label{eq:revlvt_simp}
\end{equation}
with 
\begin{equation}
c_{1} = \frac{1}{\gamma^{2}}\big\{ -(B+\mu)K^{o}+(K^{o}+A)\mu \big\}.
\label{eq:c1_revlvt}
\end{equation}
Unlike $\langle v_{L}^{*}(\mathbf{q},\omega)v_{L}(\mathbf{q},\omega) \rangle$ and $\langle v_{T}^{*}(\mathbf{q},\omega)v_{T}(\mathbf{q},\omega) \rangle$, the real and imaginary parts of $\langle v_{L}^{*}(\mathbf{q},\omega)v_{T}(\mathbf{q},\omega) \rangle$ are not necessarily positive. Therefore, we identify real and positive $\omega$ that maximizes $\omega^{2}q^{4}/[ (\omega^{2}-c_{2}q^{4})^{2}+c_{3}\omega^{2}q^{4} ]$, which is Eq.~(\ref{eq:revlvt_simp}) without the constant factor of $c_1$. This gives the maximum of Re$[\langle v_{L}^{*}(\mathbf{q},\omega)v_{T}(\mathbf{q},\omega) \rangle]$ if $c_1$ is positive, and the minimum if $c_1$ is negative. The extremum of $\text{Re}[\langle v_{L}^{*}(\mathbf{q},\omega)v_{T}(\mathbf{q},\omega) \rangle]$ occurs at
\begin{equation}
\omega = \sqrt{c_{2}}q^{2},
\end{equation}
which is real and positive for the elastic moduli values we use. However, $\text{Re}[\langle v_{L}^{*}(\mathbf{q},\omega)v_{T}(\mathbf{q},\omega) \rangle]$ is the quantity for which the continuum limit approximation conducted here does not work. Substituting Eq.~(\ref{eq:moduli_simp}) into Eq.~(\ref{eq:c1_revlvt}) gives $c_{1}=0$, which leads to $\text{Re}[\langle v_{L}^{*}(\mathbf{q},\omega)v_{T}(\mathbf{q},\omega) \rangle]=0$ according to Eq.~(\ref{eq:revlvt_simp}). This is the result from the continuum limit, which only describes the long-wavelength behavior close to $q=0$ or the $\Gamma$ point in the first Brillouin zone. Therefore, in order to find the corresponding condition for maximizing $\text{Re}[\langle v_{L}^{*}(\mathbf{q},\omega)v_{T}(\mathbf{q},\omega) \rangle]$ analytically, we should instead use the elements of the dynamic matrix for the discrete lattice case described in Eqs.~(\ref{eq:Mab})-(\ref{eq:Mltab}) and solve for $\omega$, which is much more complicated.

For the imaginary part of the cross velocity correlation function in Eq.~(\ref{eq:vlvt}) with white noise, we get
\begin{equation}
\text{Im}[\langle v_{L}^{*}(\mathbf{q},\omega)v_{T}(\mathbf{q},\omega) \rangle] = \frac{\omega^{2}\Big\{ \omega M_{TL}(\mathbf{q}) - \omega M_{LT}(\mathbf{q}) \Big\}}{\lvert \det(-i\omega\mathbb{I}-M(\mathbf{q})) \rvert^{2}}
= \frac{c_{1}\omega^{3}q^{4}}{(\omega^{2}-c_{2}q^{4})^{2}+c_{3}\omega^{2}q^{4}},
\label{eq:imvlvt_simp}
\end{equation}
with 
\begin{equation}
c_{1} = \frac{1}{\gamma}\big\{ A + 2K^{o} \big\}. \label{eq:C1_Im}
\end{equation}
We then look for $\omega$ that maximizes $\omega^{3}q^{4}/[ (\omega^{2}-c_{2}q^{4})^{2}+c_{3}\omega^{2}q^{4} ]$. The corresponding solution for the extremum is
\begin{equation}
\omega^{2} = \frac{-(2c_{2}-c_{3})+\sqrt{(2c_{2}-c_{3})^{2}+12c_{2}^{2}}}{2}q^{4},
\end{equation}
which is guaranteed to be real and positive regardless of the sign of $2c_{2}-c_{3}$. 
Similar to the above case for the real part of cross correlation $\langle v_{L}^{*}(\mathbf{q},\omega)v_{T}(\mathbf{q},\omega) \rangle$, to obtain a nontrivial condition for maximizing $\text{Im}[\langle v_{L}^{*}(\mathbf{q},\omega)v_{T}(\mathbf{q},\omega) \rangle]$ one needs to work with the expressions for the discrete lattice case.

Through the analytic calculations for the velocity correlation functions in the continuum limit instead of the current correlation functions, we have identified the approximate criteria for the occurrence of noise-driven odd elastic waves. As the expression for each velocity correlation function is different, we obtain different criteria corresponding to different modes. When the magnitude of transverse force increases (i.e., when $k_T$ becomes larger), the transverse mode of the elastic wave characterized by the transverse velocity correlation is predicted to emerge first (Eq.~(\ref{eq:alpha_TT})), and then the additional longitudinal mode at even stronger transverse force or larger $k_T$ (Eq.~(\ref{eq:alpha_LL})). We expect this trend to hold for the results from the current correlation function. The verification of the order of emergence of different modes in further analytical calculations or numerical simulations as well as possible experiments is left for the future work. With this future work, we expect to be able to draw a more detailed phase diagram than the one presented in Fig.~2 of the main text. Although the calculations shown here have some limitations, we have demonstrated a viable method to generate the persistent odd elastic waves.

\section{Cross sections of the phase diagram}
\label{sec:phasediag_slice}

In the main text, a phase diagram is presented in Fig. 2, showing four phase regimes: Regime \RomanNumeralCaps{1} with no wave behavior, regime \RomanNumeralCaps{2} with damped wave, regime \RomanNumeralCaps{3} with persistent noise-driven odd elastic wave, and regime \RomanNumeralCaps{4} with melting. This section demonstrates how this phase diagram is calculated by identifying the behavior of the system at certain values of $\alpha=k_{T}/k_{L}$, the ratio between the effective transverse and longitudinal spring constants, across different noise strengths. 
We also conducted the calculations with varying $\alpha$ at some fixed values of noise strength $v_\sigma$.
We have explored multiple different values of $\alpha$ to obtain the phase diagram shown in Fig.~2 of the main text, but here we mainly present the results at $\alpha=\alpha_{1}\simeq0.79$ and $\alpha = \alpha_{2}\simeq2.63$, as representative examples for the transition from regime \RomanNumeralCaps{2} to \RomanNumeralCaps{4} (at $\alpha_{1}$) and from regime \RomanNumeralCaps{2} to \RomanNumeralCaps{3} to \RomanNumeralCaps{4} (at $\alpha_{2}$). 

We use the stochastic odd elastic model described in Eq. (2) of the main text. As before, the parameters for the longitudinal force are taken from the experimental results of starfish embryos \cite{tan2022odd}, for which the nondimensionalized values are $\tilde{F}_{\text{st}}=53.7$ and $\tilde{f}_{\text{rep}}=785.1$ (see SM $\S$\ref{sec:starfish_model}). With these values, we obtain the effective longitudinal spring constant $k_{L} \sim 1.9$ after linearizing the longitudinal force \cite{tan2022odd} (see SM $\S$\ref{sec:starfish_model} for the explicit expression of the force). To obtain different values of the ratio $\alpha = k_{T}/k_{L}$, we adjust the value of the effective transverse spring constant $k_{T}$. For convenience, the self-spinning frequency is fixed to be $\bar{\omega}_{0}=1.0$ for all simulations. The transverse force is varied by setting different values of the coefficient $f_{0}$. Note that here we use a slightly different equation for $k_{T}$ as compared to Ref.~\cite{tan2022odd} because the self-spinning frequency is not updated in our simulations (see SM $\S$\ref{sec:wupdate}). The linearization of the transverse force without torque balance leads to 
\begin{equation}
k_{T} = \frac{2R\omega f_{0}}{r_{0}-2R},
\end{equation}
where $R$ is the radius of the particle (or embryo), $\omega = \bar{\omega}_{0}$ is the self-spinning frequency, and $r_{0}$ is the equilibrium lattice spacing. For the case of $\alpha=\alpha_{1}$, we used $f_{0} = 0.3$, which leads to $k_{T} = 1.5$ and hence $\alpha_{1} \simeq 0.79$. For the case of $\alpha_{2}$, we used $f_{0}=1.0$, which is the value used to generate the data for Fig. 1 of the main text, leading to $k_{T}=5.0$ and hence $\alpha_{2} \simeq 2.63$.
The noise strength is set by standard deviation $v_{\sigma}$ of the Gaussian white noise in the self-circling term $v_{0}(t) = \bar{v}_{0} + \xi_{v_{0}}(t)$, with $\langle \xi_{v_{0}}(t)\xi_{v_{0}}(t') \rangle = v_{\sigma}^{2}\delta (t-t')$. When varying the noise strength $v_{\sigma}$, we keep the ratio between $v_{\sigma}$ and $\bar{v}_{0}$ unchanged. The noise in the self-circling is assumed to be related to the activity level of embryos. As the activity of an embryo increases, the embryo will collide with its neighbors more frequently and experience more fluctuations, leading to larger value of $v_{\sigma}$. The increase in activity will also lead to a larger degree of motion and thus larger $\bar{v}_{0}$. As we do not have access to the experimental evidence for the relation between the changes in $v_{\sigma}$ and $\bar{v}_{0}$, we assume them to be linearly dependent with each other. In this particular test, we keep $v_{\sigma}/\bar{v}_{0}=10$ since it is the ratio used to generate the result for Fig. 1 in the main text. 

\begin{figure}[!htb]
    \includegraphics[width=0.7\columnwidth]{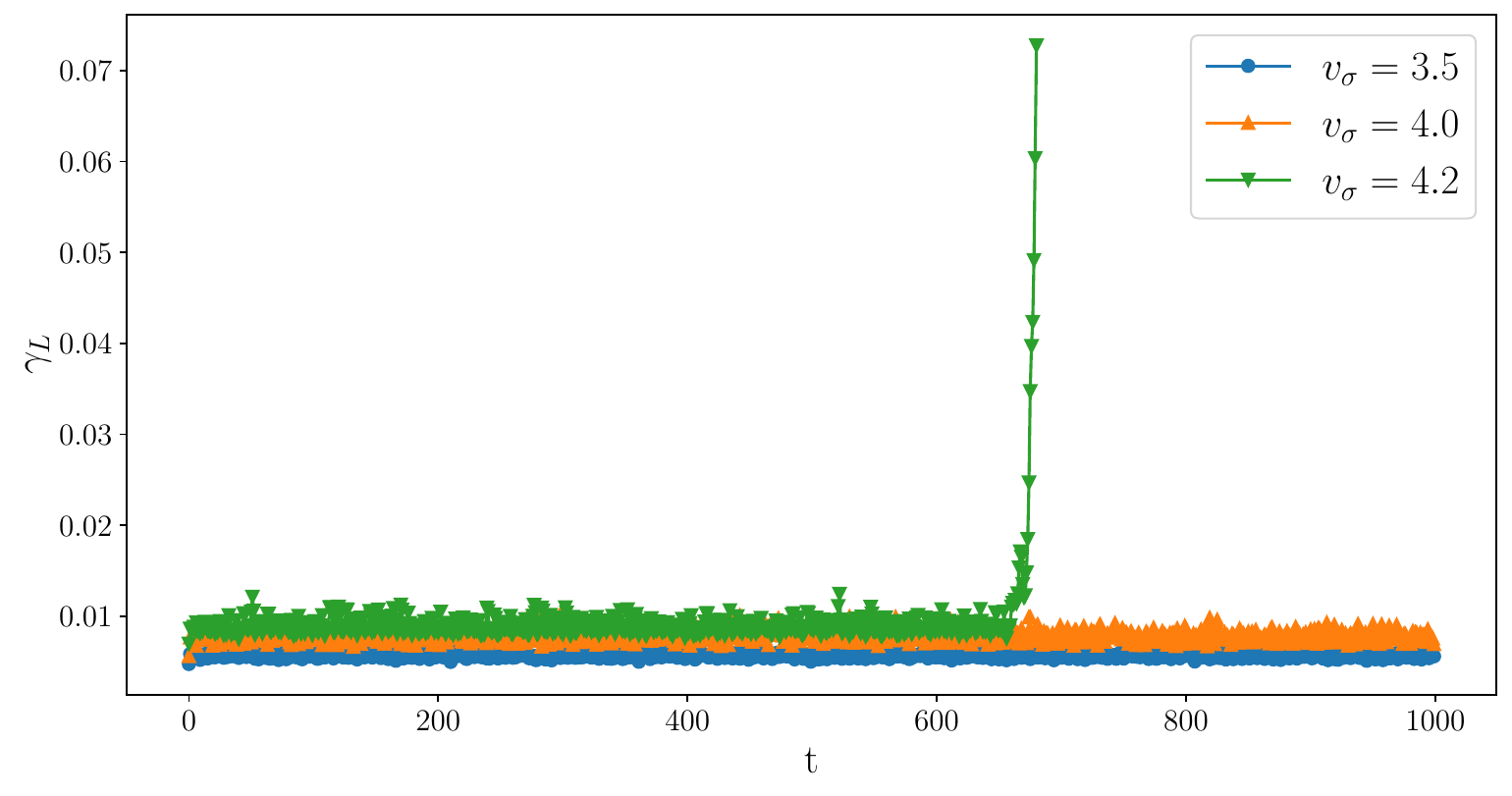}
   \caption{The time series of the dynamic Lindemann parameter $\gamma_{L}$ for different values of noise strength $v_\sigma$ at $\alpha = \alpha_{1}\simeq 0.79$. The simulation has been conducted with time step $dt = 0.001$, and the data were recorded every 100 time steps (i.e., $\Delta t=0.1$).} 
  \label{fig:dynamicLindemann_w1f03}
\end{figure}

\begin{figure}[!htb]
   \includegraphics[width=0.7\columnwidth]{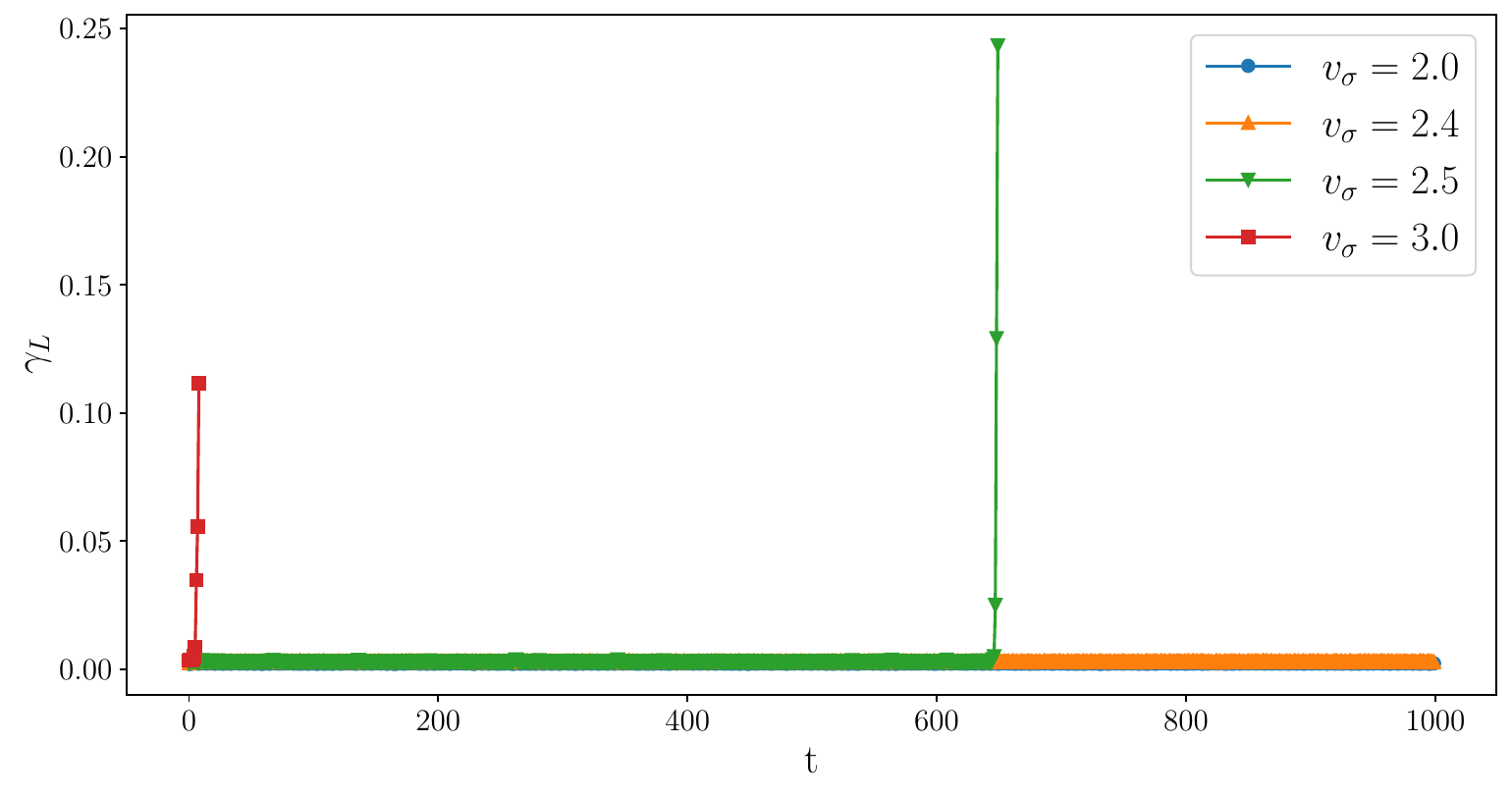}
   \caption{The time series of the dynamic Lindemann parameter $\gamma_{L}$ for different noise strengths at $\alpha = \alpha_{2}\simeq 2.63$. The simulation has been conducted with time step $dt = 0.001$, and the data were recorded every 100 time steps (i.e., $\Delta t=0.1$).} 
  \label{fig:dynamicLindemann_w1f1}
\end{figure}

The crystal melting is examined by measuring the dynamic Lindemann parameter \cite{bedanov1985modified,zahn1999two,zahn2000dynamic}, as the usual Lindemann criterion \cite{lindemann1910ueber} is only applicable to three-dimensional crystals while our system is two-dimensional and active. The dynamic Lindemann parameter $\gamma_{L}(t)$ is defined as
\begin{equation}
\gamma_{L}(t) = \frac{1}{2a^{2}}\langle\lvert \Delta \mathbf{u}_{i}(t)-\Delta \mathbf{u}_{j}(t) \rvert^{2}\rangle,
\end{equation}
where $a$ is the lattice spacing and $\Delta\mathbf{u}_{i}(t)\equiv\mathbf{u}_{i}(t)-\mathbf{u}_{i}(0)$ with $\mathbf{u}_{i}(t)$ the displacement of the $i^{th}$ particle from the equilibrium position at time $t$. $\lvert \Delta \mathbf{u}_{i}(t)-\Delta \mathbf{u}_{j}(t) \rvert^{2}$ is calculated only for the nearest neighbors (labeled by the subscripts $i$ and $j$), and $\langle\cdots\rangle$ represents the average over the population. The critical value for melting is known to be $\gamma_{L}^{c}=0.033$ and the value of $\gamma_{L}$ diverges in the melting regime (or the liquid phase) \cite{zahn1999two}.
Figures~\ref{fig:dynamicLindemann_w1f03} and \ref{fig:dynamicLindemann_w1f1} show the time evolution of $\gamma_{L}$ under different noise strengths. After the noise strength exceeds a certain value, it becomes easier for the living crystal to melt, which is represented by the divergence of $\gamma_{L}$ in the time series. Note that the simulations were aborted at the time of divergence, and thus the diverging curves do not last up to the final time step of simulation ($10^{6}$ time steps with $dt = 0.001$, so that the final time $t=1000$). It is also notable that, while it is known that $\gamma_{L}^{c}=0.033$, the onset of divergence occurs at a smaller value near 0.01. The self-circling motion of the agents (embryos in the case of our model) may have facilitated the melting. To deduce the phase diagram, we identify the value of $v_{\sigma}$ from which the onset of divergence of $\gamma_{L}$ occurs before $t=1000$, which is represented by the red circle point in Fig. 2 of the main text. Connecting these data points leads to the phase boundary between regimes \RomanNumeralCaps{2} and \RomanNumeralCaps{4} or between \RomanNumeralCaps{3} and \RomanNumeralCaps{4} in the phase diagram. 

\begin{figure}[!htb]
   \includegraphics[width=\columnwidth]{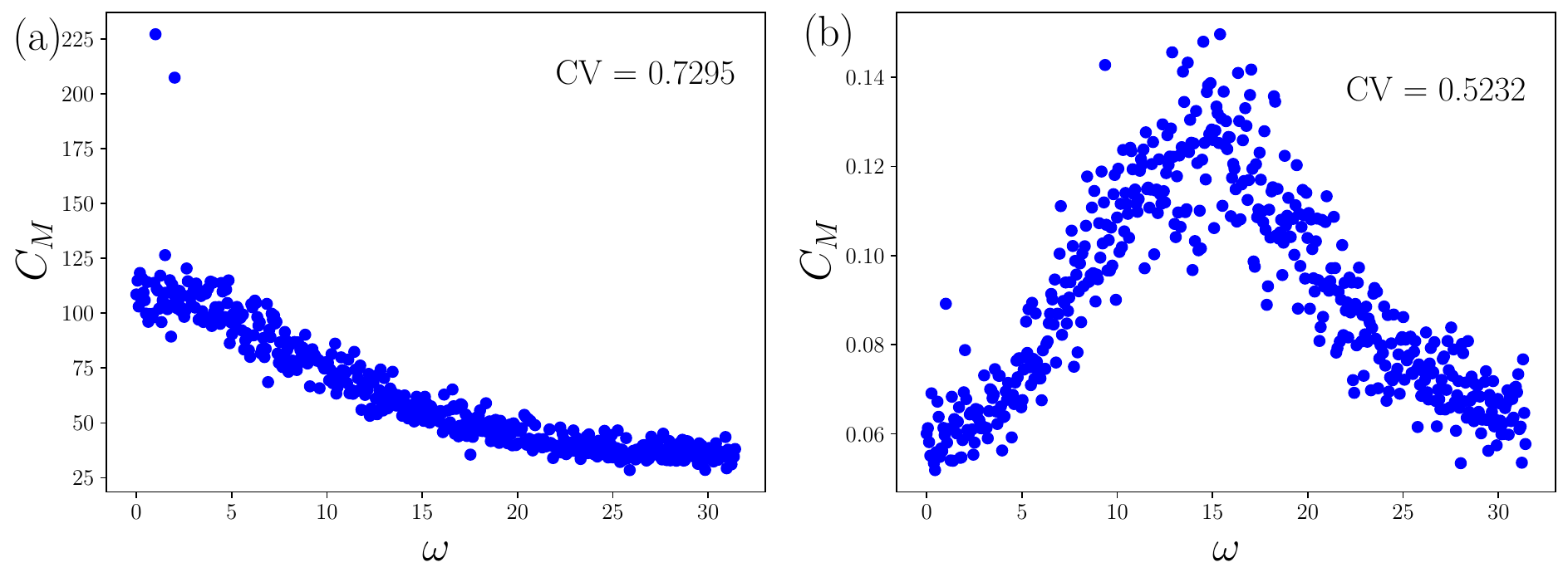}
   \caption{The current correlation function $C_{M}$ at $M$ point of the first Brillouin zone as a function of frequency $\omega$ at (a) $\alpha=\alpha_{1} \sim 0.79$ and $v_{\sigma}=1.0$ and (b) $\alpha = \alpha_{2}\sim2.63$ and $v_{\sigma}=0.02$. The simulations have been conducted with time step $dt=0.01$, and the data were recorded every 10 time steps (i.e., $\Delta t=0.1$) and averaged over 100 realizations.}
  \label{fig:CM_example}
\end{figure}

\begin{figure}[!htb]
   \includegraphics[width=\columnwidth]{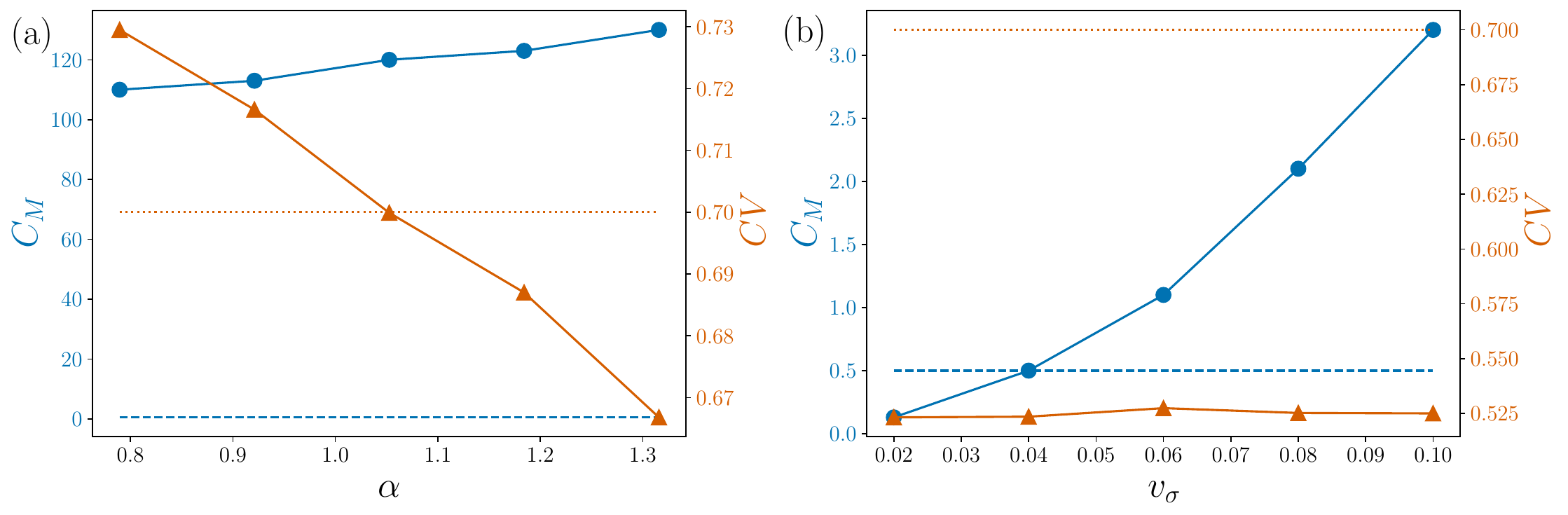}
   \caption{The peak value of the current correlation function $C_{M}$ at the $M$ point of the first Brillouin zone (blue circle) and the coefficient of variation $CV$ (vermilion triangle), for (a) varying $\alpha$ values with a fixed $v_{\sigma}=1.0$ and (b) varying $v_{\sigma}$ values with a fixed $\alpha=\alpha_{2}\sim2.63$. The simulations have been conducted with time step $dt = 0.01$, and the data were recorded every 10 time steps (i.e., $\Delta t=0.1$) and averaged over 100 realizations. The blue dashed line denotes the threshold $C_{M}^{\text{thres}}=0.5$, and the vermilion dotted line denotes the threshold $CV^{\text{thres}}=0.7$.}
  \label{fig:CM_wave}
\end{figure}

The existence of persistent elastic waves is determined by examining the dispersion relation calculated from the current correlation functions. However, the transition from the damped wave (regime \RomanNumeralCaps{2}) to the noise-driven persistent wave (regime \RomanNumeralCaps{3}) is not clearly defined. As regime \RomanNumeralCaps{2} is approached from regime \RomanNumeralCaps{3} by decreasing noise, the wave signal itself decreases, which makes it difficult to detect the wave. This tendency is captured in Eq.~(4) of the main text, which is also applicable to our derivation of the criterion for noise-driven odd elastic waves in SM $\S$\ref{sec:wave_condi}, where the velocity correlation function is proportional to the noise strength (SM $\S$\ref{sec:analytical_vv}). Although the dispersion relation is calculated from the current correlation functions, which are similar to but different from the velocity correlation functions, we expect a similar tendency, as confirmed by our simulation results (see Fig.~\ref{fig:CM_wave}). On the other hand, when regime \RomanNumeralCaps{2} is approached by decreasing $\alpha$, the wave becomes damped as predicted by the theory, but the damping of the wave is gradual. Therefore, to determine the wave behavior, we need to set two thresholds regarding the strength of the signal and the damping of the spectral peak respectively. For this task, we examine the current correlation function at the $M$ point of the first Brillouin zone, $C_{M}(\omega) = C_{LL}(\mathbf{q}_{M},\omega)+C_{TT}(\mathbf{q}_{M},\omega)$, as a function of frequency $\omega$. Here $\mathbf{q}_{M}$ is the wave vector corresponding to the $M$ point.
As shown in Fig.~\ref{fig:CM_example}, $C_{M}$ shows a peak at a certain value of $\omega$, which we have collected to construct the dispersion relation in Fig. 1 of the main text (also see SM $\S$\ref{sec:fitting}). We set a threshold peak value for the wave detection, $C_{M}^{\text{thres}}=0.5$, below which no wave would be observed. Setting a different value of threshold does not change the qualitative behavior of the phase diagram. Although this may be an arbitrary value of threshold here, in experiments this value would be determined by the sensitivity of the measuring apparatus. In order to determine the degree of damping of the spectral peak, we calculate the coefficient of variation ($CV$), which is used to measure how dispersed a distribution is, as defined by 
\begin{equation}
CV = \frac{\sigma}{\mu},
\end{equation}
where $\sigma$ is the standard deviation and $\mu$ is the mean of the distribution. To obtain $CV$ of $C_{M}$, we calculate $\sigma$ and $\mu$ as the weighted mean and the weighted standard deviation, with the weight being the value of $C_{M}$ at each $\omega$. We set the threshold for the wave detection as $CV^{\text{thres}}=0.7$. Again, in experiments this value would be determined by the peak detection resolution of the measuring apparatus. The wave behavior is detected if both $C_{M}>C_{M}^{\text{thres}}$ at the peak and $CV<CV^{\text{thres}}$ are satisfied. Figure~\ref{fig:CM_example} gives two examples of $C_{M}$ spectral peaks showing the lack of wave behavior.  
Figure~\ref{fig:CM_example}(a) shows $C_{M}$ measured at $\alpha=\alpha_{1}\sim0.79$. Because the noise is strong ($v_{\sigma}=1.0$), the value of $C_{M}$ at the peak is much larger than $C_{M}^{\text{thres}}=0.5$, but the $C_M$ vs $\omega$ relation is too dispersed, yielding $CV\sim0.73>CV^{\text{thres}}$. Therefore, the wave behavior is deemed not detectable. Figure~\ref{fig:CM_example}(b) shows $C_{M}$ measured at $\alpha_{2}\sim2.63$ and $v_{\sigma}=0.02$. In contrast to the case of Fig.~\ref{fig:CM_example}(a), the peak is more well-defined, with $CV\sim0.52<CV^{\text{thres}}$, but the peak value of $C_{M}$ is smaller than $C_{M}^{\text{thres}}$ because of too weak noise $v_{\sigma}$. Thus no wave would be detected in this case as well. As suggested by Fig.~\ref{fig:CM_example}, at the boundary between regimes \RomanNumeralCaps{2} and \RomanNumeralCaps{3}, damping or flattening of the spectral peak becomes a more important factor to determine the wave behavior for regions of smaller $\alpha$ values, while the signal strength becomes a more determining factor for regions of larger $\alpha$ values. 
This observation is demonstrated in Fig.~\ref{fig:CM_wave}. In Fig.~\ref{fig:CM_wave}(a), we start from $\alpha=\alpha_{1}\sim0.79$ and $v_{\sigma}=1.0$ (i.e., the case of Fig.~\ref{fig:CM_example}(a)) and increase $\alpha$. Throughout the process, $C_{M}$ is well above $C_{M}^{\text{thres}}$ because of large $v_{\sigma}$. $CV$ decreases with increasing $\alpha$ and crosses $CV^{\text{thres}}$ around $\alpha\sim1.05$, which is then taken as a boundary value between regime \RomanNumeralCaps{2} and \RomanNumeralCaps{3} and represented by the blue triangle point in the phase diagram of Fig. 2 in the main text. On the other hand, in Fig.~\ref{fig:CM_wave}(b) we start from $\alpha=\alpha_{2}\sim2.63$ and $v_{\sigma}=0.02$ (i.e., the case of Fig.~\ref{fig:CM_example}(b)) and increase $v_{\sigma}$. Throughout the process, $CV$ is below $CV^{\text{thres}}$ because of large $\alpha$. $C_{M}$ increases with increasing $v_{\sigma}$ and crosses $C_{M}^{\text{thres}}$ around $v_{\sigma}=0.04$, which is taken as the boundary value represented by the blue triangle point in Fig. 2 of the main text. Connecting these data points leads to the phase boundary between regimes \RomanNumeralCaps{2} and \RomanNumeralCaps{3} in the phase diagram.

\begin{figure}[!htb]
   \includegraphics[width=\columnwidth]{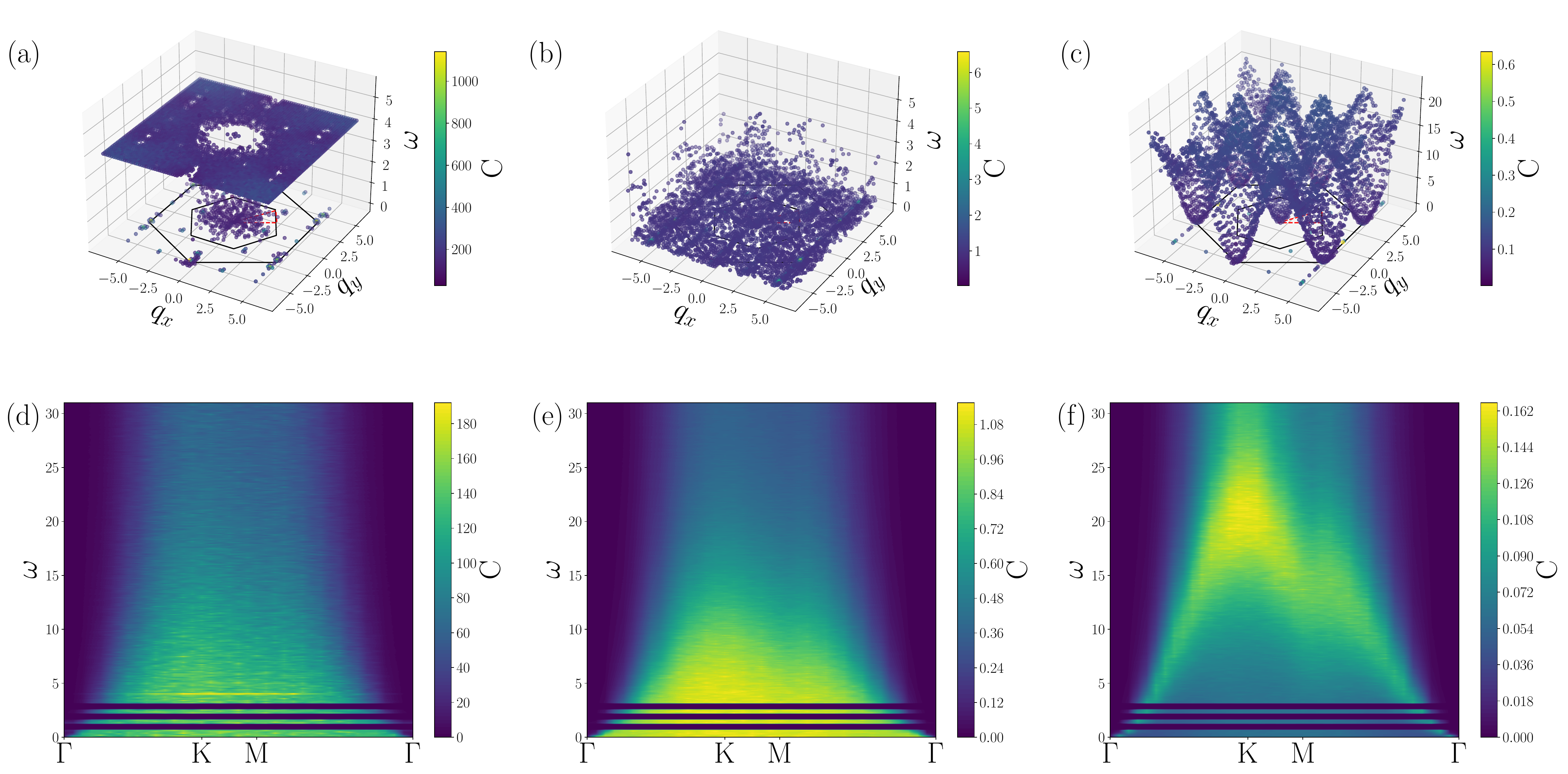}
   \caption{Dispersion results in regime \RomanNumeralCaps{2}. The first row shows the three-dimensional dispersion results for (a) $\alpha = \alpha_{1}\simeq 0.79$ and $v_{\sigma}=3.5$, (b) $\alpha = \alpha_{1}\simeq 0.79$ and $v_{\sigma}=0.1$, and (c) $\alpha = \alpha_{2}\simeq 2.63$ and $v_{\sigma}=0.02$. The second row shows the corresponding dispersion results along a path in the first Brillouin zone connecting the $\Gamma$, $K$, and $M$ points. For (a) and (d), given that the system is close to the melting point, the simulation was conducted with $dt=0.001$ up to $t=100$, and the data were recorded every 100 time steps (with $\Delta t=0.1$) and averaged over 100 realizations. For others, the simulation was conducted with $dt=0.01$ up to $t=100$, and the data were recorded every 10 time steps (with $\Delta t=0.1$) and averaged over 1000 realizations. The signals corresponding to the self-circling modes of $n=1,2,3$ have been removed to capture possible wave signals, with the corresponding locations indicated by dark lines.}
  \label{fig:nowave_comp}
\end{figure}

We note that $\alpha_{1}\simeq 0.79$ is expected to exhibit a persistent wave according to the analytic estimate of the criterion for noise-driven odd elastic waves (see Eq.~(5) in the main text and SM $\S$\ref{sec:wave_condi}), but it does not, as shown in Fig.~\ref{fig:CM_example}(a) as well as Fig.~\ref{fig:nowave_comp}(a), (b), (d) and (e). 
The system is near the melting point in Fig.~\ref{fig:nowave_comp}(a) and (d), and the signals from self-circling are very pronounced up to $n=4$, as indicated by the horizontal plane and lines in panels (a) and (d) respectively. To verify that the strong self-circling signal did not overwhelm or interfere with the wave signal, we have also examined the case with smaller self-circling signals (smaller $v_{\sigma}$ and $\bar{v}_{0}$, with the ratio between the two kept unchanged). Figures~\ref{fig:nowave_comp}(b) and (e) confirm that a persistent wave has not been generated in this case either. These observations are in contrast with what is presented in panels (c) and (f), where although the wave signal has been captured, it is weaker than the threshold value ($C_{M}^{\text{thres}}=0.5$) and thus disregarded as being undetectable. The values of the current correlation function in panels (c) and (f) are an order of magnitude smaller than those in (b) and (e), indicating that the lack of signals in (b) and (e) is not due to small noise or weakness of the wave. These simulation results contradict the approximate criterion for noise-driven odd elastic waves given in Eq.~(5) of the main text and SM $\S$\ref{sec:wave_condi}, which predicts the wave behavior for $\alpha > \sqrt{5/11} \sim 0.67$. However, this analytic prediction is an estimate based on the calculation of the velocity correlation function from the linearized equations of motion with additive white noise. On the other hand, the full model consists of nonlinear terms, and the noise is incorporated in the self-driving term. The dispersion relations are calculated from the current correlation function, which is similar to but distinct from the velocity correlation function. Therefore, some deviations from the analytic prediction of the wave criterion are expected, which does not undermine the main finding of our work that a persistent odd elastic wave can be realized in the presence of noise for large enough transverse force that is represented by the parameter $\alpha$. This is demonstrated in the case of $\alpha \simeq 2.63$ and $v_{\sigma}=0.1$ (Fig. 1 of the main text). Again, although strictly speaking the case featured in Fig.~\ref{fig:nowave_comp}(a) and (b) and the case of Fig.~\ref{fig:nowave_comp}(c) are different, as the former has spectral peaks that are too dispersed and the latter has a signal that is too weak to observe, we categorize all of them into regime \RomanNumeralCaps{2} of damped waves in the phase diagram because the occurrence of wave is predicted for both cases by the deterministic theory \cite{scheibner2020odd} but the wave is not expected to be experimentally detectable. Hence we need to use two thresholds ($C_{M}^{\text{thres}}$ and $CV^{\text{thres}}$) to determine the wave behavior, as discussed above.

\section{The experimental dispersion relations from current correlation functions}
\label{sec:expt_jj}

This section presents the full results of dispersion relations obtained from current correlation functions of the experimental data for starfish embryo living crystals observed in Ref.~\cite{tan2022odd}.

\begin{figure}[!htb]
   \includegraphics[width=0.8\columnwidth]{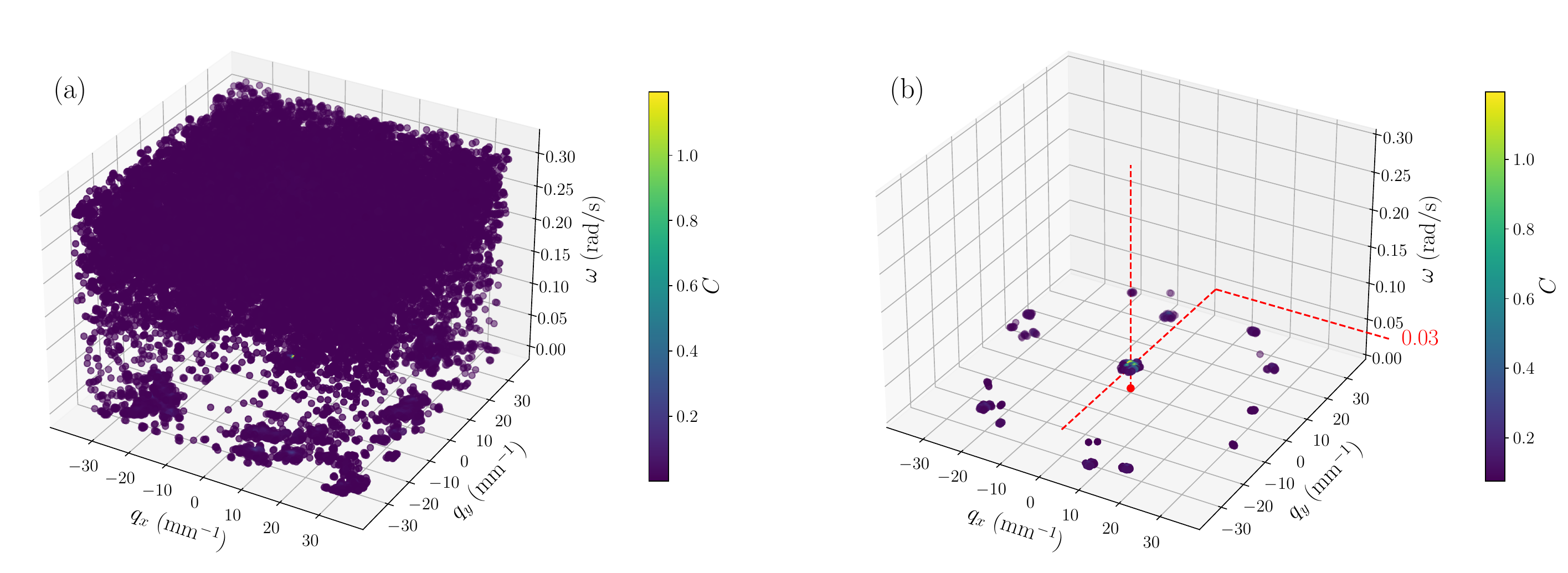}
   \caption{The dispersion results obtained from $\frac{1}{N}\langle \mathbf{J}^{*}(\mathbf{q},\omega)\cdot\mathbf{J}(\mathbf{q},\omega) \rangle = C_{LL}(\mathbf{q},\omega) + C_{TT}(\mathbf{q},\omega)$ for the starfish embryo experimental data of Ref.~\cite{tan2022odd}. Panel (a) shows the full result of calculation, and panel (b) shows only the data points for which the values of current correlation are larger than 0.075. Red dashed lines are added to indicate the location of the origin (red dot) and the frequency value $\omega = 0.03$ rad/s.}
  \label{fig:expt_Cfull_thres}
\end{figure}

Figure \ref{fig:expt_Cfull_thres}(a) shows the experimental dispersion relations obtained from the current correlation function $\frac{1}{N}\langle \mathbf{J}^{*}(\mathbf{q},\omega)\cdot\mathbf{J}(\mathbf{q},\omega) \rangle = C_{LL}(\mathbf{q},\omega) + C_{TT}(\mathbf{q},\omega)$. It is very noisy and hard to read. Therefore, we apply a threshold cutoff and plot only the data points with the current correlation value larger than 0.075. Then one can see signals around $\mathbf{q}=\mathbf{0}$ and the vertices of the reciprocal lattice at $\omega = 0.03 \text{ rad/s}$, the value very similar to the self-spinning frequency of the starfish embryos measured experimentally \cite{tan2022odd}. The result is shown in Fig.~\ref{fig:expt_Cfull_thres}(b) which is the same as Fig.~3(a) of the main text. 

\begin{figure}
   \includegraphics[width=0.62\columnwidth]{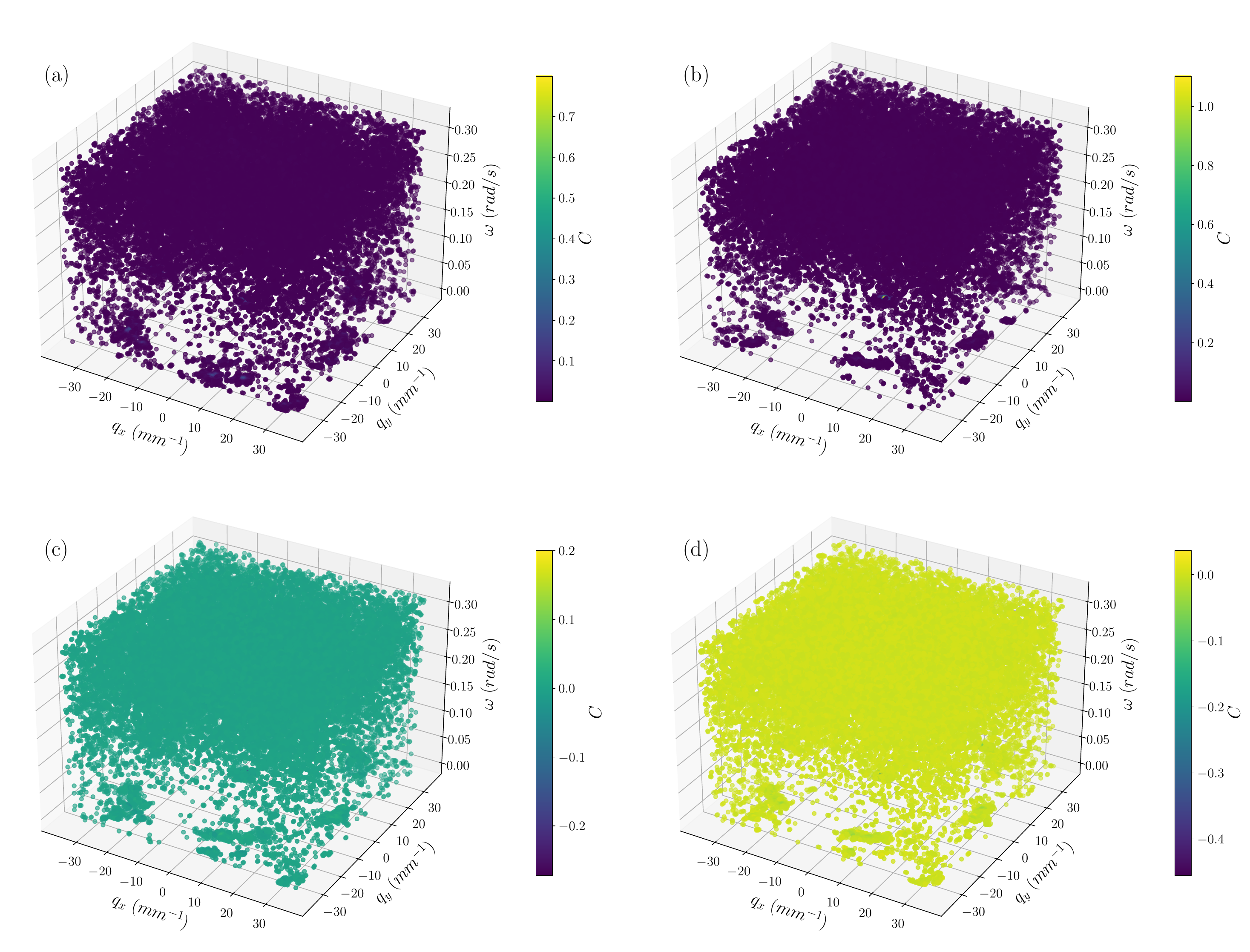}
   \caption{The dispersion results calculated from (a) $C_{LL}(\mathbf{q},\omega)$, (b) $C_{TT}(\mathbf{q},\omega)$, (c) $\text{Re}[C_{LT}(\mathbf{q},\omega)]$, and (d) $\text{Im}[C_{LT}(\mathbf{q},\omega)]$ of the current correlation for the starfish embryo experimental data. 
  \label{fig:expt_Cab}}
\end{figure}
\begin{figure}
   \includegraphics[width=0.62\columnwidth]{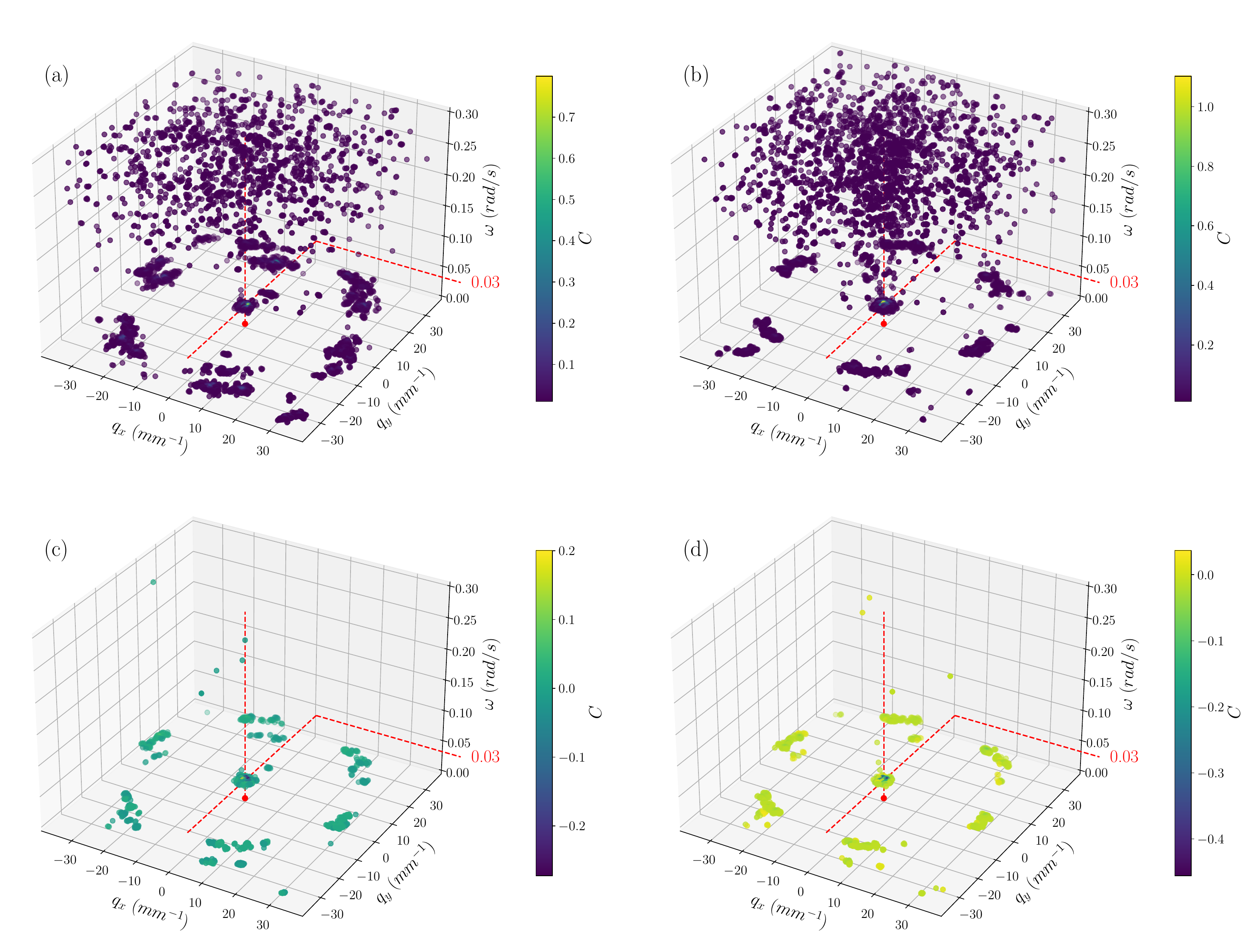}
   \caption{The dispersion results calculated from (a) $C_{LL}(\mathbf{q},\omega)$, (b) $C_{TT}(\mathbf{q},\omega)$, (c) $\text{Re}[C_{LT}](\mathbf{q},\omega)$, and (d) $\text{Im}[C_{LT}](\mathbf{q},\omega)$ for the starfish embryo experimental data, where only the points with the corresponding current correlation values larger than 0.01 are shown. Red dashed lines are reference lines to aid reading of $\mathbf{q}$ and $\omega$ of the signal. The horizontal reference lines are located at $\omega = 0.03$ rad/s, and the red dot is located at the origin.}
  \label{fig:expt_Cab_thres}
\end{figure}

Figure \ref{fig:expt_Cab} shows the experimental dispersion relations obtained from different current correlation functions, including the diagonal elements of the current correlation $C_{LL}(\mathbf{q},\omega)$ and $C_{TT}(\mathbf{q},\omega)$ along the longitudinal and transverse directions, representing the correlation of longitudinal and transverse currents [Fig.~\ref{fig:expt_Cab}(a) and (b)], and the cross current correlations [Fig.~\ref{fig:expt_Cab}(c) and (d)]. Instead of giving the dispersion results from $C_{LT}(\mathbf{q},\omega)$ and $C_{TL}(\mathbf{q},\omega)$, here we show those from $\text{Re}[C_{LT}(\mathbf{q},\omega)]$ and $\text{Im}[C_{LT}(\mathbf{q},\omega)]$, as the cross current correlation functions in Fourier space are complex and $C_{LT}(\mathbf{q},\omega)$ and $C_{TL}(\mathbf{q},\omega)$ are complex conjugates of each other.
As in Fig.~\ref{fig:expt_Cfull_thres}(a), results of Fig.~\ref{fig:expt_Cab} are noisy and hard to read. We thus apply the same cutoff threshold of 0.01 for a better illustration, which results in Fig.~\ref{fig:expt_Cab_thres}. Similar to Fig.~\ref{fig:expt_Cfull_thres}(b), Fig.~\ref{fig:expt_Cab_thres} shows that strong enough signals are detected only around $\mathbf{q}=\mathbf{0}$ and the vertices of the reciprocal lattice at frequency $\omega \approx 0.03 \text{ rad/s}$. As discussed in the main text, these dispersion results represent the overall self-circling motion of the embryos, rather than a propagating elastic wave. 
To further confirm this conclusion, SM $\S$\ref{sec:simul_sc} shows the simulation result of non-interacting self-circling particles, which reproduces the dispersion relations shown in Fig.~\ref{fig:expt_Cab_thres}. The related analytical derivation of these dispersion relations for non-interacting self-circling particles are given in SM $\S$\ref{sec:flat}.

\section{Simulations of non-interacting self-circling particles}
\label{sec:simul_sc}

In this section, to verify that the experimental dispersion relations presented above are due to self-circling, we show the result from our simulations for non-interacting self-circling particles.

The simulation was done with 900 particles on a $30\times30$ triangular lattice with periodic boundary conditions. We used the Euler method \cite{euler1768foundations} with time step $dt=0.01$ (recorded the data every 40 time steps so that the measured $\Delta t=0.4$), simulated the system up to totally 40000 time steps, and then averaged the results over 100 independent runs of simulations. The particles do not interact with each other and simply circle around the perfect lattice positions independently. The self-circling frequency is set as $\omega_{0} = 1$, and the self-propelling strength is $v_{0} = 0.01$. We choose a small value of $v_{0}$ to be consistent with our analytical derivation for the dispersion relations for self-circling particles, which uses the approximation of small $l_{0}\equiv v_{0}/\omega_{0}$ (see SM $\S$\ref{sec:flat} for more details). Physically, this choice of parameter represents the embryos' limited ability to move inside a dense cluster.

\begin{figure}[!htb]
   \includegraphics[width=\columnwidth]{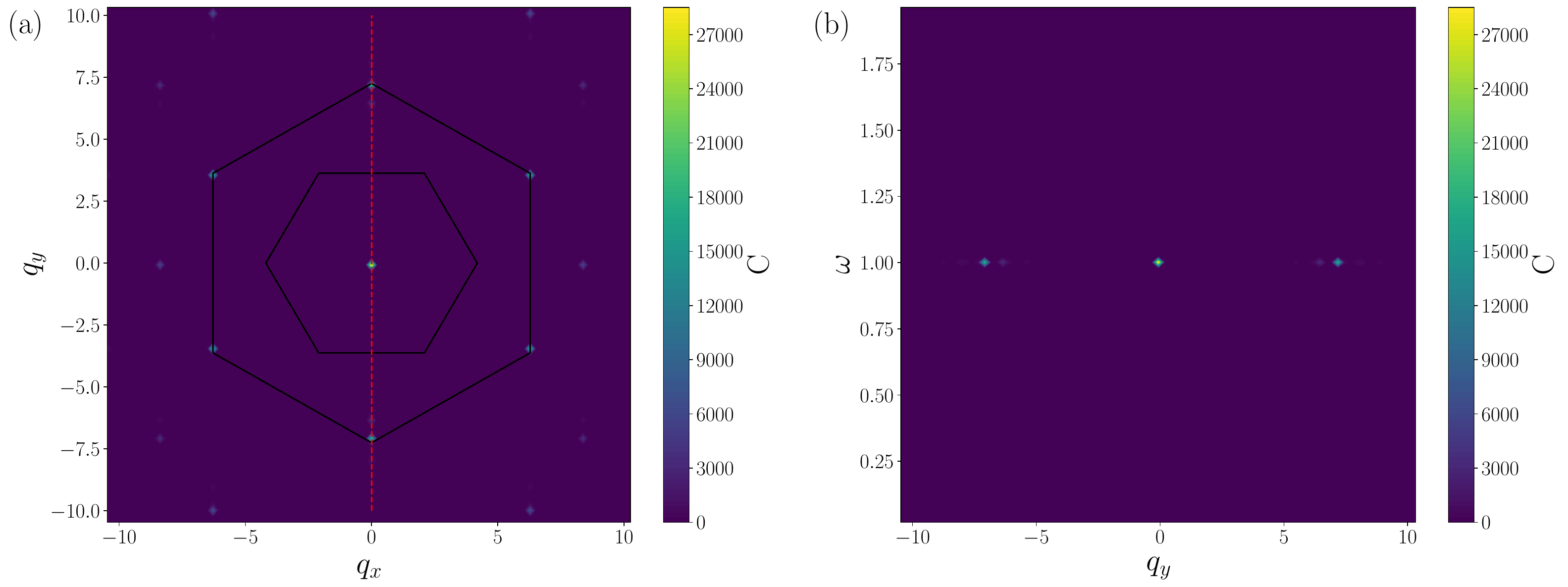}
   \caption{Results obtained from simulations of non-interacting self-circling particles, including (a) the maximum current correlation at each $\mathbf{q}$ over the whole $\mathbf{q}$-space, and (b) the dispersion result at $q_{x}=0$ (corresponding to the red dashed line in (a)).
   The signals appear only at $\mathbf{q}=\mathbf{0}$ and the vertices of the reciprocal lattice with frequency $\omega = 1.0$, which is the self-circling frequency, similar to that obtained from the experimental data. In (a), the outer hexagon represents the reciprocal lattice while the inner hexagon represents the first Brillouin zone.
  \label{fig:simple_circling}}
\end{figure}

Figure \ref{fig:simple_circling}(a) shows the color-coded value of the maximum current correlation function $\frac{1}{N}\langle \mathbf{J}^{*}(\mathbf{q},\omega)\cdot\mathbf{J}(\mathbf{q},\omega) \rangle = C_{LL}(\mathbf{q},\omega) + C_{TT}(\mathbf{q},\omega)$ at each $\mathbf{q}$ over the whole $\mathbf{q}$-space. While most of the $\mathbf{q}$-space has zero current correlation value, signals are detected at $\mathbf{q}=\mathbf{0}$ and at the vertices of the reciprocal lattice. Figure \ref{fig:simple_circling}(b) shows the dispersion result at $q_{x}=0$, corresponding to the red dashed line in Fig.~\ref{fig:simple_circling}(a). The signals appear only at $q_{y}=0$ and at the vertices of the reciprocal lattice, consistent with Fig.~\ref{fig:simple_circling}(a).
While Fig. \ref{fig:simple_circling}(a) does not show the value of $\omega$ for the signals, Figure \ref{fig:simple_circling}(b) shows the dominant signal at $\omega = 1$, which is the same as the self-circling frequency. These results agree with those from experimental data shown in Fig.~\ref{fig:expt_Cab_thres} as well as Fig.~3 of the main text, confirming that the oscillatory behavior observed in the experiment \cite{tan2022odd} is attributed to the self-circling motion of embryos.

\section{Revealing the hidden dispersion curve}
\label{sec:reveal_disp}

This section compares the dispersion curves obtained from both the simulation data of the starfish embryo model and the experimental data of Ref.~\cite{tan2022odd}, before and after the removal of the self-circling frequency part of the data. By doing so, we verify whether there exists a hidden dispersion curve in each dataset.

\begin{figure}[!htb]
   \includegraphics[width=\columnwidth]{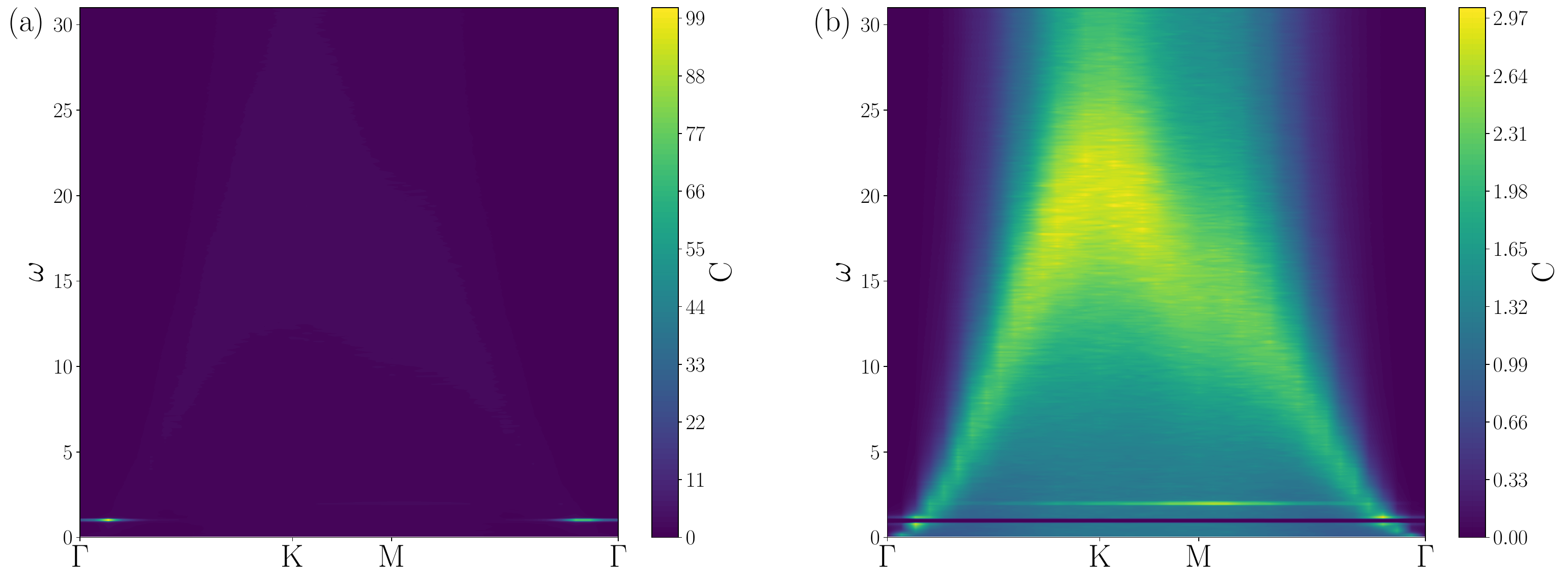}
   \caption{The dispersion result for the starfish embryo model obtained from (a) the full data and (b) the data with the self-circling frequency part removed. Panel (b) corresponds to Fig.~1(b) of the main text but without the data points for $\omega$ that maximizes the current correlation function at each $\mathbf{q}$ and without the dispersion curves identified from the analytically calculated velocity correlation functions.
  \label{fig:starfish_cutcomp}}
\end{figure}

We have shown the dispersion curves obtained from different elements of current correlation functions of the starfish embryo model in Fig.~1(b) of the main text. The figure has a dark horizontal line at frequency $\omega = 1.0$, which is the self-circling frequency used in the simulation, because we have removed the data at that frequency to make the dispersion curve visible. Figures \ref{fig:starfish_cutcomp}(a) and \ref{fig:starfish_cutcomp}(b) show the dispersion results from $\frac{1}{N}\langle \mathbf{J}^{*}(\mathbf{q},\omega)\cdot\mathbf{J}(\mathbf{q},\omega) \rangle = C_{LL}(\mathbf{q},\omega) + C_{TT}(\mathbf{q},\omega)$ before and after the removal of the self-circling frequency part from the data, respectively. Figure \ref{fig:starfish_cutcomp}(b), the result after the removal, gives the same plot as Fig.~1(b) of the main text but is reproduced here for better comparison. Before the removal (Fig.~\ref{fig:starfish_cutcomp}(a)), the dispersion curve is not visible because the signal from self-circling around the $\Gamma$ point, corresponding to $\mathbf{q}=\mathbf{0}$ and the reciprocal lattice vertices, is too strong, as the self-circling is the dominant motion of all the embryos in the simulation.

\begin{figure}[!htb]
   \includegraphics[width=\columnwidth]{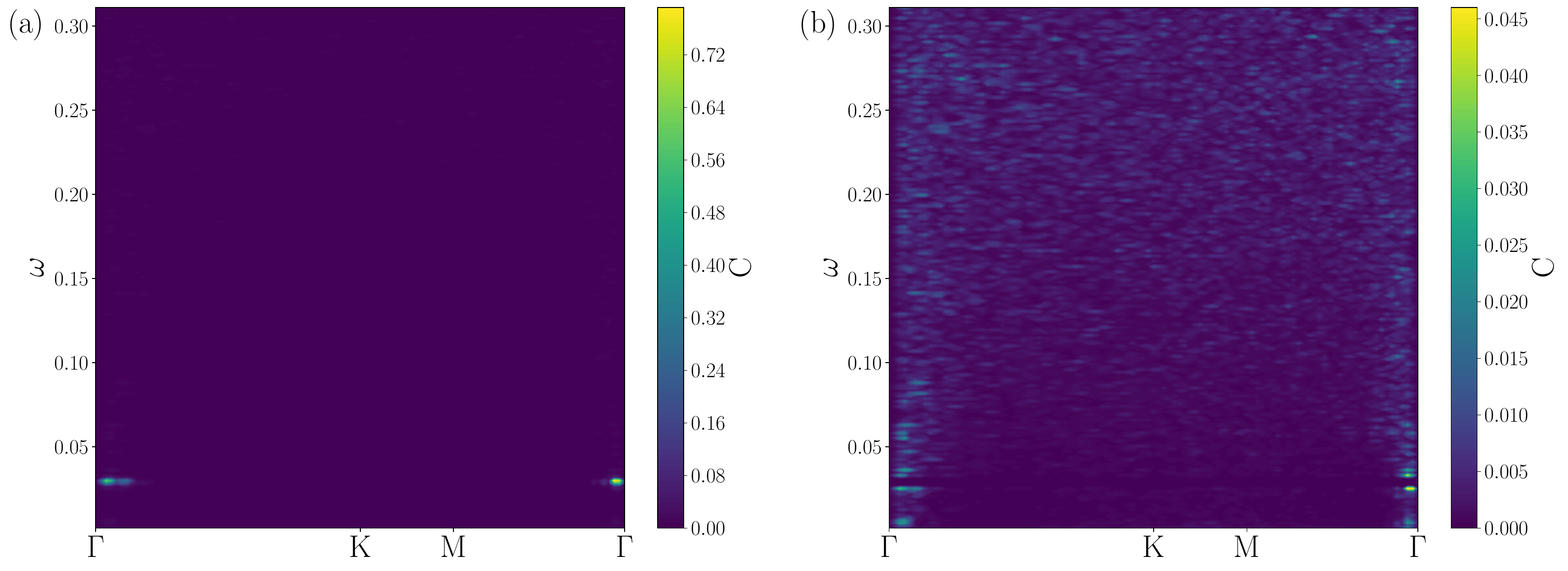}
   \caption{The dispersion result for the starfish embryo experimental data of Ref.~\cite{tan2022odd}, as obtained from (a) the full data and (b) the data with the self-circling frequency part removed. Panel (b) corresponds to Fig. 3(d) of the main text.
  \label{fig:expt_cutcomp}}
\end{figure}

This observation raises the question of whether the dispersion curve was just hidden by the strong signal from self-circling in the experimental data of Ref.~\cite{tan2022odd}, and we thus do a similar analysis with data removal. Figure \ref{fig:expt_cutcomp}(a) shows the dominant signal around the $\Gamma$ point at $\omega = 0.03 \text{ rad/s}$, which is the self-circling frequency, similar to the case in Fig.~\ref{fig:starfish_cutcomp}(a). However, unlike Fig.~\ref{fig:starfish_cutcomp}(b), Fig.~\ref{fig:expt_cutcomp}(b) does not show the hidden dispersion curve after the removal of the self-circling signal. (Fig.~\ref{fig:expt_cutcomp}(b) is the the same plot as Fig. 3(d) of the main text.) This analysis confirms that self-circling is the only signal presented in the experimental data.

\section{Noise statistics of the experimental data}
\label{sec:noise_stat_expt}

In this section, we analyze the statistics of fluctuations in the velocity of starfish embryos from the experimental data of Ref.~\cite{tan2022odd}. We use the result obtained here to simulate the starfish embryo model (Eqs. (2) and (3) of the main text) in the parameter regime that is as similar to the actual experiment as possible in the next section (SM $\S$\ref{sec:expt_emul}). 

The experimental data from Ref.~\cite{tan2022odd} are the positional trajectories of starfish embryos at each time frame separated by 10 seconds. We calculated the velocity of each embryo by subtracting its positional coordinate at a given time frame from the one at a time frame right after it and dividing the result by the time interval between the frames. This procedure is performed for the raw trajectories $\mathbf{x}(t)$ and the smoothed-out trajectories $\mathbf{\bar{x}}(t)$. The latter trajectories are obtained by averaging the position of each embryo over a time window of 50 frames, as done in Ref.~\cite{tan2022odd}. Then the fluctuation in velocity $\delta\mathbf{v}(t)$ is the difference between the raw velocity $\mathbf{v}(t)$ and the smoothed-out velocity $\mathbf{\bar{v}}(t)$, i.e.,
\begin{equation}
\mathbf{v}(t) = \frac{\mathbf{x}(t+\Delta t)-\mathbf{x}(t)}{\Delta t}, \qquad \mathbf{\bar{v}}(t) = \frac{\mathbf{\bar{x}}(t+\Delta t)-\mathbf{\bar{x}}(t)}{\Delta t}, \qquad
\delta \mathbf{v}(t) = \mathbf{v}(t)-\mathbf{\bar{v}}(t),
\label{eq:dv_def}
\end{equation}
where $\Delta t$ is the time interval (which is 10 seconds in this case).
\begin{figure}[!htb]
   \includegraphics[width=\columnwidth]{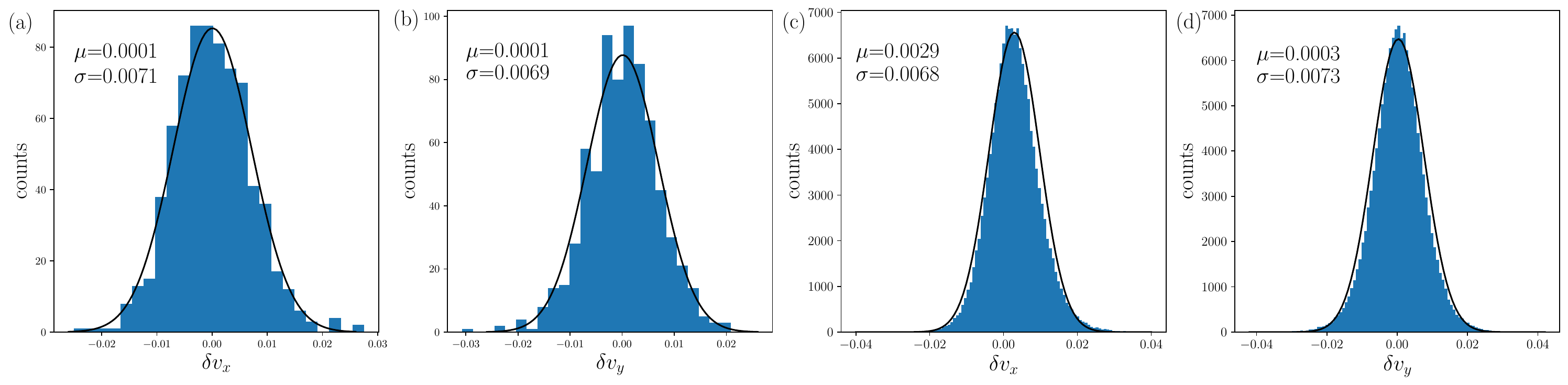}
   \caption{Histogram of the fluctuation in the velocity of starfish embryos in (a) the $x$-direction and (b) the $y$-direction, as obtained from the experimental data of Ref.~\cite{tan2022odd}, for the time frame at which the most number of embryos were detected. The corresponding histograms for all the measurement times are given in (c) for the $x$-direction and in (d) for the $y$-direction. The solid curve is the Gaussian fit. $\mu$ and $\sigma$ are the mean and standard deviation of the fitted Gaussian distribution respectively.
  \label{fig:hist_expt_dv}}
\end{figure}
Figure~\ref{fig:hist_expt_dv} shows that the histogram of $\delta \mathbf{v}(t)$ can be well-fitted to a Gaussian distribution. Since the system is two-dimensional, we show histograms corresponding to two components of the velocity along the $x$ and $y$ directions respectively. Note that we have nondimensionalized the parameters of the experimental data, in the same way as that of SM $\S$\ref{sec:starfish_model}, to obtain $\delta \mathbf{v}(t)$. The number of detected embryos in each time frame differs, as the experimental embryo tracking apparatus could lose the target. Therefore, we used the time frame at which the number of embryos detected is the largest, which is 726, for Fig.~\ref{fig:hist_expt_dv}(a) and (b). As the profile of the fitted Gaussian distribution is very similar for each time frame, we also show the histogram for $\delta \mathbf{v}$ for all the time frames together in Fig.~\ref{fig:hist_expt_dv}(c) and (d) for better statistics.

\begin{figure}[!htb]
   \includegraphics[width=\columnwidth]{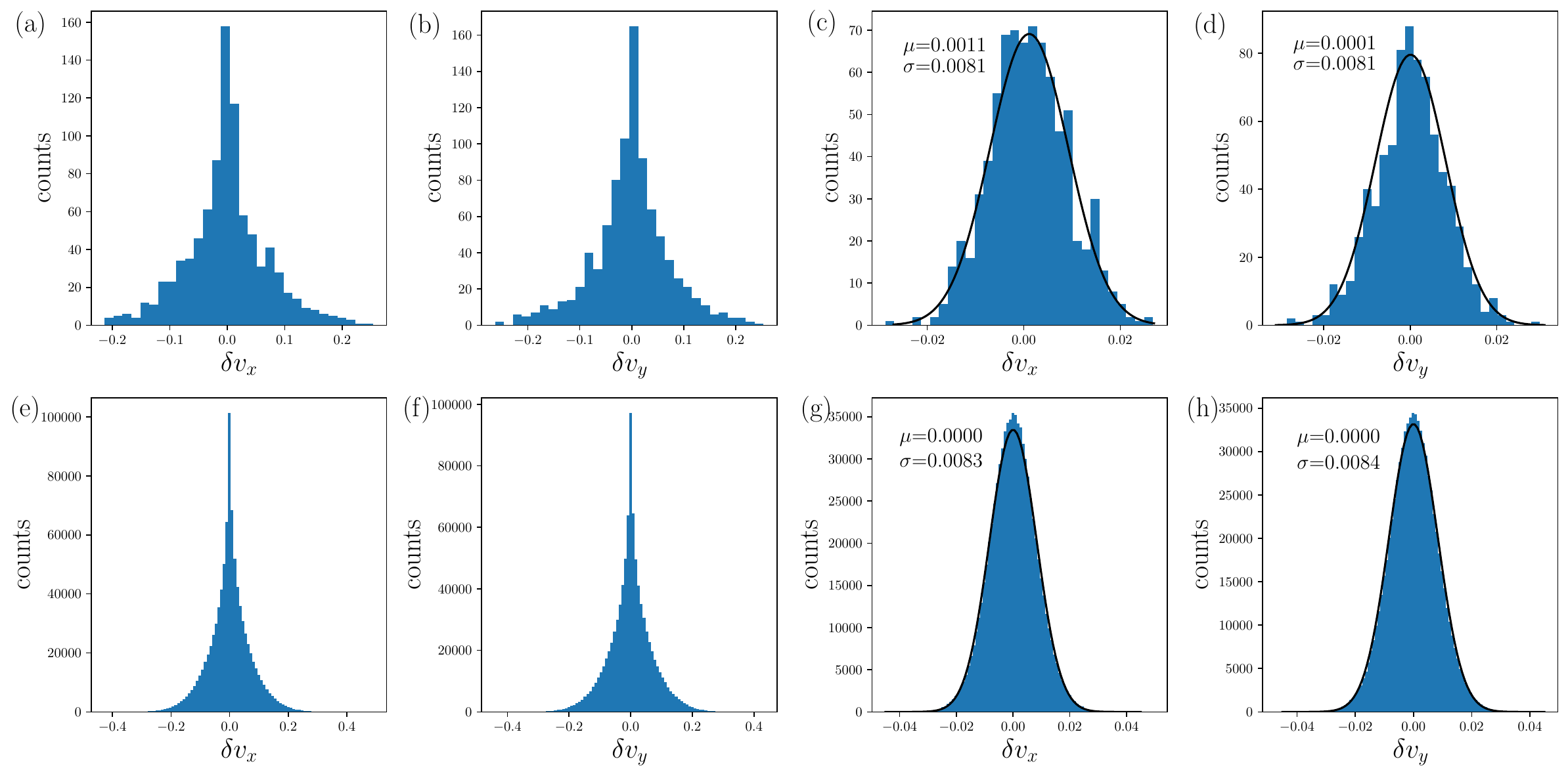}
   \caption{Histograms of the fluctuations in the velocity of starfish embryos obtained from model simulation (with the same parameter values used in the main text), including (a) instantaneous fluctuation in $v_{x}$, (b) instantaneous fluctuation in $v_{y}$, (c) averaged fluctuation in $v_{x}$, and (d) averaged fluctuation in $v_{y}$ at $t=70$ with $\Delta t=0.1$ (i.e., at the $(7\times10^{4})$th time step with $dt=0.001$ where the measurement was done per 100 time steps). (e)-(h) are the results for all the simulation times. The solid curves in (c), (d), (g) and (h) are the Gaussian fit, and $\mu$ and $\sigma$ are the mean and standard deviation of the fitted distribution. 
  \label{fig:hist_starfish_dv}}
\end{figure}

In order to replicate the experiment in our simulation, we revisit how noise is incorporated in our starfish embryo model (Eqs.~(2) and (3) of the main text) where we have introduced a self-propulsion term $v_{0}(t)\mathbf{p}_{i}(t)$. Here $v_{0}(t)$ represents the self-propulsion strength and $\mathbf{p}_{i}(t)=\big(\cos \theta_{i}(t),\sin \theta_{i}(t)\big)$ is the polarization or orientation vector of the $i^{th}$ embryo. Noise enters the system through $v_{0}(t)=\bar{v}_{0}+\xi_{v_{0}}(t)$ where $\xi_{v_{0}}(t)$ is assumed to be a Gaussian white noise with $\langle \xi_{v_{0}}(t)\xi_{v_{0}}(t') \rangle = v_{\sigma}^{2}\delta (t-t')$. Thus the noise term in our model is $\delta \mathbf{v}(t)=\xi_{v_{0}}(t)\mathbf{p}_{i}(t)$. It is nontrivial to separate $\xi_{v_{0}}(t)$ and $\mathbf{p}_{i}(t)$, and it is more convenient to measure and analyze the statistics of $\xi_{v_{0}}(t)\mathbf{p}_{i}(t)$ as a whole. Because of $\mathbf{p}_{i}(t)$, the distribution of the fluctuation in velocity at a given time instant is in fact not expected to be Gaussian. 
Figures~\ref{fig:hist_starfish_dv} shows the histograms of the velocity fluctuation $\delta\mathbf{v}(t)$ in the starfish embryo model, where we have used the same parameters as those in the main text (i.e., longitudinal attractive force $\tilde{F}_{st}=53.7$, longitudinal repulsive force $\tilde{f}_{rep}=785.1$, and transverse force $f_{0}=1.0$ with self-spinning frequency $\bar{\omega}_{0}=1.0$ for $30\times30$ agents in a triangular lattice with periodic boundary conditions). The simulation was run for $10^{5}$ time steps with $dt=0.001$, and the measurement was done every 100 time steps, corresponding to $\Delta t=0.1$ with $t_{f}=100$. 
(On the other hand, the simulation for the Hookean toy model presented in the End Matter was run for $4\times10^{4}$ time steps with $dt = 0.01$ and with measurement done every 40 time steps, so that $\Delta t=0.4$. The one for the starfish embryo model presented in the main text was run for $10^{4}$ time steps with $dt = 0.01$ and recorded every 10 time steps, so that $\Delta t=0.1$.)
Figures~\ref{fig:hist_starfish_dv}(a) and (b) confirm that the distribution of $\delta\mathbf{v}(t)$ at a given time $t$ is not Gaussian. However, while the instantaneous velocity at a given moment can be obtained in simulations through the equations of motion, the velocity calculated from the experimental data is not instantaneous. Since it is calculated by the difference between positions separated by a finite time interval, as shown in Eq.~(\ref{eq:dv_def}), it is related to the average of instantaneous velocities. Therefore, to compare with the experimental data, we need to calculate the averaged velocity over a time interval based on the positional trajectories obtained from the simulation, just as how we analyzed the experimental data. In our simulation, the positions of the agents were recorded every 100 time steps, making the measurement time interval $\Delta t=0.1$.
Then the $\delta\mathbf{v}(t)$ value calculated yields a Gaussian distribution, as shown in Fig.~\ref{fig:hist_starfish_dv}(c) and (d), which agrees with the experimental data. Since the distribution of $\delta\mathbf{v}$ for every time step is similar, we also show the histogram results for all the simulation times (Fig.~\ref{fig:hist_starfish_dv}(e)-(h)) for better statistics.

To be specific, the distribution of embryo velocity fluctuation in Fig.~\ref{fig:hist_starfish_dv}(a) and (b) is different from that in Fig.~\ref{fig:hist_starfish_dv}(c) and (d) because the former is the distribution of instantaneous velocity fluctuation while the latter is the distribution of the average of the velocity fluctuation within the time interval between two consecutive measurements of positions. In other words,
\begin{equation}
\delta\mathbf{v}_{measured}(t) = \frac{1}{\Delta t}\int_{t}^{t+\Delta t} dt' \delta\mathbf{v}_{ins}(t'),
\label{eq:dv_measured}
\end{equation}
where $\delta\mathbf{v}_{ins}(t)$ is the instantaneous velocity fluctuation of an embryo, $\delta\mathbf{v}_{measured}(t)$ is the measured velocity fluctuation calculated from the embryo trajectory data, and $\Delta t$ is the time interval between two measurements. Each $\delta\mathbf{v}_{ins}(t)=\xi_{v_{0}}(t)\mathbf{p}_{i}(t)$, and the distribution of $\delta\mathbf{v}_{measured}$ is essentially the distribution of the sample mean of $\delta\mathbf{v}_{ins}$. According to the Central Limit Theorem (CLT), the distribution of the sample mean of independent and identically distributed (i.i.d.) random variables follows the Gaussian distribution. To verify whether the CLT applies to $\delta\mathbf{v}_{measured}$, we need to check whether the $\delta\mathbf{v}_{ins}$'s are i.i.d. This can be examined from the autocorrelation function of $\delta\mathbf{v}_{ins}(t)$,
\begin{equation}
\begin{split}
\langle\delta\mathbf{v}_{ins}(t)\cdot\delta\mathbf{v}_{ins}(t')\rangle &= \langle
\xi_{v_{0}}(t)\xi_{v_{0}}(t')\big[ \cos (\theta_{i}(t))\cos(\theta_{i}(t')) + \sin(\theta_{i}(t))\sin(\theta_{i}(t')) \big]\rangle \\
&= \langle
\xi_{v_{0}}(t)\xi_{v_{0}}(t')\rangle\langle \cos (\theta_{i}(t))\cos(\theta_{i}(t')) + \sin(\theta_{i}(t))\sin(\theta_{i}(t')) \rangle \\
&= v_{\sigma}^{2}\delta(t-t')\langle \cos (\theta_{i}(t))\cos(\theta_{i}(t')) + \sin(\theta_{i}(t))\sin(\theta_{i}(t')) \rangle,
\end{split}
\label{eq:fluc_v_autocorr}
\end{equation}
which is equal to 0 for $t\neq t'$. Therefore, each instance of $\delta\mathbf{v}_{ins}(t)$ is independent of the others, and the sum of these random variables, which leads to $\delta\mathbf{v}_{measured}(t)$, is expected to be Gaussian-distributed.

\begin{figure}[!htb]
   \includegraphics[width=\columnwidth]{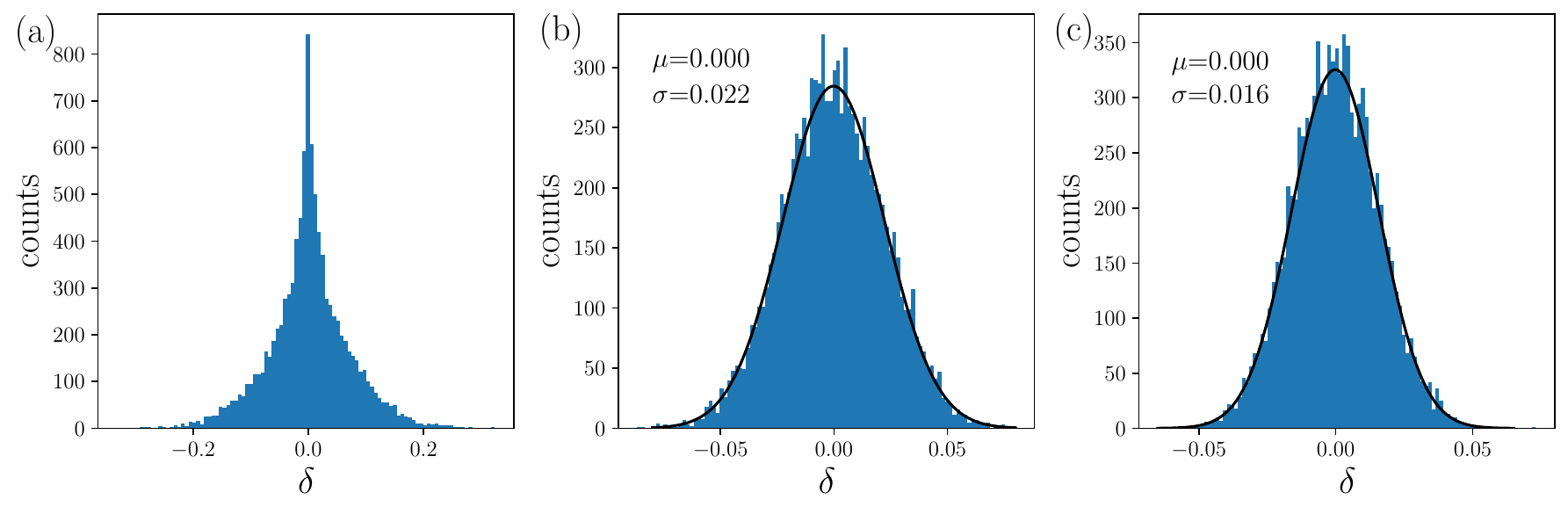}
   \caption{Histograms of the randomly generated variable $\delta = \xi p$, where $\xi$ is a Gaussian noise and $p = \cos (\omega n + \theta_{0})$ with $\omega=1.0$, $\theta_{0}$ being a uniform random variable, and time step $n$ increased monotonically. These include (a) histogram of the instantaneous random variables $\delta(n)$, (b) histogram of the random variables $\delta$ averaged over the time window of 10 time steps, and (c) histogram of the random variables $\delta$ averaged over the time window of 20 time steps. The solid curves in (b) and (c) are the Gaussian fit, and $\mu$ and $\sigma$ are the mean and standard deviation of the fitted distribution.
  \label{fig:hist_rv_test}}
\end{figure}

We verify this argument numerically by generating random variables representing the velocity fluctuation introduced in the starfish embryo model. At each time step $n$, $10^{4}$ random variables of the form $\delta(n)=\xi(n)\cos (\omega n + \theta_{0})$ are generated, with frequency $\omega=1.0$. These $10^{4}$ random variables have explicit dependence on the time step $n$ in the cosine term and implicit dependence on $n$ due to $\xi(n)$, which is a Gaussian noise with mean 0 and standard deviation 0.1 generated at each time step. The constant $\theta_{0}$ is drawn from a uniform distribution from 0 to $2\pi$ at $n=0$. First, these random variables $\delta(n)$ representing the fluctuation in velocity of $10^{4}$ embryos are generated for one time step. The resulting distribution of the random variables is shown in Fig.~\ref{fig:hist_rv_test}(a), whose shape is consistent with that of Fig.~\ref{fig:hist_starfish_dv}(a) and (b) as well as (e) and (f). We then repeat the process for 10 time steps and collect the average of each $\delta (n)$ (so we collect $10^{4}$ averaged $\delta(n)$). We also repeat this process for the time interval of 20 time steps. The resulting distribution of the averaged $\delta(n)$ is shown in Fig.~\ref{fig:hist_rv_test}(b) and (c). Unlike the result of instantaneous $\delta(n)$, the distribution of the averaged $\delta(n)$ is Gaussian. This agrees with the experimental observation shown in Fig.~\ref{fig:hist_expt_dv} and the model simulation in Fig.~\ref{fig:hist_starfish_dv}. Thus this test confirms that although the instantaneous velocity fluctuation of embryos is not expected to be Gaussian-distributed, the distribution of the measured velocity fluctuation is Gaussian, as the measured velocity is the average of instantaneous velocities over a finite time interval.

\section{Emulation of the experiment}
\label{sec:expt_emul}

In this section, we verify that the experimental system of Ref.~\cite{tan2022odd} is not in the parameter regime that yields odd elastic waves either with the absence or in the presence of noise, by emulating the experiment through simulations. As before, we use the self-propelling starfish embryo model given in Eqs. (2) and (3) of the main text. In the study of our theory, we have used stronger transverse force ($f_{0}=1.0$ and $\bar{\omega}_{0}=1.0$) than the one in Ref.~\cite{tan2022odd} ($f_{0}=0.06$ and $\omega_{0}/2\pi = 0.72$), so that the system is in the parameter regime where the elastic waves are expected to emerge. In the demonstration of this section, we use the exact values obtained from Ref.~\cite{tan2022odd}. It is difficult to extract the components of the noise term in the model from the experimental data. However, in the above SM $\S$\ref{sec:noise_stat_expt} we have demonstrated that the measured noise follows the Gaussian distribution although the noise term in the model for instantaneous velocity fluctuations is not Gaussian-distributed, as the measured velocity fluctuation is the average of instantaneous fluctuations. From the standard deviation of the experimentally measured velocity, we deduce the value of the strength of the instantaneous velocity fluctuation $v_{\sigma}$ to put in our simulation to emulate the experiment.

From Eqs.~(\ref{eq:dv_measured}) and (\ref{eq:fluc_v_autocorr}), the autocorrelation function of the measured velocity fluctuation in the $x$ direction is expressed as 
\begin{equation}
\langle \delta v_{x}^{m}(t)\delta v_{x}^{m}(t') \rangle = \frac{1}{\Delta t^{2}}\int_{t}^{t+\Delta t}dt''\int_{t'}^{t'+\Delta t}dt''' v_{\sigma}^{2}\delta (t''-t''')\langle \cos (\theta_{i}(t''))\cos(\theta_{i}(t''')) \rangle,
\end{equation}
where $\delta v_{x}^{m}$ is the measured velocity fluctuation in the $x$ direction. We calculate the $x$ and $y$ components of the fluctuation separately because $\delta v_{x}$ and $\delta v_{y}$ are measured separately. We are interested in the case of $t=t'$, for which 
\begin{equation}
\langle \delta v_{x}^{m}(t)^{2} \rangle = \frac{1}{\Delta t^{2}}\int_{t}^{t+\Delta t}dt'' v_{\sigma}^{2}\langle \cos^2(\theta_{i}(t'')) \rangle = \frac{v_{\sigma}^{2}}{2\Delta t}.
\end{equation}
Similarly, we get $\langle \delta v_{y}^{m}(t)^{2} \rangle = v_{\sigma}^{2}/2\Delta t$. The standard deviation of the measured velocity fluctuation $\sigma$ is then
\begin{equation}
\sigma = \sqrt{\langle \delta v_{x}^{m}(t)^{2} \rangle} = \sqrt{\langle \delta v_{y}^{m}(t)^{2} \rangle} = \frac{v_{\sigma}}{\sqrt{2\Delta t}},
\label{eq:extract_expt_v_sigma}
\end{equation}
such that $v_{\sigma} = \sigma\sqrt{2\Delta t}$. In the experiment, $\Delta t = 10$ s, so we use the dimensionless $\Delta t = 10$ following the non-dimensionalization convention used in SM $\S$\ref{sec:starfish_model}, and we take $\sigma\sim 0.007$ from Fig.~\ref{fig:hist_expt_dv}, giving $v_{\sigma}\sim 0.03$ for the instantaneous velocity fluctuation in the experiment. While time in the experiment is continuous, time steps in the simulation are discrete. Instead of integration as in Eq.~(\ref{eq:dv_measured}), what we get from the simulation is the arithmetic mean of instantaneous velocity fluctuations. The number of velocity fluctuations being averaged is the number of time steps $n_{interval}$ within the time interval $\Delta t$. Therefore, the measured velocity fluctuation $\sigma$ from the simulation is predicted to be
\begin{equation}
\sigma = \frac{v_{\sigma}}{\sqrt{2n_{interval}}}.
\label{eq:sig_n_interval}
\end{equation}
The relation between $\sigma$ and $n_{interval}$ in Eq.~(\ref{eq:sig_n_interval}) is verified in Fig.~\ref{fig:hist_rv_test} for $v_{\sigma}=0.1$. When $n_{interval}=10$, $\sigma=0.1/\sqrt{2\times10}\sim0.0223$ is expected from Eq.~(\ref{eq:sig_n_interval}), and Fig.~\ref{fig:hist_rv_test}(b) gives $\sigma\sim 0.022$. When $n_{interval}=20$, $\sigma=0.1/\sqrt{2\times20}\sim0.0158$ is expected, and Fig.~\ref{fig:hist_rv_test}(c) shows $\sigma\sim 0.016$.

\begin{figure}[!htb]
   \includegraphics[width=\columnwidth]{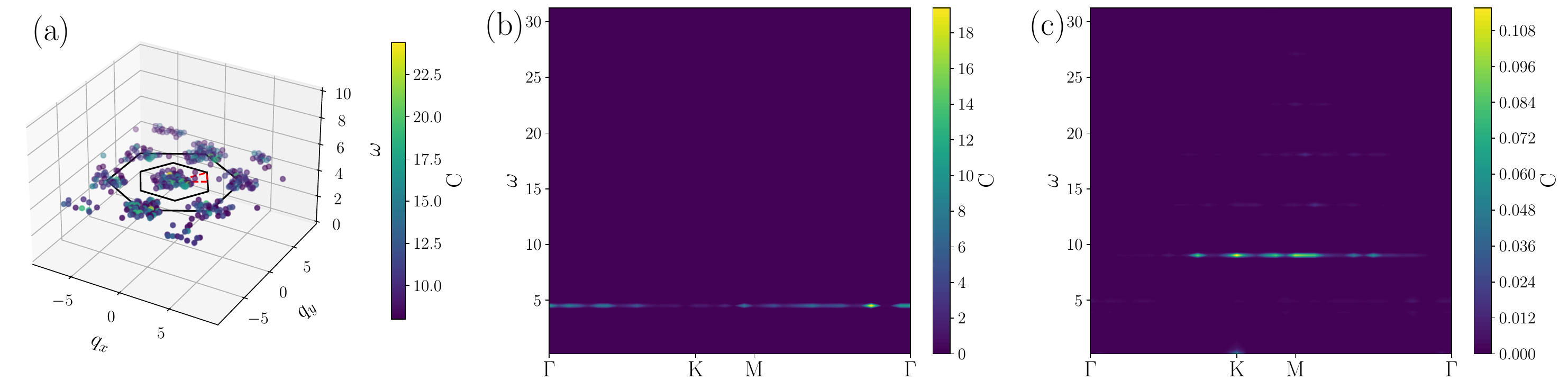}
   \caption{Dispersion results obtained from the model simulation for one realization, with the use of the same parameters as Ref.~\cite{tan2022odd}, i.e., $\tilde{F}_{st}=53.7$, $\tilde{f}_{rep}=785.1$, $f_{0}=0.06$, and $\bar{\omega}_{0}/2\pi=0.72$ in the absence of noise. (a) 3D dispersion where data points exceeding a threshold $C>8$ are shown. The outer hexagon represents the reciprocal lattice, and the inner hexagon represents the first Brillouin zone (BZ). The red dashed line inside the first BZ is the path along which the 2D dispersion results presented in (b) and (c) are obtained. The dispersion result after cutting off the signal of self-circling is shown in (c).
  \label{fig:disp_f006_w072twopi_noNoise}}
\end{figure}

\begin{figure}[!htb]
   \includegraphics[width=\columnwidth]{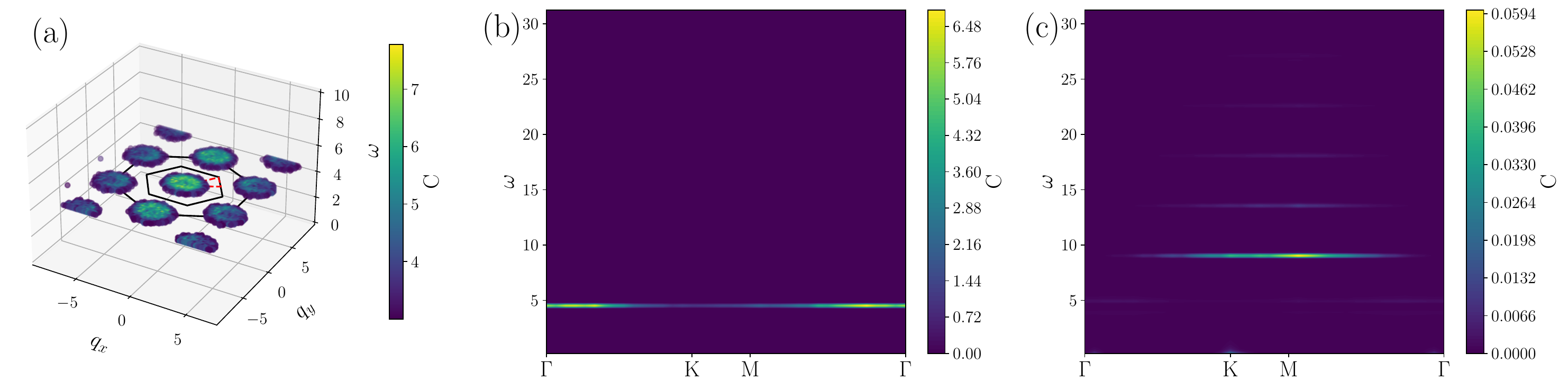}
   \caption{Dispersion results obtained from the model simulation averaged over 100 realizations, with the use of the same parameters as Ref.~\cite{tan2022odd}, i.e., $\tilde{F}_{st}=53.7$, $\tilde{f}_{rep}=785.1$, $f_{0}=0.06$, and $\bar{\omega}_{0}/2\pi=0.72$ in the absence of noise. (a) 3D dispersion where data points exceeding a threshold $C>3$ are shown. The outer hexagon represents the reciprocal lattice, and the inner hexagon represents the first Brillouin zone (BZ). The red dashed line inside the first BZ is the path along which the 2D dispersion results presented in (b) and (c) are obtained. The dispersion result after cutting off the signal of self-circling is shown in (c).
  \label{fig:disp_f006_w072twopi_noNoise_100runs}}
\end{figure}

We emulated the experiment in several versions. First, we use the exact values of the parameters presented in Ref.~\cite{tan2022odd}, which are $\tilde{F}_{st}=53.7$ for the longitudinal attractive force, $\tilde{f}_{rep}=785.1$ for the longitudinal repulsive force, $f_{0}=0.06$ for the transverse force, and $\bar{\omega}_{0}= 2\pi \times 0.72$ for the self-spinning frequency. Note that we have non-dimensionalized the equations and the values presented here are dimensionless (SM $\S$\ref{sec:starfish_model}). As before, totally 900 agents are located at a $30\times 30$ triangular lattice with periodic boundary conditions. We use the mean self-circling velocity $\bar{v}_{0}=0.01$ as in our previous simulations to study our theory since the amplitude of self-circling itself does not affect the wave behavior. To have a similar statistics as the experiment, we run the simulation for $4000$ time steps with $dt = 0.01$ and measure the position of agents every 10 time steps so that $n_{interval}=10$, $\Delta t=0.1$ and the number of measured time steps is 400, which is the same as the number of time frames used for the experimental data analysis. (The total number of experimental time frames was 800, while we used the later half of the data.) To be as similar to the experiment as possible, we first analyze the data from just one realization of the simulation (Fig.~\ref{fig:disp_f006_w072twopi_noNoise}) and then analyze the data from the average of 100 realizations (Fig.~\ref{fig:disp_f006_w072twopi_noNoise_100runs}) for better statistics. In this version, we do not include any noise. As shown in Fig.~\ref{fig:disp_f006_w072twopi_noNoise}(a) and Fig.~\ref{fig:disp_f006_w072twopi_noNoise_100runs}(a), the signals are localized at $\mathbf{q}=0$ and the reciprocal lattice vertices, all at $\omega = \bar{\omega}_{0}$. The reference lines representing the reciprocal lattice and the first Brillouin zone (BZ) have been drawn at $\omega = \bar{\omega}_{0}$ to indicate the location of the signals. While we show strong enough signals for $C>8$ in Fig.~\ref{fig:disp_f006_w072twopi_noNoise}(a) and for $C>3$ in Fig.~\ref{fig:disp_f006_w072twopi_noNoise_100runs}(a), the weaker signals below the threshold also exist only on the $(q_{x},q_{y})$ planes located at $\omega=n\bar{\omega}_{0}$ with $n=1,2,...$, as demonstrated in Fig.~\ref{fig:disp_f006_w072twopi_noNoise}(b) and (c) as well as Fig.~\ref{fig:disp_f006_w072twopi_noNoise_100runs}(b) and (c) which show the dispersion results along the path in the first BZ (red dashed lines in Fig.~\ref{fig:disp_f006_w072twopi_noNoise}(a) and Fig.~\ref{fig:disp_f006_w072twopi_noNoise_100runs}(a)). This observation agrees with our prediction given in SM $\S$\ref{sec:flat}, confirming that the signal is due to self-circling motion, instead of a wave. Since the signal at $\omega=\bar{\omega}_{0}$ is the strongest, we see the corresponding single horizontal line in Fig.~\ref{fig:disp_f006_w072twopi_noNoise}(b) and Fig.~\ref{fig:disp_f006_w072twopi_noNoise_100runs}(b). To verify whether the wave signal has been overwhelmed by the self-circling signal, as in SM $\S$\ref{sec:reveal_disp} we removed the self-circling signal, leading to Fig.~\ref{fig:disp_f006_w072twopi_noNoise}(c) and Fig.~\ref{fig:disp_f006_w072twopi_noNoise_100runs}(c). Unlike Fig.~\ref{fig:starfish_cutcomp} where the wave signal is present, Fig.~\ref{fig:disp_f006_w072twopi_noNoise}(c) and Fig.~\ref{fig:disp_f006_w072twopi_noNoise_100runs}(c) do not reveal any dispersion curve, other than showing the self-circling signals at $\omega=n\bar{\omega}_{0}$ for $n\geq 2$. 

The second version of the emulation uses the same parameter values as those depicted above but includes the noise in the self-propulsion term of our theory. Here we use $v_{\sigma}=0.03$ deduced from the experimental data. First, we check whether the distribution of noise in this system still follows the behavior discussed in SM $\S$\ref{sec:noise_stat_expt}.
\begin{figure}[!htb]
   \includegraphics[width=\columnwidth]{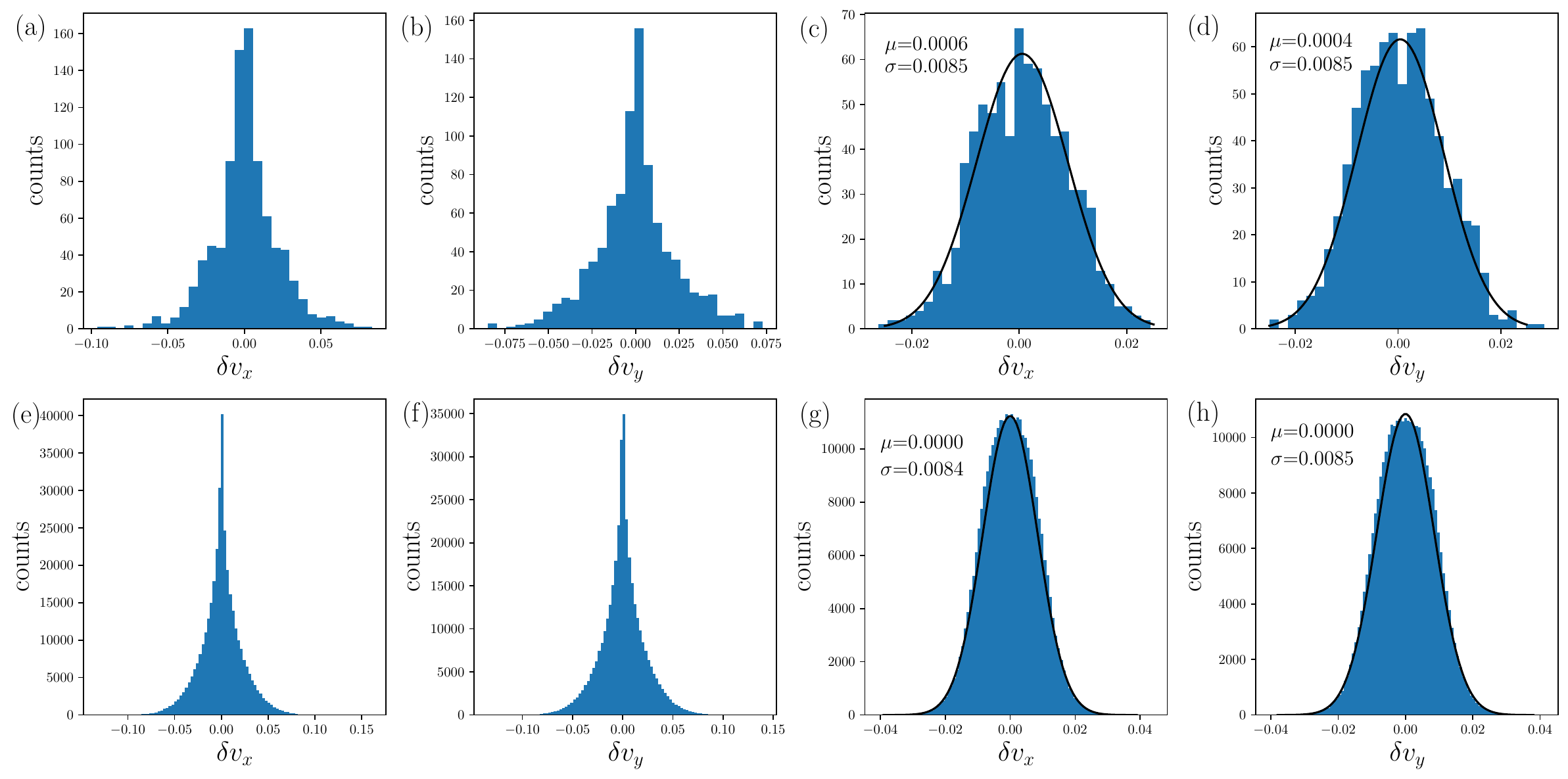}
   \caption{Histograms of the velocity fluctuation from the model simulation for one realization, with the use of the same parameters as Ref.~\cite{tan2022odd}, i.e., $\tilde{F}_{st}=53.7$, $\tilde{f}_{rep}=785.1$, $f_{0}=0.06$, and $\bar{\omega}_{0}/2\pi=0.72$, and the noisy self-propulsion term with $\bar{v}_{0}=0.01$ and $v_{\sigma}=0.03$ as deduced from the experimental data. (a)--(d) are obtained at $t=12$ with $\Delta t=0.1$ (i.e., at the $1200$th time step with $dt=0.01$ where the measurement was done per 10 time steps). (e)--(h) are obtained from the whole simulation time. (a) and (e) are for the instantaneous velocity fluctuation in the $x$ direction, and (b) and (f) are for the instantaneous velocity fluctuation in the $y$ direction. Panels (c), (d), (g) and (h) are for the averaged velocity fluctuations obtained from the positional coordinates in the $x$ and $y$ directions. The solid curves in (c), (d), (g) and (h) are the Gaussian fit, and $\mu$ and $\sigma$ are the mean and standard deviation of the fitted distribution.
  \label{fig:histdv_f006_w072twopi_noise}}
\end{figure}
Figure~\ref{fig:histdv_f006_w072twopi_noise} shows the histogram of each component of both the instantaneous velocity fluctuations and the measured (or averaged) velocity fluctuations. We show the result from one randomly chosen time point (Fig.~\ref{fig:histdv_f006_w072twopi_noise}(a)-(d)) and from the whole simulation time (Fig.~\ref{fig:histdv_f006_w072twopi_noise}(e)-(h)) for better statistics. Since the time interval between two consecutive measurements is $n_{interval}=10$ time steps, we expect the measured standard deviation to be $\sigma\sim0.007$ according to Eq.~(\ref{eq:sig_n_interval}), but the actual measured standard deviation turns out to be $\sigma\sim0.0085$ (Fig.~\ref{fig:histdv_f006_w072twopi_noise}(c),(d),(g) and (h)). This disagreement can be attributed to the fact that the relation between $\sigma$ and $\Delta t$ (Eq.~(\ref{eq:extract_expt_v_sigma})) as well as $n_{interval}$ (Eq.~(\ref{eq:sig_n_interval})) has been derived under the assumption that the velocity fluctuation $\delta\mathbf{v}$ does not have contribution from interparticle interaction terms. However, in general, the contributions from the interaction terms in $\mathbf{v}(t)$ and $\bar{\mathbf{v}}(t)$ in Eq.~(\ref{eq:dv_def}) do not completely cancel with each other. Therefore, $\delta\mathbf{v}(t)$ ends up including the remaining contribution from interparticle interactions. However, this contribution and hence the disagreement between the prediction and the measurement is small enough that the measured $\sigma\sim0.0085$ is close to the predicted $\sigma\sim0.007$, and we proceed with this result without further fine-tuning of the model and the analysis method.

\begin{figure}[!htb]
   \includegraphics[width=\columnwidth]{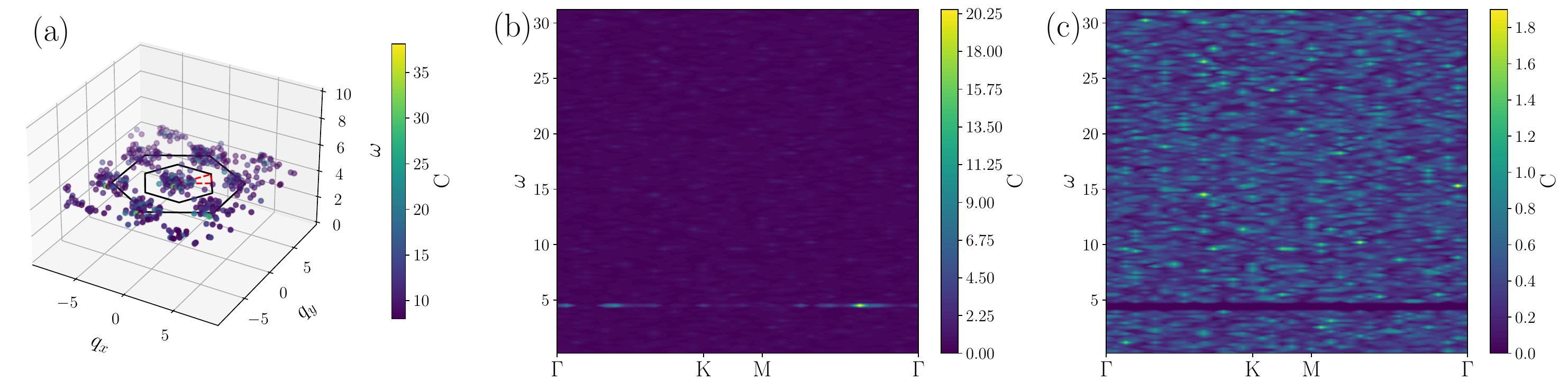}
   \caption{Dispersion results from one realization of the model simulation, with the use of same parameters as Ref.~\cite{tan2022odd}, i.e., $\tilde{F}_{st}=53.7$, $\tilde{f}_{rep}=785.1$, $f_{0}=0.06$, and $\bar{\omega}_{0}/2\pi=0.72$, and the noisy self-propulsion term with $\bar{v}_{0}=0.01$ and $v_{\sigma}=0.03$ as deduced from the experimental data. (a) 3D dispersion where data points exceeding a threshold $C>8$ are shown. The outer hexagon represents the reciprocal lattice, and the inner hexagon represents the first BZ. The red dashed line inside the first BZ is the path along which the 2D dispersion results presented in (b) and (c) are obtained. The dispersion result after cutting off the signal of self-circling is shown in (c).
  \label{fig:disp_f006_w072twopi_noise_1run}}
\end{figure}

\begin{figure}[!htb]
   \includegraphics[width=\columnwidth]{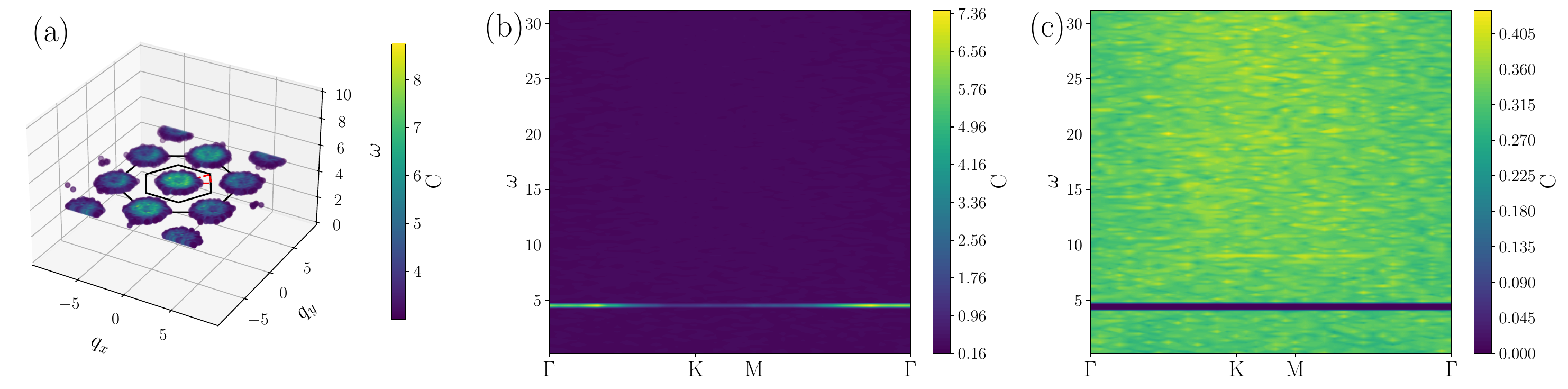}
   \caption{Dispersion results from the average of 100 realizations of the model simulation, with the use of same parameters as Ref.~\cite{tan2022odd}, i.e., $\tilde{F}_{st}=53.7$, $\tilde{f}_{rep}=785.1$, $f_{0}=0.06$, and $\bar{\omega}_{0}/2\pi=0.72$, and the noisy self-propulsion term with $\bar{v}_{0}=0.01$ and $v_{\sigma}=0.03$ as deduced from the experimental data. (a) 3D dispersion where data points exceeding a threshold $C>3$ are shown. The outer hexagon represents the reciprocal lattice, and the inner hexagon represents the first BZ. The red dashed line inside the first BZ is the path along which the 2D dispersion results presented in (b) and (c) are obtained. The dispersion result after cutting off the signal of self-circling is shown in (c).
  \label{fig:disp_f006_w072twopi_noise_100runs}}
\end{figure}

Figure \ref{fig:disp_f006_w072twopi_noise_1run} shows the dispersion result for one realization of this simulation. We analyze the result for one realization to mimic the experiment, and also analyze the result from the average of 100 realizations (Fig.~\ref{fig:disp_f006_w072twopi_noise_100runs}) for better statistics. These two figures look very similar to the cases shown in Fig.~\ref{fig:disp_f006_w072twopi_noNoise} and Fig.~\ref{fig:disp_f006_w072twopi_noNoise_100runs}  in the absence of noise. The signals are localized at $\mathbf{q}=0$ and the reciprocal lattice vertices, at $\omega=\bar{\omega}_{0}$ which is the self-circling frequency. Removal of the self-circling signal does not reveal the wave signal, which confirms that no wave behavior is detected. Because of the presence of noise, the result after removing the self-circling signal is more noisy in Fig.~\ref{fig:disp_f006_w072twopi_noise_1run}(c) and Fig.~\ref{fig:disp_f006_w072twopi_noise_100runs}(c) than in Fig.~\ref{fig:disp_f006_w072twopi_noNoise}(c) and Fig.~\ref{fig:disp_f006_w072twopi_noNoise_100runs}(c). The noisy result in Fig.~\ref{fig:disp_f006_w072twopi_noise_1run}(c) is comparable to the experimental result shown in Fig.~\ref{fig:expt_cutcomp} after cutting off the self-circling signal.

\begin{figure}[!htb]
   \includegraphics[width=\columnwidth]{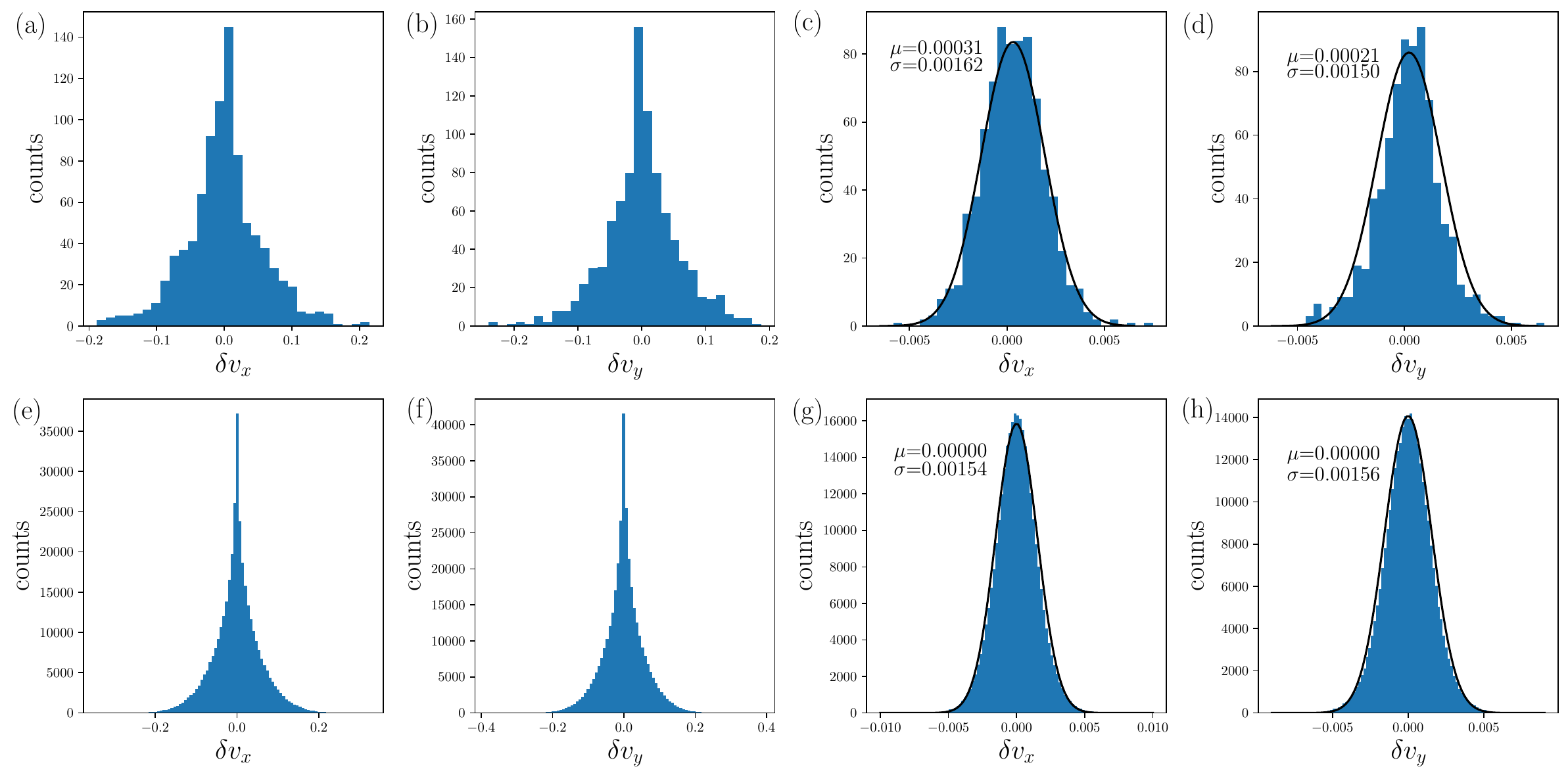}
   \caption{Histograms of the velocity fluctuation from the model simulation for one realization, with the use of $\tilde{F}_{st}=53.7$, $\tilde{f}_{rep}=785.1$, $f_{0}=0.06$, and $\bar{\omega}_{0}=0.03$, and the noisy self-propulsion term with $\bar{v}_{0}=0.06$ and $v_{\sigma}=0.08$ to match the experimental observation. (a)-(d) are obtained at $t=300$ with $\Delta t=2$ ($60000^{th}$ time step with $dt=0.01$ where measurement was done every 200 time steps). (e)-(h) are obtained from the whole simulation time. (a) and (e) are the instantaneous velocity fluctuation in the $x$ direction, (b) and (f) are instantaneous velocity fluctuation in the $y$ direction. Panels (c), (d), (g) and (h) are for the averaged velocity fluctuations obtained from the positional coordinates in the $x$ and $y$ directions. The solid curves in (g) and (h) are the Gaussian fit, and $\mu$ and $\sigma$ are the mean and standard deviation of the fitted distribution.
  \label{fig:histdv_f006_w003_noise}}
\end{figure}

\begin{figure}[!htb]
   \includegraphics[width=\columnwidth]{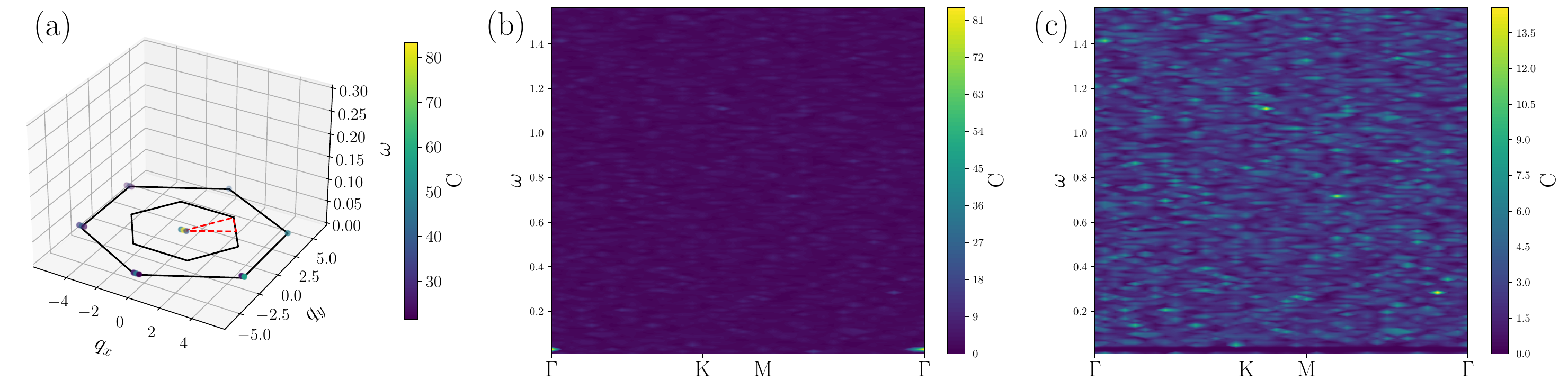}
   \caption{Dispersion results obtained from one realization of the model simulation, with the use of $\tilde{F}_{st}=53.7$, $\tilde{f}_{rep}=785.1$, $f_{0}=0.06$, and $\bar{\omega}_{0}=0.03$, and the noisy self-propulsion term with $\bar{v}_{0}=0.06$ and $v_{\sigma}=0.08$ to match to the experimental data. (a) 3D dispersion where data points exceeding a threshold $C>20$ are shown. The outer hexagon represents the reciprocal lattice, and the inner hexagon represents the first BZ. The red dashed line inside the first BZ is the path along which the 2D dispersion results presented in (b) and (c) are obtained. The dispersion result after cutting off the signal of self-circling is shown in (c).
  \label{fig:disp_f006_w003_noise_1run}}
\end{figure}

The third version of the emulation uses a smaller value of the self-spinning frequency, $\bar{\omega}_{0}=0.03$. The value used in the simulations presented in Ref.~\cite{tan2022odd}, which is $\omega_{0}/2\pi=0.72$, is the self-spinning frequency of an isolated single embryo. The spinning of embryos becomes slower inside a cluster\cite{tan2022odd}, and thus Ref.~\cite{tan2022odd} used the torque balance in their simulation to reflect this. As the constant self-spinning frequency is used in our modeling here (for reasons elaborated in SM $\S$\ref{sec:wupdate}), we need to directly use the value of the self-spinning frequency inside a cluster in our simulations. This value has been identified in our analysis of dispersion in the experimental data (Fig. 3 in the main text) as well as the mode chirality analysis in Ref.~\cite{tan2022odd}. Since the expected frequency signal is weaker with much smaller $\bar{\omega}_{0}$, we use $n_{interval}=200$ with $dt=0.01$ so that $\Delta t=2$ which leads to finer $\omega$ resolution than the other two versions of the emulation described above. We keep the number of observed time frames to be 400 by running the simulation for 80000 time steps. 
We should keep $v_{\sigma}$ of the experiment consistent since it is the microscopic parameter directly entering the equations of motion. However, as discussed in the second version of the emulation, the velocity fluctuation $\delta\mathbf{v}$ actually contains the contribution from interparticle interactions, which leads to discrepancy between the predicted $\sigma$ and the observed $\sigma$. In this version of the emulation, instead of directly using the deduced value of $v_{\sigma}$ from Eq.~(\ref{eq:extract_expt_v_sigma}), we calculate $\sigma$ predicted by Eq.~(\ref{eq:sig_n_interval}) given $n_{interval}$ and $v_{\sigma}$ deduced by Eq.~(\ref{eq:extract_expt_v_sigma}), find the values of $v_{\sigma}$ and $\bar{v}_{0}$ that give this value of $\sigma$ in the simulation and use them for the further analyses. The reasoning is that, since Eq.~(\ref{eq:extract_expt_v_sigma}) does not consider nonlinear effects or contribution from interaction terms, $v_{\sigma}$ deduced from this equation is likely not the true value of this microscopic parameter. However, we assume that the relation between the statistics from the continuous time and that from the discrete time still holds. In other words, from Eq.~(\ref{eq:extract_expt_v_sigma}) and Eq.~(\ref{eq:sig_n_interval}),
\begin{equation}
v_{\sigma}=\sigma_{ex}\sqrt{2\Delta t} = \sigma_{s}\sqrt{2n_{interval}}, 
\qquad
\sigma_{s} = \sigma_{ex}\sqrt{\frac{\Delta t}{n_{interval}}}
\end{equation}
where the subscript $ex$ represents ``experiment" and the subscript $s$ represents ``simulation," and we assume that the above relation between $\sigma_{ex}$ and $\sigma_{s}$ holds even if $v_{\sigma}$ deduced this way is not the true value. From the experimental value of $\sigma_{ex}=0.007$ with $\Delta t=10$, we expect $\sigma_{s} \sim 0.00156$ for $n_{interval}=200$.
By adjusting the microscopic noise strength to be $v_{\sigma}=0.08$, we obtained the expected noise, as shown in Fig.~\ref{fig:histdv_f006_w003_noise}. We also adjusted the mean self-circling velocity to be $\bar{v}_{0}=0.06$ so that the self-circling signal is visible in the dispersion result (Fig.~\ref{fig:disp_f006_w003_noise_1run}). The value of $\bar{v}_{0}$ does not affect the noise profile shown in Fig.~\ref{fig:histdv_f006_w003_noise}.

\begin{figure}[!htb]
   \includegraphics[width=\columnwidth]{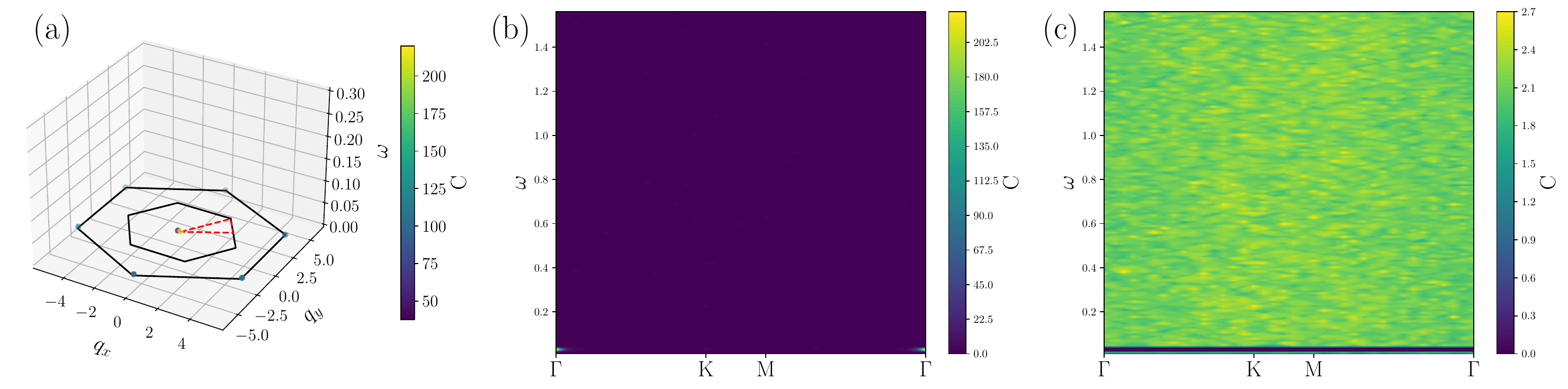}
   \caption{Dispersion results obtained from the average of 100 realizations of the model simulation, with the use of $\tilde{F}_{st}=53.7$, $\tilde{f}_{rep}=785.1$, $f_{0}=0.06$, and $\bar{\omega}_{0}=0.03$, and the noisy self-propulsion term with $\bar{v}_{0}=0.06$ and $v_{\sigma}=0.08$ to match to the experimental data. (a) 3D dispersion where data points exceeding a threshold $C>30$ are shown. The outer hexagon represents the reciprocal lattice, and the inner hexagon represents the first BZ. The red dashed line inside the first BZ is the path along which the 2D dispersion results presented in (b) and (c) are obtained. The dispersion result after cutting off the signal of self-circling is shown in (c).
  \label{fig:disp_f006_w003_noise_100runs}}
\end{figure}

Figure~\ref{fig:disp_f006_w003_noise_1run} shows the dispersion result for one realization of this simulation, and Fig.~\ref{fig:disp_f006_w003_noise_100runs} shows the result averaged over 100 realizations. With much smaller $\bar{\omega}_{0}$, the signals in Fig.~\ref{fig:disp_f006_w003_noise_1run}(a) as well as in Fig.~\ref{fig:disp_f006_w003_noise_100runs}(a) are more localized than those in Figs.~\ref{fig:disp_f006_w072twopi_noNoise}(a), \ref{fig:disp_f006_w072twopi_noNoise_100runs}(a), \ref{fig:disp_f006_w072twopi_noise_1run}(a), and \ref{fig:disp_f006_w072twopi_noise_100runs}(a), occupying smaller areas in the $(q_{x}, q_{y})$ plane, while the observation that the signals are located at $\mathbf{q}=0$ and the vertices of the reciprocal lattice at $\omega=\bar{\omega}_{0}$ remains the same. As the signals are occupying a smaller range on the $(q_{x}, q_{y})$ plane, they are shown as a point signal at the $\Gamma$ point in Fig.~\ref{fig:disp_f006_w003_noise_1run}(b) and Fig.~\ref{fig:disp_f006_w003_noise_100runs}(b) rather than a horizontal line. Similar to Fig.~\ref{fig:disp_f006_w072twopi_noise_1run}(c) and Fig.~\ref{fig:disp_f006_w072twopi_noise_100runs}(c), removing the signal at $\omega=\bar{\omega}_{0}$ does not reveal a dispersion curve in Fig.~\ref{fig:disp_f006_w003_noise_1run}(c) and Fig.~\ref{fig:disp_f006_w003_noise_100runs}(c), indicating the absence of wave behavior. The result demonstrated in Fig.~\ref{fig:disp_f006_w003_noise_1run} is very similar to the experimental result shown in Fig.~\ref{fig:expt_cutcomp}, confirming that the parameter set in the experimental system does not yield persistent odd elastic waves.

\section{Two measurements of elastic moduli}

In Ref.~\cite{tan2022odd} the elastic moduli of the living crystal of starfish embryos were calculated via two methods. The first one was through the microscopic experimental parameters, and the second one was through the local strains identified from topological defects. While the values of the elastic moduli obtained from the former do not satisfy the deterministic elastic wave condition \cite{scheibner2020odd}, those from the latter do. This section explicitly shows this observation. In the main paper, we have derived a new criterion for noise-driven odd elastic waves. Here we also show that the disagreement on whether or not the experimental parameters are in the wave regime remains the same for the new criterion.

The elastic moduli calculated from the experimental parameters and the linearized equation of motion for starfish embryos \cite{tan2022odd} are given by 
\begin{equation}
A \approx 1.9 ~s^{-1}, 
\qquad
K^{o} \approx 0.8 ~s^{-1},
\qquad
B \approx 7.0 ~s^{-1},
\qquad
\mu \approx 3.5 ~s^{-1}.
\label{eq:parameters_microscopic}
\end{equation}
Plugging these values into the deterministic wave condition \cite{scheibner2020odd} gives
\begin{equation}
\bigg(\frac{B}{2}\bigg)^{2}-K^{o}A-(K^{o})^{2}\approx10.09 ~s^{-2} > 0,
\end{equation}
which means that the system is not in the wave regime, as the wave behavior is predicted if the above expression is negative. The quantities listed in Eq.~(\ref{eq:parameters_microscopic}) are dimensional and have been calculated using the self-spinning frequency of an isolated single embryo \cite{tan2022odd}, which is of larger value than that of an embryo in a cluster. If linearizing the non-dimensionalized equations of motion (SM $\S$\ref{sec:starfish_model}) without torque balance, we obtain 
\begin{align}
k_{L} = \tilde{f}_{\text{rep}}\frac{39}{2^{10}\tilde{r}_{0}^{14}}-\tilde{F}_{\text{st}}\frac{\tilde{r}_{0}^{2}-0.5}{4\pi(\tilde{r}_{0}^{2}+1)^{2}},
&&
k_{T} = \frac{2R\omega f_{0}}{\tilde{r}_{0}-2R},
&&
F_{T}^{0} = 2R\omega f_{0}\ln\bigg(\frac{d_{c}}{\tilde{r}_{0}-2R}\bigg),
\end{align}
where $k_{L}$ and $k_{T}$ are effective spring constants in the longitudinal and transverse directions respectively, and $F_{T}^{0}$ is the zeroth order term of the transverse force. In the following we omit torque balance and directly use the self-spinning frequency of an embryo inside a cluster, $\omega=0.03$. As before, $\tilde{F}_{\text{st}}=53.7$, $\tilde{f}_{\text{rep}}=785.1$, $f_{0}=0.06$, and $2R=1$. In our modeling, we have used $\tilde{r}_{0}=1.2$ and $d_{c}=R$ to facilitate the simulations, but here we use the values presented in the experimental work \cite{tan2022odd}, which are $\tilde{r}_{0}=1.1$ and $d_{c}=0.5R$. This leads to $k_{L}\approx7.46$, $k_{T}\approx0.018$, and $F_{T}^{0}\approx0.0016$. Using Eq.~(\ref{eq:elastic_moduli}) (with $F_{L}^{0}=0$), we get
\begin{equation}
A \approx 0.017, 
\qquad
K^{o} \approx 0.0071,
\qquad
B \approx 6.46,
\qquad
\mu \approx 3.23.
\label{eq:parameters_microscopic_v2}
\end{equation}
Note that here the values of $A$ and $K^{o}$ are very small because we use small experimental value of $\omega$ for the case of embryo cluster. With these values, 
\begin{equation}
\bigg(\frac{B}{2}\bigg)^{2}-K^{o}A-(K^{o})^{2}\approx10.42 > 0,
\end{equation}
showing again that the system is not in the wave regime.
On the other hand, the strains determined from topological defects provide an estimate of the elastic moduli as \cite{tan2022odd}
\begin{equation}
A/\mu = 7.7 \pm 0.61, 
\qquad
K^{o}/\mu = 7.1 \pm 0.59,
\label{eq:parameters_defects}
\end{equation}
with $B=2\mu$ given by the linear elasticity theory for a triangular spring network with nearest-neighbor interactions \cite{scheibner2020odd,braverman2021topological}. The corresponding expression for the deterministic wave condition is then
\begin{equation}
\bigg(\frac{B}{2}\bigg)^{2}-K^{o}A-(K^{o})^{2}\approx-104.08 \mu^{2} < 0,
\end{equation}
which means that the system is in the wave regime. 

Substituting Eq.~(\ref{eq:parameters_microscopic}) into the approximate criterion for noise-driven odd elastic waves that is derived in this work, we have
\begin{equation}
(K^{o})^{2}+(K^{o}+A)^{2}+4K^{o}(K^{o}+A)-\mu^{2}-(B+\mu)^{2}\approx -105.93~s^{-2}<0,
\end{equation}
indicating that the system is not in the noise-driven wave regime. Using Eq.~(\ref{eq:parameters_microscopic_v2}) leads to
\begin{equation}
(K^{o})^{2}+(K^{o}+A)^{2}+4K^{o}(K^{o}+A)-\mu^{2}-(B+\mu)^{2}\approx -104.24<0,
\end{equation} 
giving the same conclusion.
On the other hand, using Eq.~(\ref{eq:parameters_defects}) and $B=2\mu$ gives
\begin{equation}
(K^{o})^{2}+(K^{o}+A)^{2}+4K^{o}(K^{o}+A)-\mu^{2}-(B+\mu)^{2}\approx 679.77\mu^{2}>0,
\end{equation}
which means that the system is in the noise-driven wave regime. It is not surprising that the disagreement persists for the new criterion, given that the moduli related to odd elasicity ($A$ and $K^{o}$) are smaller than $B$ and $\mu$ in the first case of measurement (Eq.~(\ref{eq:parameters_microscopic}) or Eq.~(\ref{eq:parameters_microscopic_v2})) while they are larger than $B$ and $\mu$ in the second case (Eq.~(\ref{eq:parameters_defects})).

Those two measurements yield not only different numerical values of elastic moduli but also different conclusions regarding whether the system satisfies the deterministic wave condition and the criterion for the noise-driven waves. Therefore, a method that does not depend on the measured values of the elastic moduli becomes essential to identify the elastic wave behavior, which is described in this work through the dispersion relations determined from current correlation functions.

\medskip
\begin{acknowledgments}
We thank Tzer Han Tan and Nikta Fakhri for providing the experimental data of Ref.~\cite{tan2022odd} for the trajectories of starfish embryos. We gratefully acknowledge valuable discussions with Tzer Han Tan, Alexander Mietke, J{\"o}rn Dunkel, Vincenzo Vitelli and Ken Elder.
Z.-F.H. acknowledges support from the National Science Foundation under Grant No. DMR-2006446.
\end{acknowledgments}

%